\documentclass[fleqn,usenatbib]{mnras}

\usepackage{newtxtext,newtxmath}
\usepackage{comment} %Needs to be deleted in the end
\usepackage[T1]{fontenc}

\DeclareRobustCommand{\VAN}[3]{#2}
\let\VANthebibliography\thebibliography
\def\thebibliography{\DeclareRobustCommand{\VAN}[3]{##3}\VANthebibliography}

\usepackage{graphicx}	% Including figure files
\usepackage{amsmath}	% Advanced maths commands
\usepackage{subcaption} 
\newcommand\hdist{2}
\newcommand\vdist{2}
\newcommand\centpara{8}
\newcommand\x{0.51}
\newcommand\y{0.8}
\newcommand\z{2}

\newcommand\m{1.0}

\newcommand\hdistTF{2}

\title[Optimizing high redshift galaxy surveys for environmental information]{Optimizing high redshift galaxy surveys for environmental information}

\author[Tobias J. Looser et al.]{Tobias J. Looser,$^{1,2,3,}\thanks{E-mail: tjl54@cam.ac.uk}$
Simon J. Lilly,$^{1}$
Larry P. T. Sin,$^{1}$
Bruno M. B. Henriques,$^{1}$
\newauthor
Roberto Maiolino,$^{2,3,4}$
Michele Cirasuolo$^{5}$
\\
$^{1}$Department of Physics, ETH Zurich, Wolfgang-Pauli-Strasse 27, CH-8093 Zurich, Switzerland\\
$^{2}$Cavendish Laboratory, University of Cambridge, 19 J. J. Thomson Ave., Cambridge CB3 0HE, UK\\
$^{3}$Kavli Institute for Cosmology, University of Cambridge, Madingley Road, Cambridge CB3 0HA, UK\\
$^{4}$Department of Physics and Astronomy, University College London, Gower Street, London WC1E 6BT, UK\\
$^{5}$European Southern Observatory, Karl-Schwarzschild-Strasse 2, D-85748 Garching bei Muenchen, Germany\\
}

\date{Accepted 2021 April 13. Received 2021 March 15; in original form 2020 December 03}

\pubyear{2020}

\begin{document}
\label{firstpage}
\pagerange{\pageref{firstpage}--\pageref{lastpage}}
\maketitle

\begin{abstract}
We investigate the performance of group finding algorithms that reconstruct galaxy groups from the positional information of tracer galaxies that are observed in redshift surveys carried out with multiplexed spectrographs.  We use mock light-cones produced by the L-Galaxies semi-analytic model of galaxy evolution in which the underlying reality is known. We particularly focus on the performance at high redshift, and how this is affected by choices of the mass of the tracer galaxies (largely equivalent to their co-moving number density) and the (assumed random) sampling rate of these tracers.  We first however compare two different approaches to group finding as applied at low redshift, and conclude that these are broadly comparable. For simplicity we adopt just one of these, "Friends-of-Friends" (FoF) as the basis for our study at high redshift.
We introduce 12 science metrics that are designed to quantify the performance of the group-finder as relevant for a wide range of science investigations with a group catalogue. These metrics examine the quality of the recovered group catalogue, the median halo masses of different richness structures, the scatter in dark matter halo mass and how successful the group-finder classifies singletons, centrals and satellites. We analyze how these metrics vary with the limiting stellar mass and random sampling rate of the tracer galaxies, allowing quantification of the various trade-offs between different possible survey designs. Finally, we look at the impact of these same design parameters on the relative "costs" in observation time of the survey using as an example the potential MOONRISE survey using the MOONS instrument.
\end{abstract}

\begin{keywords}
galaxies: groups: general -- catalogues -- surveys -- galaxies: high-redshift -- dark matter  -- large-scale structure of Universe
\end{keywords}

%%%%%%%%%%%%%%%%%%%%%%%%%%%%%%%%%%%%%%%%%%%%%%%%%%

%%%%%%%%%%%%%%%%% BODY OF PAPER %%%%%%%%%%%%%%%%%%

\section{Introduction} \label{Introduction}
Large scale galaxy redshift surveys carried out with efficient multi-object spectrographs (MOS) allow us to investigate the evolution of galaxies over cosmic time. While the earliest of these simply established the broad characteristics of the evolving galaxy population \citep[e.g.][]{Colless1990,Lilly1995,Broadhurst1998}
in terms of the luminosity (or mass) functions, in the last two decades more extensive redshift surveys have enabled the study of the role of environment in driving this evolution, even for galaxies in quite typical environments, both at low redshift \citep[e.g.][]{Colless2001} and at much greater look-back times \citep[e.g.][]{Davis2003,LeFevre2005,Gunn2006,Lilly2009}. 
Since the underlying and dominant dark matter structures are largely undetectable directly, these structures must be traced by other methods, such as the presence of extended hot gas (which is impractical on many scales of interest) or the distribution of the galaxies themselves, which is in principle accessible through highly-multiplexed redshift-surveys of large numbers of galaxies.

The identification of galaxy "groups", i.e. a set of galaxies populating the same dark matter halo, from galaxy redshift catalogues enables the environments of galaxies to be characterized down to quite low halo masses. It is widely understood that the mass of the host halo of a galaxy, and whether the galaxy is the "central" or a "satellite" within that halo, are the dominant determinants of the evolution of the galaxy. The determination of these properties as accurately as possible is a pre-requisite for a physical understanding of the properties of galaxies. Even if one is interested in further "second-order" effects, such as the growth history of the halo \citep[e.g.][]{Forbes2013}, or the location of the halo within the filamentary structure of the cosmic web \citep[e.g.][]{Alonso2015}, or effects such as "galactic conformity" \citep[e.g.][]{Kauffmann2013,Knobel2015,Sin2017}, one must understand as well as possible the immediate halo environment in order to remove the effects of this from any more detailed analysis.
 
Classifying the group environment of galaxies allows many physical investigations, including for instance the relationships between stellar mass, environment and star formation rate (SFR). Previous analyses of the SFR in galaxies in relation to their (group) environment have led to the isolation of distinct processes of "mass quenching", and "environment quenching" \citep{Peng2010}, the latter particularly affecting satellite galaxies. Differentiating centrals (which we operationally define to be the most massive member of the group, independent of its actual spatial position within the group) from satellites has established that the SFR of satellites is effected by their environment:  When former central galaxies fall into larger dark matter haloes and become satellites, they are likely to quench their star formation and become quiescent, a process known as "satellite quenching" \citep[e.g.][]{van_den_Bosch2008,Peng2012,Wetzel2012}. 

In planning an observational campaign to construct a large redshift catalogue of distant galaxies using a multi-object spectrograph (MOS), there are a number of considerations in the design of the survey. These can have a major impact on the amount of observing time that is required. These design decisions include, straightforwardly, the total number of galaxies to be observed and the amount of observing time required to secure a redshift for each, which will itself depend on parameters such as the stellar masses, SFR or other selection criteria. The projected number density of target galaxies on the sky, which will also be determined by the selection criteria of the galaxies and by the desired redshift range, will also determine how many individual "configurations" of the MOS are required in each field. It may also affect how efficiently the multi-object capability of the instrument can be used, since this will generally degrade as the choice of targets within each field becomes more and more constrained.  This may depend in detail on the design of the instrument.  Another key parameter is the required sampling rate, i.e. what fraction of the available tracer targets are to be observed (including what fraction of those should have a successful redshift measurement).  A high required sampling rate will generally produce a more constrained set-up and this will lower the efficiency. Finally, how the set of tracer galaxies is defined, and in particular their number density, will certainly affect, along with the sampling rate, the range of halo masses that is probed, and the accuracy with which the halo environment of each galaxies can be reconstructed. While it is obvious that these design decisions for the survey all have a potential impact on the resulting group catalogue, their quantitative impact is not trivially assessed.    

The aim of this paper is therefore to quantitatively examine some of these issues. This invebfstigation was carried out in the context of the design of the future MOONRISE \citep{Maiolino2020} %\footnote{http://www.eso.org/sci/publications/messenger/archive/no.180-jun20/messenger-no180-24-29.pdf}
survey which will be undertaken using the MOONS MOS \citep{Cirasuolo2020}
%\footnote{Multi Object Optical and Near-infrared Spectrograph; https://www.eso.org/sci/facilities/develop/instruments/MOONS.html}
that will soon be commissioned at the VLT\footnote{Very Large Telescope; https://www.eso.org/public/teles-instr/paranal-observatory/vlt/?lang}.
Our approach however is to understand the main effects by idealizing the problem. MOONS-specific issues are as far as possible minimized or isolated, and the main results should be relevant for any similar survey. The point of this paper is not to design a particular survey with a particular instrument, but rather to explore the trade-offs in survey design that should be relevant for almost any multi-object spectroscopic survey. Of course, for any individual future survey a more detailed survey strategy considering the real data, the real instrument and the real fibre-positioning or other target selection software needs to be defined.

Semi-analytic models (SAMs) of the evolving galaxy population are now good enough that the basic properties of the galaxy population produced by the models, such as the mass- or luminosity- functions, well match the real galaxy population at different redshifts. This is especially true for those SAMs that use Markov Chain Monte Carlo (MCMC) methods to explicitly tune the main parameters of the SAM so as to best match the main distribution functions of the galaxy population. Likewise, the success of the standard model of cosmology means that the properties of the underlying population of dark matter haloes in such models are likely to closely approximate reality. The mock "light-cones" from these models should therefore be quite a good match to the observed sky in the most relevant aspects. Such mock catalogues therefore enable us to apply exactly the same selection criteria as in a potential survey and then compare the properties of the reconstructed group catalogue to the underlying known "reality". This enables us to assess the actual performance of the group-finder in the proposed survey using quantifiable metrics.  At the very least, the differential performance of different survey designs can be quite reliably examined.

No group-finder based on galaxy positions can hope to perfectly assign galaxies to haloes because of the limited information available to the astronomer (essentially two projected spatial positions and a third radial velocity measurement). It should also be appreciated that even within the simulated Universe the association of galaxies to parent haloes is not without some ambiguity. The isolation of individual dark matter haloes is not perfectly defined as they continually accreate new material. Further difficulties can arise in unambiguously associating galaxies to be satellites within larger haloes, especially during the early phases of virialization.  A gravitationally bound galaxy (sub-halo) may still be found at some distance from the halo even after a first passage through it - the so-called "splash-back" effect. 

An important part of the current paper will be to construct quantitative metrics for assessing the performance of a group finding algorithm. These metrics go beyond the simple statistical concepts of purity and completeness and aim to be relevant for a wide range of different scientific applications of the output from the group-finder.

We will concentrate on the relative performance of the group-finder as the design of the survey is varied. We will focus especially on the two major design criteria for the selection of the tracer galaxies. For simplicity, we will assume that the primary selection criteria for the survey is the limiting stellar mass of the target galaxies, $M_{\mathrm{stellar}}$. While previous deep spectroscopic surveys (e.g. CFRS, VVDS, zCOSMOS, etc.) used flux-limited samples, photometric redshifts are now good enough in well-studied fields like COSMOS that the stellar mass uncertainties are quite modest, allowing mass-selection of targets in future deep spectroscopic survey designs.  As an example, the scatter in the observed H-band mass-light ratio for fixed rest-frame colour at high redshift $z \approx 1.5$ is about 0.1 dex ($1 \sigma$). Additionally, the ability to vary the exposure times for different targets allow an efficient observation of mass-selected samples. For instance, the planned MOONS GTO survey MOONRISE will be selecting targets based on their photometrically estimated stellar masses as well as on their (AB) H magnitudes, see \citet{Maiolino2020}.

The limiting stellar mass of the target galaxies translates directly into their volume number density in the Universe. Clearly, the volume number density of the tracers is one of the most important parameters in their ability to define structure in the Universe. To first order, any set of galaxies selected by some other criteria (e.g. the luminosity in some band) could be converted, via their volume number density, to an "equivalent" $M_{\mathrm{stellar}}$ limit.   Of course, any particular selection of the tracers (including the straight stellar mass) will unavoidably introduce some biases into the galaxy content of the resulting identified groups. These biases in galaxy-content, arising from using other tracers than stellar mass, will not be considered in this paper.

The other major design parameter aside from the mass (or number density) of the tracers is the required sampling rate, $s$. This is defined as the ratio of the number of tracer galaxies with a usable redshift to the total number of such tracers.  Incomplete sampling may arise from either not observing a galaxy at all, or in failing to secure a spectroscopic redshift  - we do not distinguish between these in this investigation.  Choice of lower $s$ may reflect a simple desire to maximise the area of sky covered by the survey, or to maximise the multiplexing efficiency of the instrument due to technical constraints from the instrument design.

Non-random incompleteness, in the sense that the observations may fail to secure a redshift for some targets with particular properties, may be especially a problem at high redshifts. While this can be mitigated by adaptively varying the exposure time until a redshift is obtained, there will inevitable be some remaining incompleteness.  Similarly, there will likely be a range of reliabilities of the redshifts, which may again introduce biases into the content of recovered structures.

Furthermore, the incompleteness may not be completely spatially random if local projected over-densities or particular geometries of nearby objects, lead to a reduction in the fraction of targets that can be observed. However, if the survey covers a large enough range in radial distance, the projected surface density of tracers is not strongly correlated with the volume number density associated with individual structures due to the high number of foreground and background objects (we examine this question later in the paper).  Multiplex spectrographs like MOONS allow variable fibre-placement densities by using overlapping patrol fields of individual fibres, and the effects of target geometry can be reduced by multiple passes over each part of the sky. Nonetheless, technical constraints might potentially prevent the proper sampling of clustered targets. Even though the effect is small, this issue must be addressed in any final survey design. In the current work, we will assume for simplicity that the incompleteness is completely random across the adopted set of tracers. The fraction of randomly observed objects is therefore described by a random sampling rate $s$.

There are several approaches in the literature to constructing group catalogues from galaxy redshift catalogues. These span a range of philosophies, depending on how much astrophysical information is assumed. On the one extreme are algorithms such as "Friends-of-Friends" (FoF) \citep[e.g.][]{Huchra_Geller1982,Merchan2002,Merchan2005,Robotham2011,Eke2004,Gerke2005,Berlind2006,Knobel2009} and similar approaches such as that based on the Voronoi–Delaunay tesselation \citep{Marinoni2002}. These rely only on the spatial location of galaxies in projected space and in velocity to associate galaxies together.  In order to optimize the group finding, the FoF method uses three free parameters that control the galaxy-galaxy "linking-lengths" perpendicular to, and along, the line of sight. These three free parameters can be optimized by applying the group-finder to simulations, for which we know the underlying dark matter haloes and the "true" groups populating them, and optimizing the parameters to recover the "true" catalogue as well as possible.  Once structures are identified in this way, their dark matter masses must be estimated through the application of suitable algorithms, e.g. using the richness, the integrated stellar mass, estimates of the size or velocity dispersion, or some combination of these. These mass-estimators may be calibrated against the mock catalogues, or by using other approaches such as abundance matching.  FoF and related techniques have been extensively used at high redshift \citep[e.g.][]{Gerke2005,Knobel2009}.  

Particularly at low redshifts, e.g. for the SDSS \footnote{Sloan Digital Sky Survey; https://www.sdss.org/}, some other, more refined, approaches have been developed. These assign membership of galaxies to groups in an iterative scheme that builds up the population of dark matter haloes around galaxies based on assumptions about the sizes of haloes (in projected space and velocity). These then assign membership of galaxies to individual groups based on a probabilistic approach. These approaches generally have estimates of the dark matter mass built in to the algorithm. This approach has been used by e.g. \citet{Yang2007} and \citet{Tinker2011}. Again, such algorithms generally still have tunable parameters that should be optimized.

Of course, in principle, galaxies for which only photometric (but no spectroscopic) redshifts are available could also be included in any sample of galaxies. This was done for example in \citet{Kovac2009} to define the density field out to $z=1$ or in \citet{Wang2020} for group finding. For simplicity, we do not include these galaxies in the group catalogues used in this work.

In this work, we will first review in Section
\ref{group_finding} the basic approaches to group finding used in this paper, review the most basic ideas of purity and completeness and how these may be used to optimize the tunable parameters of the group finding algorithms.  We also introduce the mock light-cones used throughout the paper.  We describe a new implementation of a FoF group-finder (in this paper) and of what we call a "halo-based" group-finder that has also been re-implemented in Sin et al. (in prep.).

Then in Section \ref{science_metrics} we construct a set of twelve metrics to quantitatively assess the performance of group finding algorithms in a scientific context.
In Section \ref{Tinker_vs_FoF_main} we use these metrics to first examine the relative performance of the FoF and "halo-based" algorithms when they are applied to an SDSS-like mock catalogue at low redshift.  
We will use for this a fixed stellar mass cut of $10^{9.0}$ solar masses for the tracers, but examine a wide range of sampling rates, $s$, so as to better understand the strengths and short-comings of these two approaches as the sampling rate is reduced, concluding that their performance is comparable.

We will then apply, in Section \ref{Science_metrics_M_s_space}, the FoF group-finder (alone) to three high redshift ranges within the overall redshift range $0.9<z<2.6$, and examine how these metrics change as we vary both the stellar mass cut in the range $10^{9.1-10.5}$ solar masses and the sampling rate $s$.  This leads to a more quantitative understanding of the impact of the choices of $M_{\mathrm{stellar}}$ and $s$ on the size and quality of the recovered group catalogues and on their usability for various scientific investigations. 

In order to assess the recovered FoF catalogues, and how they change with $M_{\mathrm{stellar}}$, $s$ and $z$, we present numerous quality metrics and introduce various science metrics designed to capture a wide range of environmental information about the recovered galaxy group catalogues.   

As a realistic science example, we also explore the degree to which imperfections in the recovered group catalogue perturb a simple science measurement: the quenched fraction of central and satellite galaxies as a function of their host halo mass. We explore how the metrics can be used to construct simple corrections to the observed quenched fractions to best recover the correct values.

Finally, we present an analysis of the "costs" of possible survey designs in terms of observing time. This highlights the trade-offs between the selections in stellar mass $M_{\mathrm{stellar}}$ and the choice of sampling rate $s$ in terms of the total telescope observing time and the efficiency with which the multiplexing capability of the instrument is utilized.  As part of this we present a general formalism for consideration of the survey costs, in which the instrument specific issues are all concentrated in just one of the four terms, enabling a relatively straightforward consideration of these trade-offs.  We then summarize the conclusions of the paper in Section \ref{summary}. Throughout the paper we assume a $\Lambda \textrm{CDM}$ cosmology with the following parameters: $H_0 = 69.6$ km $\textrm{s}^{-1}$/ Mpc, $ \Omega_{\textrm{M}} = 0.286$ and $\Omega_{\Lambda} = 0.714$.

\section{Galaxy group finding}
\label{group_finding}

Galaxy groups are the set of galaxies populating the same dark matter halo. However, dark matter haloes are not directly observable and so, in practical terms, galaxy groups must be recovered from catalogues of the observable positions (RA, DEC and redshift) and possibly other quantities (e.g. stellar mass) of the galaxies observed in any large scale galaxy redshift survey.  Fortunately, mock catalogues from "light-cones" based on realistic models of galaxy formation and evolution are now available in which the group membership of all galaxies are, at least in principle, known from the underlying model.  These mock catalogues can therefore provide an underlying "truth" against which a recovered or reconstructed group catalogue may be compared. 

In this section, we first briefly describe the L-Galaxies Munich Galaxy Model that was used as the (simulated) mock reality throughout this work.  This mock reality is used firstly to optimize the parameters of different group finding algorithms and also then to quantitatively compare their performance, both relative to each other and also, for a given group-finder, as the tracer selection (in terms of stellar mass) and sampling rate are varied.  

We then briefly review the basic concepts of how real and recovered groups can be matched and the quantitative definitions of completeness and purity of the recovered group catalogue.  We also briefly review the operation of the two representative group-finders examined in this work, which, although described in previous papers are here slightly modified: The first is the Sin et al. re-implementation (Sin et al., in prep.) of a group-finder based on the work of the \citet{Tinker2011}. The second is a multi-run implementation of a FoF group-finder \citep[e.g.][]{Huchra_Geller1982,Gerke2005,Knobel2009}.  
We again briefly show how the parameters of these algorithms can be optimized (generally by balancing purity and completeness).  Finally, we construct an empirical mass-estimator for the FoF algorithm. 

\subsection{The L-Galaxies SAM}
The L-Galaxies semi-analytical model (SAM) \footnote{https://wwwmpa.mpa-garching.mpg.de/galform/virgo/millennium/} \citep{Henriques2015} of galaxy formation and evolution is used throughout this paper as the "mock" reality. 
L-Galaxies is built onto the Millennium \citep{Springel2005} and Millennium II \citep{BoylanKolchin2008} dark matter simulations, which are $\Lambda CDM$ cosmological N-body simulations, using $10^{10}$ particles, predicting the evolution of dark matter haloes over cosmic time.  Such simulations retain the merger history of individual dark matter haloes and the evolution of individual "sub-haloes" within the larger haloes.  The L-Galaxies SAM then models the evolution of the baryonic matter component within these dark matter haloes, based on prescriptions for the physical processes that affect galaxy formation and evolution such as gas cooling, star formation, supernova feedback, the formation and growth of black holes and the feedback from these, as well as galaxy interactions and mergers.

In this paper, we use the mock 'pencil-beam' "light-cones" produced by the L-Galaxies SAM using M05 stellar populations.  These mimic the distant (model) universe as observed by us.  These light-cones include information about the mass and location of dark matter haloes with redshift, and of the simulated galaxies within them. In what follows, these simulated galaxy groups are regarded to represent "reality", and we will refer to them as the "true galaxy groups". Each true galaxy group is defined to be the set of galaxies populating the same dark matter halo in the simulation. The ensemble set of true galaxy groups forms a true galaxy group catalogue.  The Richness $N$ is defined as the number of member tracer galaxies. Some galaxies may be in a group of just one galaxy, and these will be referred to as singletons.

\begin{figure}
\includegraphics[width=1.\columnwidth]{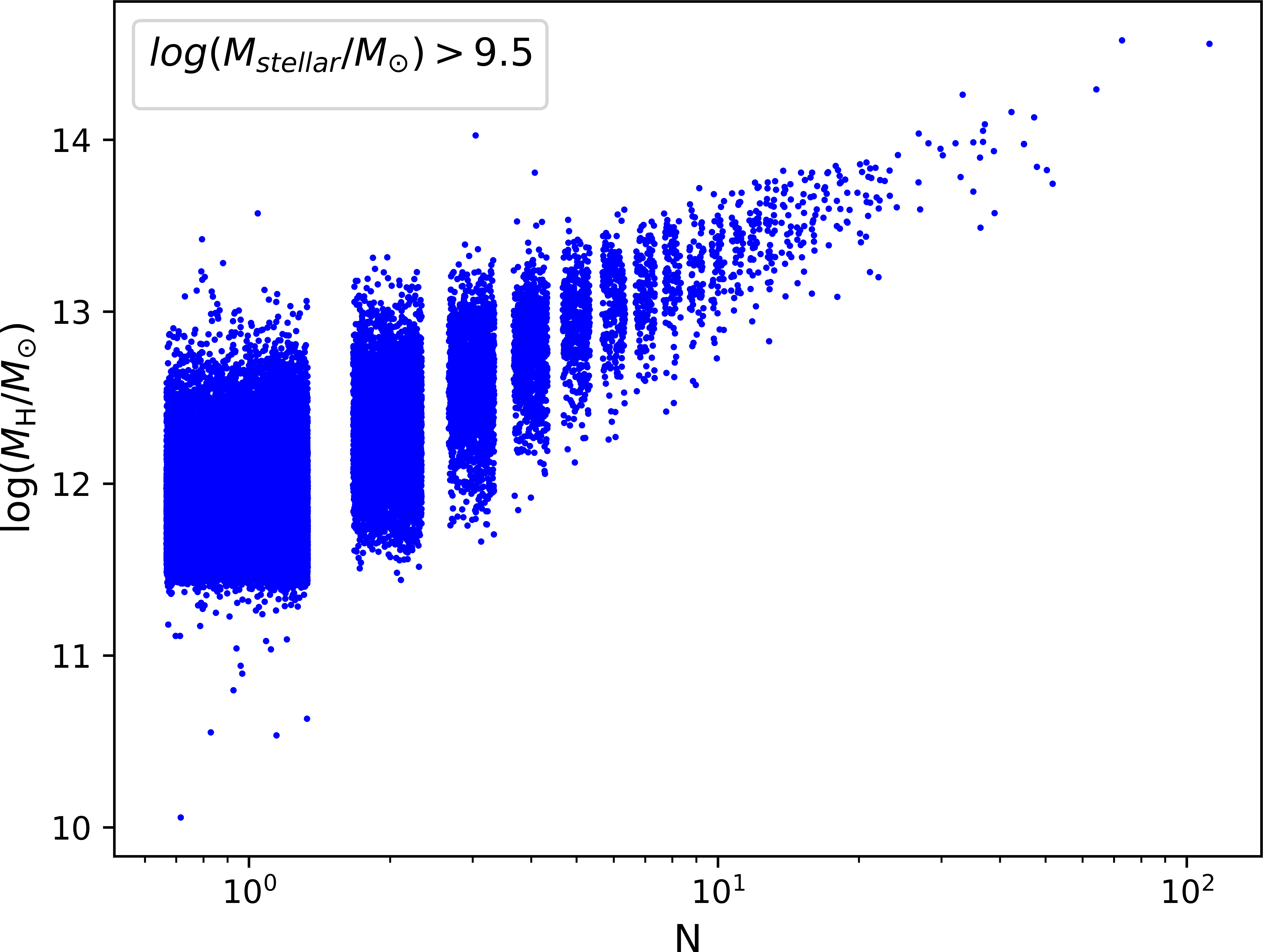}
\caption{The distribution of dark matter halo mass as function of true group richness; a basic property of the L-Galaxies SAM illustrative true group catalogue for $\textrm{log}_{10}(M_{\mathrm{stellar}}/M_{\odot}) > 9.5$ within the redshift range $1.2 < z < 1.7$. An artificial random scatter is added to the richness values for readability.} \label{Illustrative_true_gr_cat_1}
\end{figure}

\begin{figure}
\includegraphics[width=1.\columnwidth]{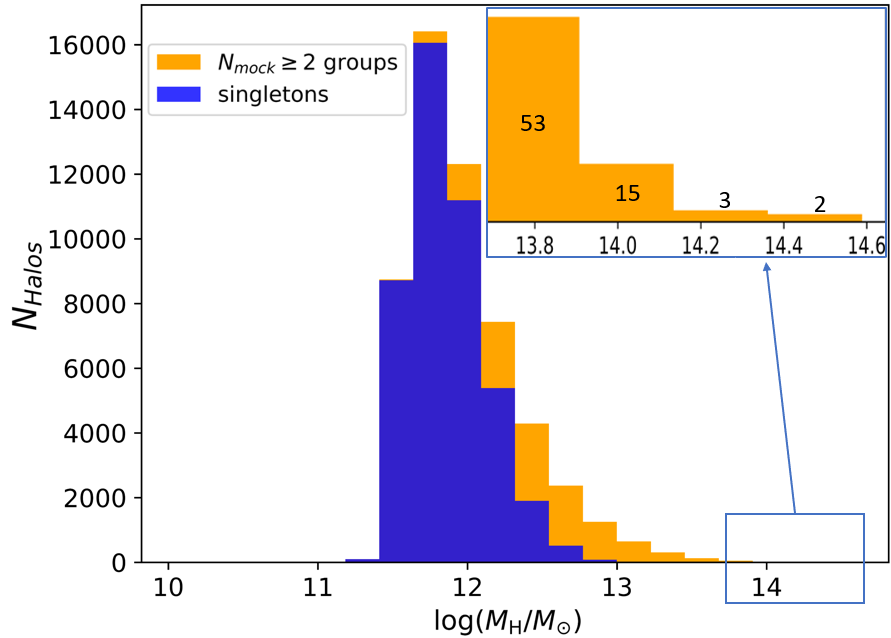}
\caption{The number of dark matter haloes populated by singletons and richer structures as a function of dark matter halo mass; a basic property of the L-Galaxies SAM illustrative true group catalogue for $\textrm{log}_{10}(M_{\mathrm{stellar}}/M_{\odot}) > 9.5$ within the redshift range $1.2 < z < 1.7$.} \label{Illustrative_true_gr_cat_2}
\end{figure}

For orientation, it is worthwhile to briefly review the true group catalogue from the L-Galaxies model. As an example, we consider the catalogue that is obtained using tracers of stellar mass $\textrm{log}_{10}(M_{\mathrm{stellar}}/M_{\odot}) > 9.5$ within a light-cone of 2deg x 2deg on the sky and in the redshift range $1.2 < z < 1.7$. This volume contains 75,838 galaxies above the stellar mass limit.  Two basic properties, the halo mass $M_{\mathrm{H}}$ versus richness $N$ distribution and the number density of groups as function of halo mass, are shown in Fig.~\ref{Illustrative_true_gr_cat_1} and Fig.~\ref{Illustrative_true_gr_cat_2}, respectively.

\begin{figure}
\includegraphics[width=\columnwidth]
{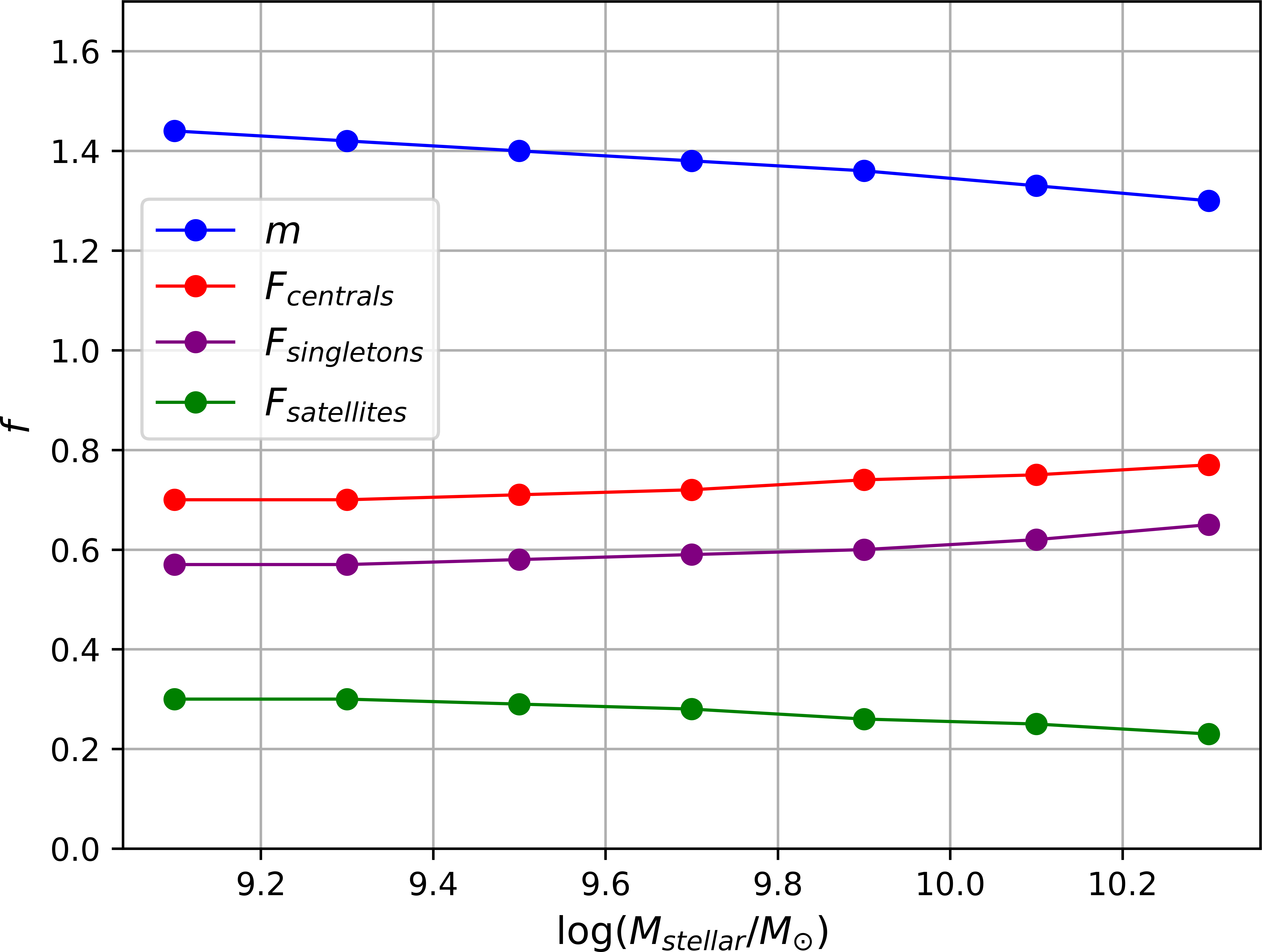}
\caption{The basic characteristics of the true group catalogue for different limiting mass cuts $M_{\mathrm{stellar}}$ in the redshift regime $1.2 < z < 1.7$. The multiplicity $m$ is the average number of members per group. Note that the fraction of centrals  $F_{\mathrm{centrals}}$ includes the singletons.  Most centrals are in fact singletons.} \label{True_Catalog_bfs}
\end{figure}

The effects of varying the stellar mass selection limit $M_{\mathrm{stellar}}$ on the characteristic properties of the group catalogue are shown in Fig.~\ref{True_Catalog_bfs}. With a more restrictive (higher) mass cut, the multiplicity $m$, which is the average number of members per group, decreases, as does the fraction of galaxies that are satellites $F_{\mathrm{satellites}}$. The fraction of galaxies that are centrals (including the singletons) $F_{\mathrm{centrals}}$ and the fraction of singleton galaxies themselves $F_{\mathrm{singletons}}$, on the other hand, both increase. The choice of limiting stellar mass selections systematically shifts the basic properties of the resulting galaxy group catalogue.

In this analysis we use several of the 24 available L-Galaxies pencil-beam 2deg x 2deg light-cones (always using the M05 stellar models). We use one light-cone to train the algorithms by optimizing the free parameters (e.g. b, R for the FoF group-finding algorithm, see Subsection \ref{FoF_group_finder} and to fit the polynomial coefficients in our halo mass estimator, see Subsection \ref{M_halo_est}).  We then apply the algorithms to a different light-cone in order to obtain the results in the Sections \ref{science_metrics}, \ref{Tinker_vs_FoF_main}, \ref{Science_metrics_M_s_space} and \ref{cost_analysis}. The single individual light-cones contain large enough numbers of galaxies and groups to allow statistically meaningful conclusions.   As a detail, the light-cones have sharp edges and hence cut through some galaxy groups. The galaxies in these truncated groups have not been excluded in this analysis. This might slightly bias the sample of galaxy groups, but the fraction of such groups is extremely small and hence the biases introduced by this effect are negligible.

\subsection{The fidelity of recovered galaxy groups}
In this section, we review the most basic statistical quantities that may be used to assess the quality of the recovered group catalogue produced by a given group-finder applied to a given set of tracer galaxies.
\subsubsection{Matching of real and recovered groups}

\begin{figure*}
\centering
\includegraphics[width=1.8\columnwidth]
{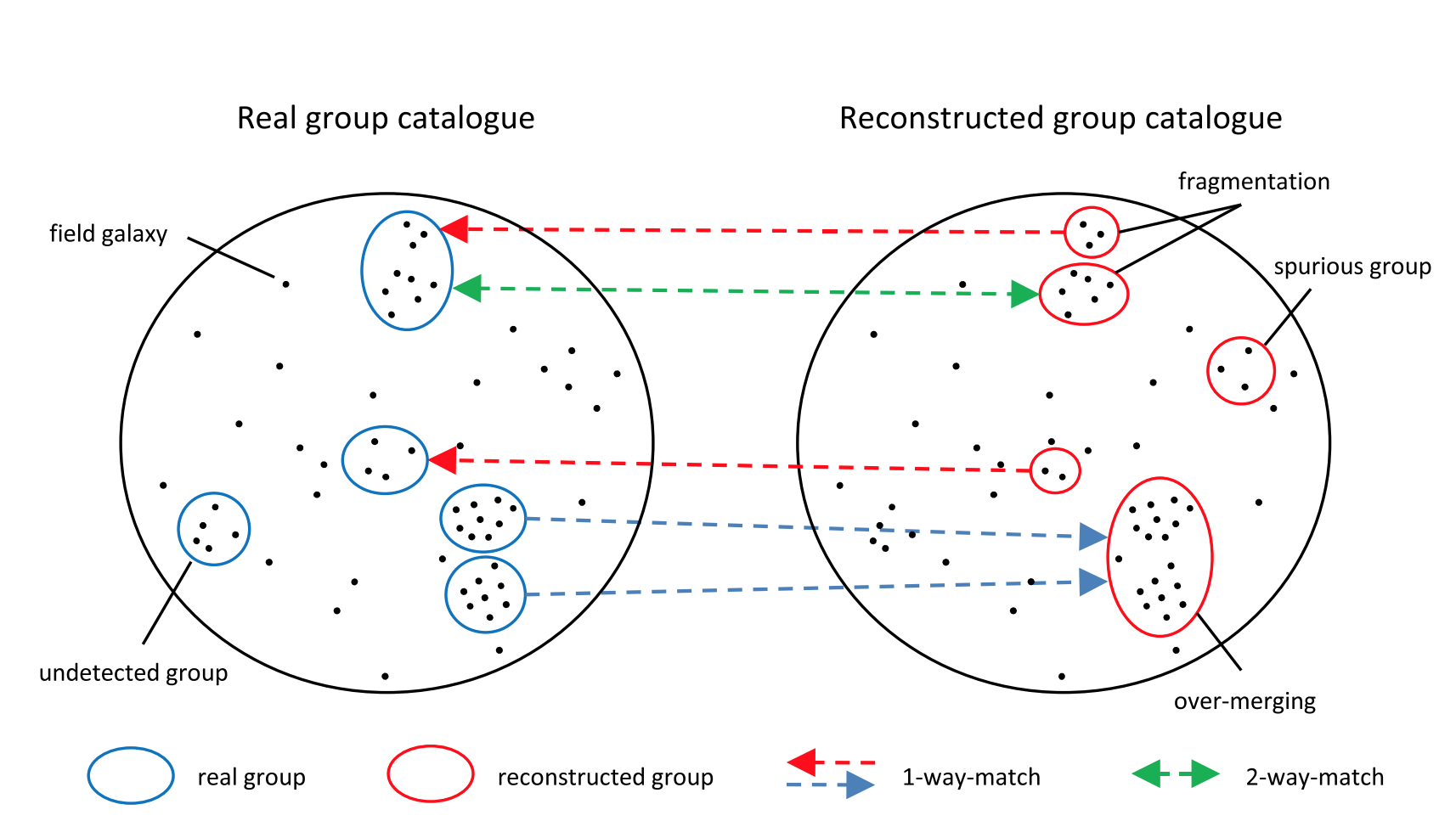}
\caption{Diagrammatic illustration \citep{Knobel2009} of the matching between the "real" groups (from the simulation) and the reconstructed  groups. See text for details.} \label{Comp Rec Re Groups}
\end{figure*}

The performance of any group-finder can be assessed by comparing the recovered groups (from the group-finder) and those in the true group catalogue (from the mock simulation). Although we will later define more science-oriented metrics, these most basic quantities are used to optimize the parameters of the group-finder, and also serve as the basis for some of the later metrics.

Following \citet{Knobel2009} and \citet{Gerke2005} the matching procedure is conceptually illustrated in Fig.~\ref{Comp Rec Re Groups}. Each point represents a galaxy and we compare the two orderings (the true group catalogue and the reconstructed group catalogue) of this identical set of points. The precise definitions used are as follows:
\begin{itemize}
\item Match: A group $i$ is matched to another group
$j$ if group $j$ contains a fraction $f > f_M$ of
the members of group $i$. For this association to
be unique, it must hold that $f_M \geq 0.5$. Throughout this
paper, we use this minimum $f_M = 0.5$, as in \citet{Knobel2009} and \citet{Gerke2005}. Hence, for a match, more than $50 \%$ of the members of group $i$ have to be found in group j. For example, if group i is a $N = 4$ member group, group j must contain at least 3 of those 4 galaxies as members, in order to be matched. Similarly, both members of an $N = 2$ group must be present for a match.
\item 1-way match: The case where group i is associated to group j, but
group j is not associated to group i (illustrated by
a one-way arrow from group i to group j). 1-way matches can occur from real to recovered groups or vice versa.
\item 2-way match: The case where group i is associated to group j and group j is also associated back to group i (illustrated by a double-arrow). 
\end{itemize}
While each group can by definition only have a single, unique associated group (i.e. an arrow pointing away), it might well happen that a certain group is the matched group for
two or more other groups (i.e. have two or more arrows pointing inwards to
it). We therefore use the following further terminology:
\begin{itemize}
\item Over-merged group: If more than one real group is matched
to the same reconstructed group.
\item Fragmented group: If more than one reconstructed group
is matched to the same real group.
\item Spurious group: A reconstructed group which is not matched to any
real group.
\item Undetected group: A real group which has no matched
reconstructed group.
\item Singleton: A galaxy that is not associated to any group in the catalogue.
\end{itemize}
An important concept is that of "2-way matching" between real and recovered groups.  For these groups at least $50 \%$ of the members of a single real group are found within a single recovered group, and at least $50 \%$ of the members of that single recovered group are also members of the original real group. 2-way matches are by construction unique, in the sense that each real or recovered group can have at most a single 2-way match.  They are the most reliably reconstructed groups.

\subsubsection{Completeness and purity} \label{completeness_purity}

In order to quantify the basic fidelity of the group-finder, we again follow \citet{Knobel2009} in defining the completeness $c$ and purity $p$ for assessing the overall performance of the group-finder. The 1-way completeness is defined as
\begin{equation}
c_1 \left(N \right) = \frac{\mathcal{A} \left[ N^{ \tiny \textsf{mock}}_{ \tiny \textsf{gr}}\! \left(N \right) \rightarrow  N^{\tiny \textsf{rec}}_{ \tiny \textsf{gr}} \left(> \! N/2 \right) \right]}{ N^{\tiny\textsf{mock}}_{ \tiny \textsf{gr}} \! \left(N \right)},\end{equation}\label{c_1}
where $\mathcal{A} \left[ N^{ \tiny \textsf{mock}}_{ \tiny \textsf{gr}}\! \left(N \right) \rightarrow  N^{\tiny \textsf{rec}}_{ \tiny \textsf{gr}} \left(> \! N/2 \right) \right]$ counts the number of true groups $ N^{ \tiny \textsf{mock}}_{ \tiny \textsf{gr}}\! \left(N \right)$ with a population of $N$ members, that can be matched to a recovered group, by either a 1-way or 2-way match. $c_1(N)$ therefore evaluates the fraction of true $N$-member groups that are 1-way or 2-way matched. In the same way, 
\begin{equation}
c_2 \left(N \right) = \frac{\mathcal{A} \left[ N^{ \tiny \textsf{mock}}_{ \tiny \textsf{gr}}\! \left(N \right) \leftrightarrow  N^{\tiny \textsf{rec}}_{ \tiny \textsf{gr}} \left(> \! N/2 \right) \right]}{ N^{\tiny\textsf{mock}}_{ \tiny \textsf{gr}} \! \left(N \right)} ,
\end{equation} \label{c_2_formula}
evaluates the fraction of groups that are matched using the more restrictive 2-way matches only. Analogously we define the 1-way purity
\begin{equation}
p_1 \left(N \right) = \frac{\mathcal{A} \left[ N^{ \tiny \textsf{rec}}_{ \tiny \textsf{gr}}\! \left(N \right) \rightarrow  N^{\tiny \textsf{mock}}_{ \tiny \textsf{gr}} \left(>\! N/2 \right) \right]}{ N^{\tiny\textsf{rec}}_{ \tiny \textsf{gr}} \! \left(N \right)},  
\end{equation} \label{p_1}
and the 2-way purity
\begin{equation}
p_2 \left(N \right) = \frac{\mathcal{A} \left[ N^{ \tiny \textsf{rec}}_{ \tiny \textsf{gr}}\! \left(N \right) \leftrightarrow  N^{\tiny \textsf{mock}}_{ \tiny \textsf{gr}} \left(> \! N/2 \right) \right]}{ N^{\tiny\textsf{rec}}_{ \tiny \textsf{gr}} \! \left(N \right)},
\end{equation} \label{p_2_formula}
of the recovered group catalogue. 
The purity evaluates the fraction of recovered groups with a population of $N$ members which have a match to a true group.

The quality of a group-finder can then be assessed by these two performance metrics, completeness and purity, which respectively measure the fraction of true groups that are recovered and the fraction of recovered groups that have an associated true group. Ideally, we would want the group-finder to exhibit both high completeness and purity, but these are not independent and, generally, one expects a trade-off between the two.

If the group-finder is set up to easily find structures it will generally exhibit high completeness at the cost of low purity, due to over-merging and the formation of spurious groups. Likewise, if the group-finder is set up so that there are restrictions in identifying structures, then the purity will be high at the cost of completeness, as true groups will be in danger of being fragmented and there will be undetected groups. 

Depending on the scientific goals, one might wish to value purity over completeness, or vice versa. However, usually a group-finder should represent a balance between completeness and purity. Hence, again following \citet{Knobel2009}, we quantify the "quality" of the recovered group catalogue by a parameter
\begin{equation}
g_1 = \sqrt{\left(1-c_1 \right)^2 + \left(1-p_1 \right)^2},\label{g_1}
\end{equation} 
which values completeness and purity equally. The parameter $g_1$ gauges the deviation from a "perfect" group catalogue which would have $\left(c_1,p_1 \right) = (1,1)$. Note that $g_1$ should therefore be as small as possible.

\subsection{FoF group-finder} \label{FoF_group_finder}
Our implementation of the FoF group-finder is closely related to the FoF algorithm used in \citet{Knobel2009}, which itself was based on the FoF implementation used for the DEEP2 group catalogue by Gerke et al. \citet{Gerke2005}. The group finding approach of the FoF method is to "link" nearby galaxies, and subsequently form groups of all those galaxies that are linked together, either directly or indirectly via other galaxies. The linking of galaxies depends on their transverse (spatial) and radial (velocity) separations. The maximum allowed transverse separation, in order to be linked, is governed by the perpendicular linking length parameter $l_{\perp}$, and the maximum allowed radial velocity separation is governed by the parallel linking length $l_{\parallel}$.  These two parameters $l_{\perp}$ and $l_{\parallel}$ are controlled by the three free parameters of the FoF group-finder $b$, $R$ and $L_{\textrm{max}}$ as follows:

\begin{itemize}
\item The basic linking length parameter $b$ is the main parameter controlling the perpendicular linking length $l_{\perp}$. $b$ links $l_{\perp}$ to the mean observed spatial galaxy number density $\bar{n}$, which may (slowly) vary as a function of redshift if required. The purpose of relating $l_{\perp}$ to $\bar{n}$ is to allow an increase of the linking length as the number density of tracers decreases.  
\item As $\bar{n}$ decreases, however, $l_{\perp}$ may increase so much as to be larger than the physical scale of any structures of interest. Hence, $L_{\textrm{max}}$ is introduced to limit the maximum linking length. The perpendicular linking length is therefore given by
\begin{equation}
l_\perp = \textbf{min} \left[\frac{b}{\bar{n}^{1/3}} , L_{\small \textrm{max}}\left(1 + z\right) \right].
\end{equation} \label{l_perp}
\item The transverse and radial linking lengths are related by $R$. The need for this arises due to the peculiar velocities of the galaxies. These local velocities shift the measured redshifts, and hence considerably increase the apparent comoving radial distances between galaxies based on redshift alone, the famous "Fingers of God" effect. $R$ is therefore introduced in order to allow a larger radial linking tolerance. $R$ is defined as the ratio between $l_{\perp}$ and $l_{\parallel}$,
\begin{equation}
l_\parallel = R \ l_\perp.
\end{equation} \label{l_par}
\end{itemize}
$l_{\perp}$ and $l_{\parallel}$ then govern the linking of galaxies in the following way: Two galaxies $i$ and $j$ with apparent comoving radial distances from the observer $d_i$ and $d_j$, as observed from their redshifts, and an intermediate angle $\theta_{ij}$ are considered linked if the difference of their apparent comoving radial distances fulfills the inequality 
\begin{equation}
\vert d_i - d_j \vert \leq \frac{l_{\parallel,i} + l_{\parallel,j}}{2},
\end{equation} \label{d_i-j}
and, at the same time, their intermediate angle satisfies
\begin{equation}
\theta_{ij} \leq \frac{1}{2} \left(\frac{l_{\perp,i}}{d_i}  +   \frac{l_{\perp,j}}{d_j} \right).
\end{equation} \label{theta_ij}
These two inequalities are designed to impose adequate limits, dependant on the observed redshifts, on the maximal transverse and radial separations of two linked galaxies.

The choice of the free parameters $b$, $R$ and $L_{\small \textrm{max}}$ in the FoF group finding algorithm involves a trade off between completeness and purity: A large linking length improves the completeness at the cost of purity, and vice versa. 

In order to obtain an optimal balance between completeness and purity, we define the optimal free parameters as those one which minimize $g_1$, defined above in equation~(\ref{g_1}), in the resulting recovered group catalogue.

Within a large enough volume of the Universe, there will be a wide range of dark matter halo masses and richnesses, from singletons up to groups of fifty or more members. Rich groups and small groups may have quite different densities in projected transverse and radial velocity space.  As a result, the optimal parameters for identifying rich groups may well differ from those for smaller groups. 

Again following \citet{Knobel2009},
we therefore apply a "multi-run" scheme in the FoF group finding algorithm.   The optimal linking parameters are determined across the range of richnesses of the real groups. The full recovered group catalogue is then constructed by first applying the group-finder with parameters optimized for some high bin of (true) richness. All recovered groups within this same richness bin are then selected, and all their member galaxies are removed from the tracer population.  The next run uses the optimal linking parameters for the next lower richness bin, more groups are identified and the members are again removed.  This is repeated all the way down to the final 2-member bin. The galaxies which then still remain are classified as singletons.\footnote{As an aside, we investigated a variation of this multi-run method, in which the linking parameters for each richness bin are re-optimized on the remaining galaxies at each step. It was found that this approach performed only slightly better on the test samples and slightly worse on average on the validation samples, and was therefore discarded.}

\begin{figure}
\includegraphics[width=1.\columnwidth]{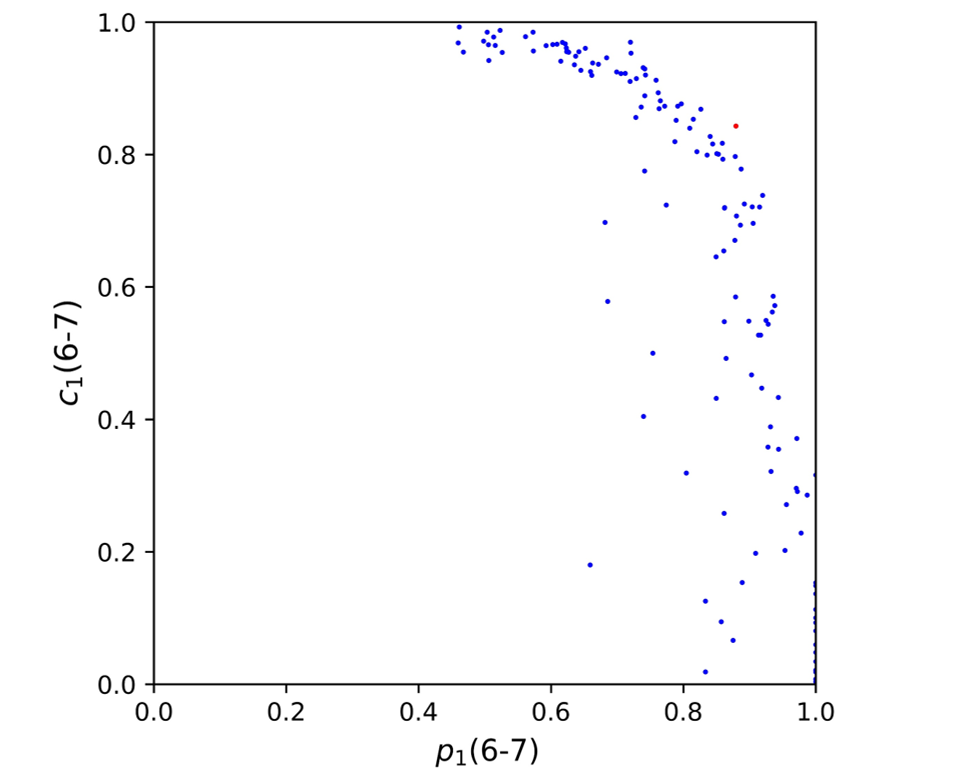}
\caption{FoF $(b,R)$-parameter optimization with respect to $g_1$ in the  $1.2 < z < 1.7$ redshift range for a stellar mass cut of $\textrm{log}_{10}(M_{\mathrm{stellar}}/M_{\odot}) = 9.5$ and a sampling rate of $s = 0.8$ in the $N =6-7$ richness bin. The deviation of the optimal recovered FoF group catalogue, with corresponding optimized parameters $b=0.06$ and $R=40$, from the true group catalogue is $g_1=0.198$, as illustrated by the red dot. The expected trade-off between completeness and purity as the FoF parameters are adjusted is clearly seen.}
\label{b_R}
\end{figure}

For illustration, Fig.~\ref{b_R} shows the parameter optimization in the $N =6-7$ richness bin in the $1.2 < z < 1.7$ redshift range for a stellar mass cut of $\textrm{log}_{10}(M_{\mathrm{stellar}}/M_{\odot}) = 9.5$ and a sampling rate of $s = 0.8$. As the spatial galaxy density $\bar{n}$ is reasonably high for these survey parameters, we can omit $L_{\mathrm{max}}$ and optimize for $b$ and $R$ only. To optimize the FoF group finding method we apply a ($b,R$) grid-search for every combination of limiting stellar mass and random sampling rate ($M_{\mathrm{stellar}},s$) in each of the three redshift ranges.

We observe that while the main parameter $b$ is rather stable over all richness's, $R$ alters substantially, presumably because of the changing effects of peculiar velocities. However, $g_1$ is more responsive to changing $b$ than it is to changing $R$.
When re-running the multi-run procedure multiple times, for the same mass-selection and sampling rates, but with a different random galaxy sampling each time, we observe the same: Over all richness ranges, the $b$ parameter is stable, while $R$ varies a good deal. Despite this, $g_1$ is largely stable for an extended range of $R$, meaning that the precise value of $R$ is not decisive for the overall performance of the multi-run group-finder.

\begin{figure}
\includegraphics[width=1.\columnwidth]{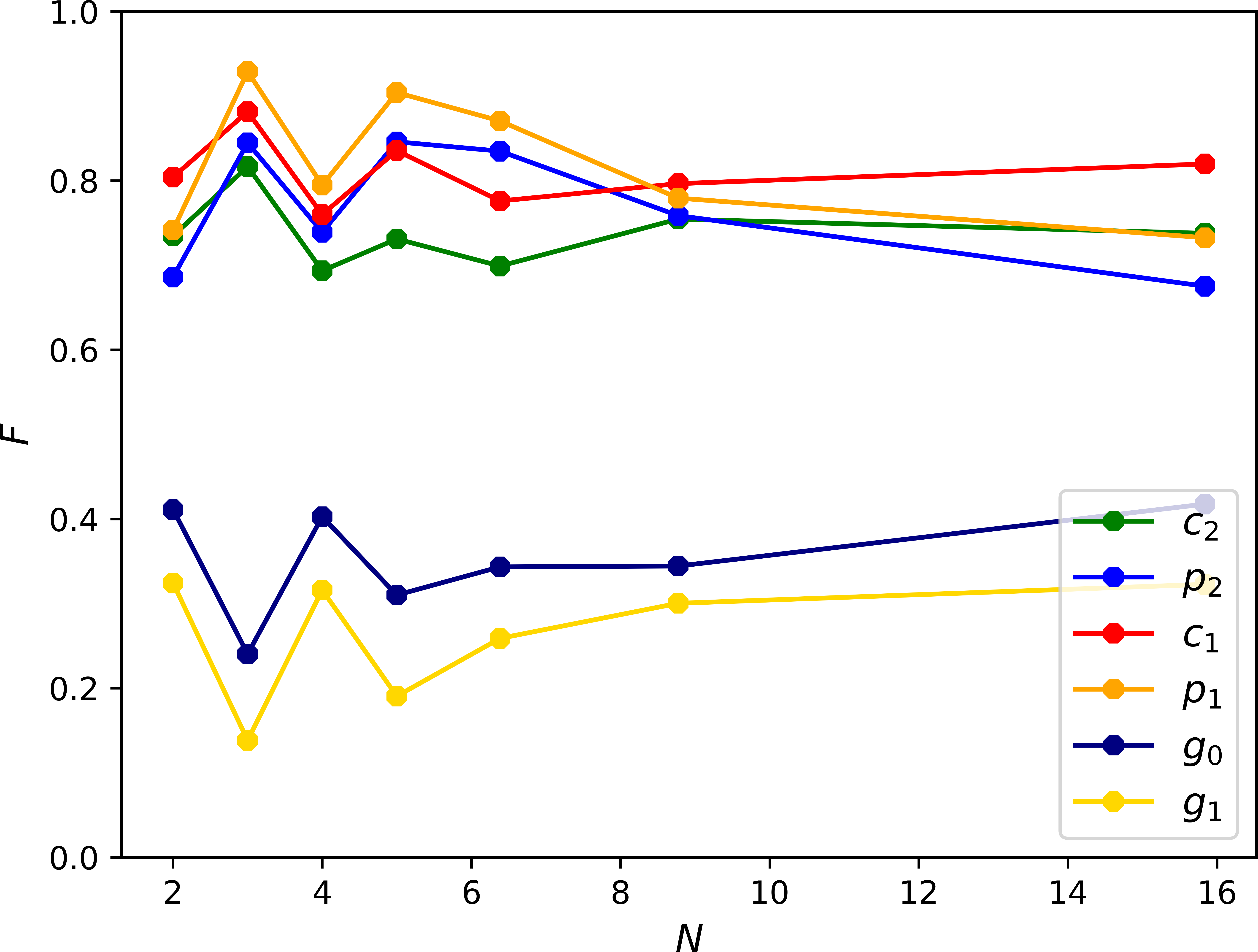}
\caption{FoF 1-way and 2-way completeness and purity, and $g_1$ and $g_0$ as function of group richness $N$ in the  $1.2 < z < 1.7$ redshift range and for a stellar mass cut of $\textrm{log}_{10}(M_{\mathrm{stellar}}/M_{\odot}) = 9.5$ and a sampling rate of $s = 0.8$. It should be noted that the closer $p_1$, $p_2$, $c_1$ and $c_2$ are to 1, and thus $g_0$ and $g_1$ to 0, the better the quality of the recovered group catalogue.}
\label{per_bins}
\end{figure}

In Fig.~\ref{per_bins} the quality of the recovered group catalogue, in terms of 1-way and 2-way completeness and purity, as function of richness is shown. Over all richness bins, we observe a total 1-way completeness of $c_1=0.82$ and a 1-way purity of $p_1=0.79$, with an overall quality of $g_1=0.28$ and $g_0=0.37$, where
\begin{equation}
g_0 = \sqrt{\left(1-c_2 \right)^2 + \left(1-p_2 \right)^2}
\end{equation} assesses the relative frequency of failures in finding 2-way matches for true and recovered groups. 

\subsection{"H\lowercase{alo-based"} galaxy group-finder}

The group finding algorithm that we call the "halo-based" method is heavily based on the methodology developed by \citet{Tinker2011} as implemented in Sin et.al. (in prep.) and we refer to that paper for details of the implementation. 

The "halo-based" methodology works as follows:
\begin{enumerate}
\item Initially, every galaxy is defined to be the sole member of its group and a dark matter halo mass is assigned to each of these "groups" via its stellar mass, using a calibration based on sub-halo abundance matching (or, conceptually, any other method).

\item Subsequently, the local matter density $P_M$ (defined below) is evaluated and, starting from the highest $P_M$, galaxies with $P_M > B$  are put together as members of a single group and masked from further group assignment (for this iteration).  The parameter $B$ is tunable.

\item Dark matter halo masses are then re-assigned to these new groups again using the total stellar masses of the recovered groups.

\item The steps (ii) and (iii) are repeated iteratively until the group membership converges, i.e remains unchanged between two iterations. 
\end{enumerate}

The local matter density $P_M$ is defined as 
\begin{equation}
P_M=\frac{c}{H_0} \Sigma( R_{\rm{proj.}}, \alpha r_s ) p( \Delta z, \beta \sigma_v ),
\end{equation}
where $\Sigma$ is the projected NFW profile of a halo with scale radius $r_s$ evaluated at projected radius $R_{\rm{proj.}}$, and $p$ is the normalized Gaussian function of a halo with velocity distribution $\sigma_v$ evaluated at redshift-space separation $\Delta z$, relative to the respective centres of that halo. The formulae for these two terms are given in \citet{Tinker2011}, with the difference that in the Sin et.al. (in prep.) implementation, two additional tunable parameters $\alpha$ and $\beta$ are introduced, allowing further degrees of freedom in optimizing the group finder.

As a detail, the "halo-based" method used in this work is optimized for the balanced pairwise accuracy metric $\Pi$ introduced in Sin et.al. (in prep.),
\begin{equation}
\Pi = \sqrt{f_{\rm{correct}}^2 - (f_{\rm{frag.}}^2 + f_{\rm{merge}}^2)},
\end{equation}
where $f_{\rm{correct}}$, $f_{\rm{frag.}}$, and $f_{\rm{merge}}$ are respectively the fraction of pairs which are correctly classified, fragmented (i.e. same-halo pairs which are misclassified as different-halo), and merged (i.e. different-halo pairs which are misclassified as same-halo), when one considers all pairs which are separated by less than $1\,\rm{Mpc}$ in projected separation, and $500\,\rm{km\, s^{-1}}$ in apparent velocity difference.

Relative to the FoF algorithm, the "halo-based" approach is more refined, in the sense that it uses more astrophysical and cosmological "knowledge".   Of course, if that knowledge is correct, then this will likely produce a more accurate result.  The trade-off is the risk of inaccuracies if that knowledge is incomplete or not correct.  One of the goals of the current work was to investigate in particular how the relative performance of the refined "halo-based" approach compared with the more basic FoF approach changes as the information on the tracer population is degraded, especially through a reduction in the tracer sampling rate $s$. 

\subsection{$M_{\mathrm{H}}$-estimator} \label{M_halo_est}

In this section, we introduce a dark matter halo mass $M_{\mathrm{H}}$-estimator for the recovered galaxy groups.   An estimate of the dark matter mass is required for many science applications and the scatter in recovered halo mass will be one of the science-based metrics that we introduce in Section \ref{science_metrics}.  Whereas the "halo-based" approach yields estimated masses directly, by construction, the masses of FoF groups are estimated separately, {\it post facto}.  One could of course adopt abundance matching methods, as in the "halo-based" approach, but one can also try to calibrate directly from the mock universe.  We here construct an "empirical" mass estimator for FoF groups based on the total stellar mass of the members of the galaxy groups $\sum M_{\mathrm{stellar}}$, the RMS radial velocity dispersion of the groups $\sigma_v$ and the projected RMS sizes of the groups $\sigma_r$.  This is constructed by examining the true groups in the mock "reality" introduced earlier in this paper.

For each recovered galaxy group we define a $\sigma_v$ as
\begin{equation}
\sigma_v = \sqrt{\frac{1}{N_{\mathrm{gal}} - 1} \sum_{i = 0}^{N_{\mathrm{gal}}} \left(c \ \frac{z_i - \bar{z}}{1 + \bar{z}} \right)^2},
\end{equation}
where $N_{\mathrm{gal}}$ is the number of galaxies within the group. The factor $\left(N_{\mathrm{gal}} - 1 \right)^{-1}$ compensates for the fact, that one degree of freedom is used in order to estimate $\bar{z}$.  Analogously, we define a $\sigma_r$ as
\begin{equation}
\sigma_r = d_{\mathrm{comdist}} \left(\bar{z}\right) \ \sqrt{\frac{1}{N_{\mathrm{gal}} - 1} \sum_{i = 0}^{N_{\mathrm{gal}}} \left( \left(\theta_i - \bar{\theta}\right)^2 + \cos^2{\bar{\theta}}\left(\phi_i - \bar{\phi}\right)^2 \right)},
\end{equation}
where $d_{\mathrm{comdist}} \left(\bar{z}\right)$ is the comoving distance corresponding to the average redshift of the group $\bar{z}$. 

\begin{figure}
\includegraphics[width=1.\columnwidth]{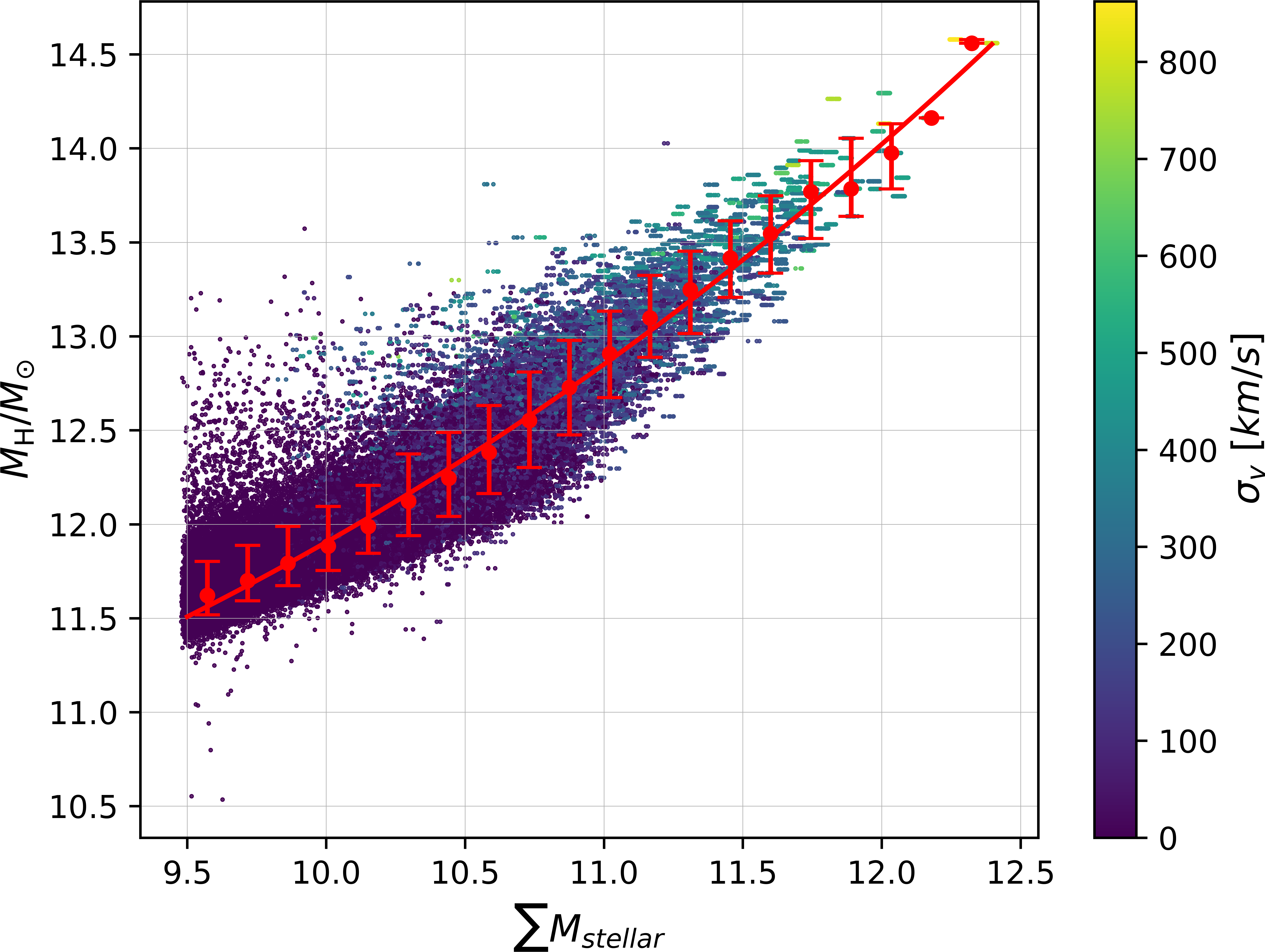}
\caption{The correlation of halo mass $M_{\mathrm{H}}$ (plotted in log stellar mass) to $\sum M_{\mathrm{stellar}}$ and $\sigma_v$ in the $1.2 < z < 1.7$ redshift regime, for $s=0.8$ and $\textrm{log}_{10}(M_{\mathrm{stellar}}/M_{\odot}) = 9.5$. The red line shows an $M_{\mathrm{H}}$-estimator based on $\sum M_{\mathrm{stellar}}$ only.}
\label{M_H_cc_sigma_v}
\end{figure}

Not surprisingly, the dominant determinant of $M_{\mathrm{H}}$ is $\sum M_{\mathrm{stellar}}$. $\sigma_v$ can be used to improve the $M_{\mathrm{H}}$-estimator further: In Fig.~\ref{M_H_cc_sigma_v} we show that galaxies with high $\sigma_v$ tend to lie above the red line, representing a halo mass estimate based on the total stellar mass only. 
For $\sigma_r$, there is little empirical evidence that it improves the halo mass estimate further over  $\sigma_v$ or $\sum M_{\mathrm{stellar}}$. We therefore implemented an $M_{\mathrm{H}}$-estimator calculating $M_{\mathrm{H,est}}$ as a polynomial function of $\sum M_{\mathrm{stellar}}$ and $\sigma_v$ only. Because it returns empirically the best results we adopted a third degree polynomial function,
\begin{multline}
\textrm{log}_{10}(M_{\mathrm{H,est}}/M_{\odot})=\sum_{\alpha + \beta \leqslant 3} c_{\alpha \beta} \ \times \\\times \left( \textrm{log}_{10} \left(\sum M_{\mathrm{stellar}}/M_{\odot} \right) \right)^{\alpha} \left(\textrm{log}_{10}((\sigma_v/(\textrm{km/s})) \right)^{\beta},
\end{multline}
where $c_{\alpha \beta}$ are the coefficients of the polynomial function. Note that the coefficients $c_{\alpha \beta}$ are dependant on the considered redshift regime, the limiting stellar mass and the sampling rate, and hence have to be calibrated for each individual ($z,M_{\mathrm{stellar}},s$)-bin separately. Further, the galaxy-based and the group based coefficients are slightly different; while in the galaxy-based analysis we consider each galaxy in each halo individually (hence the estimator is weighted by the number of galaxies in each halo), in the group-based analysis we consider haloes as a whole \footnote{To give an example, in the $1.2 < z < 1.7$ redshift regime with $\textrm{log}_{10}(M_{\mathrm{stellar}}/M_{\odot}) > 9.5$ and $s=0.8$, the coefficients of the galaxy-based estimator are $c_{00} = 9.61899242e+00$, $c_{10} = 2.97574209e-15$, $c_{20} = 8.52193554e-15$, $c_{30} = 2.27243612e-03$,
 $c_{01} = 5.46838695e-23$, $c_{02} = 2.77947535e-25$, $c_{03} = 2.38635562e-02$, , $c_{11} = 4.39220440e-23$, $c_{21} = 1.43856458e-23$, $c_{12} = 6.03824430e-30$.}.

\begin{figure}
\includegraphics[width=1.\columnwidth]{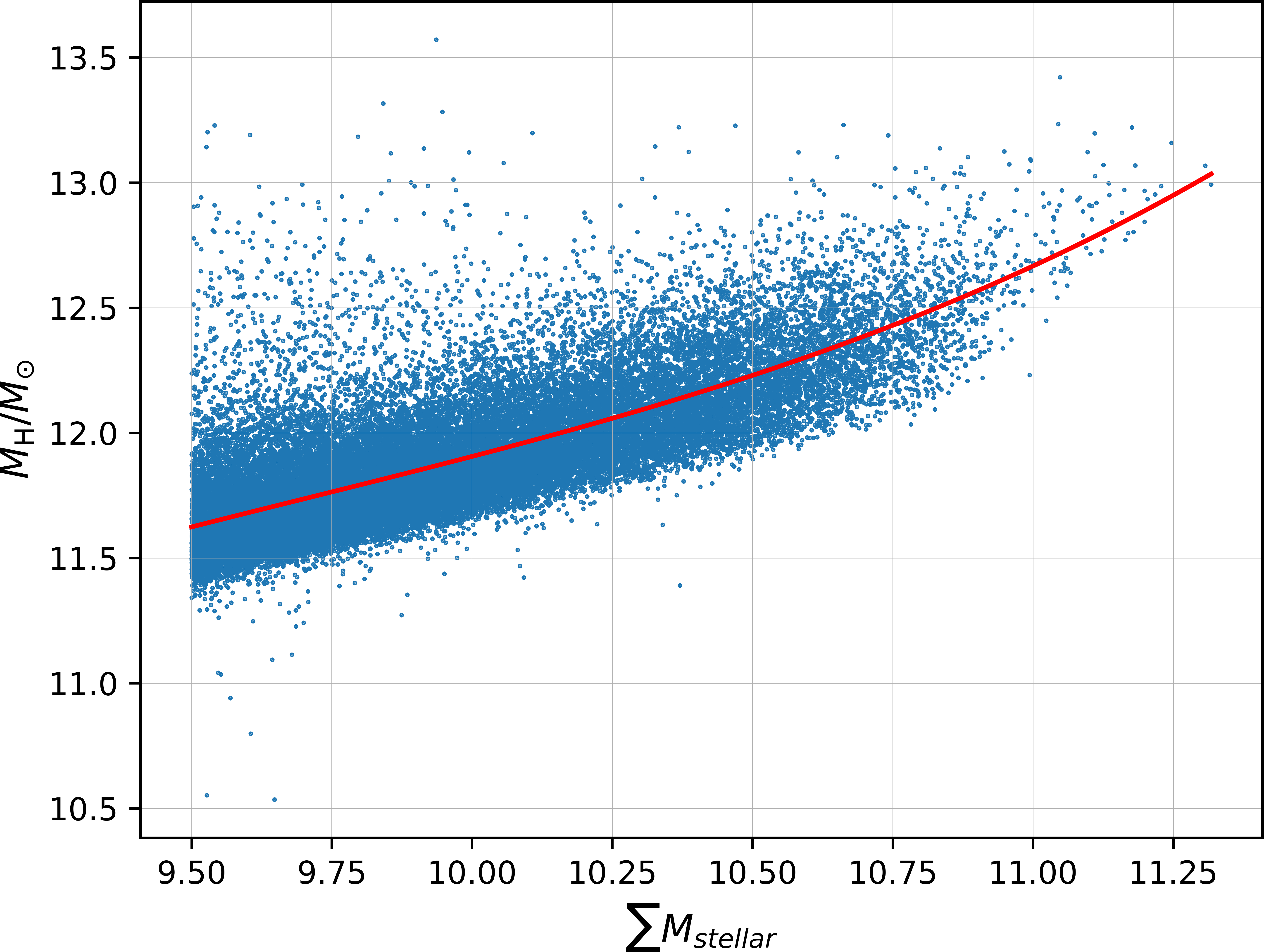}
\caption{$M_{\mathrm{H}}$-estimator for singletons in the $1.2 < z < 1.7$ redshift regime, for $s=0.8$ and $\textrm{log}_{10}(M_{\mathrm{stellar}}/M_{\odot}) = 9.5$.}
\label{M_halo_est_sin}
\end{figure}

 For singletons, the RMS velocity dispersion and the RMS size are not defined, hence we implement for them a third degree polynomial $M_{\mathrm{H}}$-estimator based on $\sum M_{\mathrm{stellar}}$, i.e. $M_{\mathrm{stellar}}$ only, as illustrated in Fig.~\ref{M_halo_est_sin}.
 
The halo mass estimator is a tool that will be used below to assign masses to recovered groups in order to calculate the scatter in the real and recovered $M_{\mathrm{H}}$.  It is important to appreciate that this scatter will therefore include both the intrinsic scatter in the $M_{\mathrm{H}}(\sum M_{\mathrm{stellar}},\sigma_v)$ relation in the real (mock) universe, plus the additional effects arising from any infidelities of the group catalogue.
 
\section{Science metrics for performance estimation}\label{science_metrics}

The purity and completeness reviewed above (see Section \ref{completeness_purity}) are operationally well-defined and, as explained there, are ideal for optimizing the values of the adjustable parameters of any group finding algorithm.  However, it is not immediately clear how the completeness and purity actually map over to a quantitative degradation of the scientifically useful information that is in principle contained within a group catalogue.

Therefore, in this section, we will construct several "science metrics". Each of these is designed to assess quantitatively the performance of a given galaxy group-finder (with optimized parameters) from the perspective of an end-user scientific investigator interested in one or other scientific questions.  

We stress that the purpose of constructing these different metrics is not to to try to identify one or two metrics that are somehow "the best", but rather to explore quite a large number of different metrics that are relevant for the wide and diverse range of scientific investigations that can be based on a (recovered) group catalogue from a given redshift survey.   Different end-users may be interested in different combinations of these metrics, according to their particular scientific goals.

For definiteness of illustration, we will also present in this section the quantitative values of each science metric for the particular case of a recovered FoF group catalogue in the $1.2 < z < 1.7$ redshift range, obtained with a tracer mass cut of $\textrm{log}_{10}(M_{\mathrm{stellar}}/M_{\odot}) > 9.5$ and a random sampling rate of $ s = 0.8$. Later in the paper, we will look at how these metrics differ between FoF and "halo-based" group-finders when applied at low redshift, as we vary the sampling rate (alone).  We will then examine how the FoF performance varies at high redshifts as the limiting stellar mass cut and sampling rate are varied, and how this changes with redshift.

\subsection{Masses of the recovered groups on a galaxy basis}\label{gal_basis}

One defining property of a galaxy group is the mass of the associated dark matter halo.  The mass is a fundamental quantity of any gravitationally bound structure, and halo mass may well also be one of the most important drivers of galaxy evolution.

We first distinguish between metrics that are constructed from the properties of the parent haloes that are assigned to each individual galaxy, and those that refer to the groups themselves. Imperfect fidelity of group reconstruction may affect these two in quite different ways.  The former may be most relevant, for example, in studies of galaxy evolution because one needs to know the underlying total parent halo mass of each individual galaxy. The latter will be more relevant for studies of the X-ray emitting gas and the like. In this subsection, we will construct metrics describing the performance of the group-finder in accurately describing the parent halo of each individual galaxy.  We refer to these as "galaxy-basis" metrics, as opposed to the "group-basis metrics" considered later.

\begin{figure}
\includegraphics[width=1.\columnwidth]{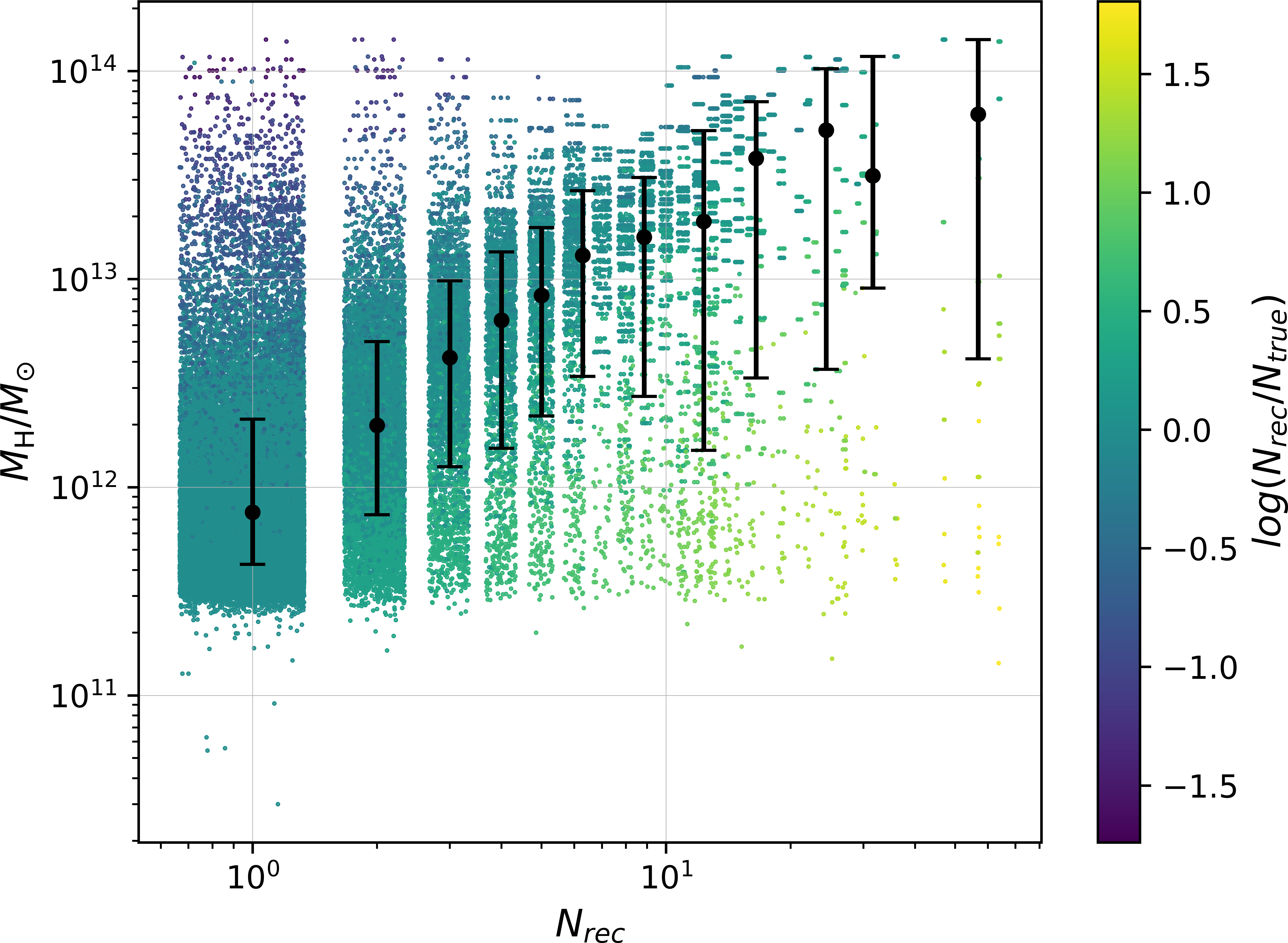}
\caption{True parent halo masses of galaxies in recovered groups as a function of the richness of the recovered group, plotted on a galaxy basis. Every colored dot represents a galaxy. The black dots stand for the median halo masses $M_{H,m}$ in each (recovered) richness bin.  The error bars indicate the  $16 \%$ and $84 \%$ in true halo mass dispersion.}
\label{M_halo_gal}
\end{figure}

 Fig.~\ref{M_halo_gal} shows the distribution of the true parent halo mass for every galaxy in the sample, as a function of the richness of the recovered group that contains that galaxy. In other words, it looks at the range of real parent halo masses for the galaxies that are in the reconstructed groups of a particular richness. Each dot therefore represents a single galaxy.  At low richnesses the (integer) values of $N_{\mathrm{rec}}$ have been artificially horizontally broadened for clarity. It should be noted that the L-Galaxies simulation will already have an intrinsic scatter between the dark matter halo mass $M_{\mathrm{H}}$ and the galaxy group richness $N_{\mathrm{true}}$. This will be further increased in Fig.~\ref{M_halo_gal} by the imperfect assignment of galaxies to groups bringing into the (recovered) group interlopers who have quite different (true) halo masses. 
 The color-coding in this figure represents the ratio between the assigned (recovered) richness of the group that contains the galaxy in question, $N_{\mathrm{rec}}$, and the true richness $N_{\mathrm{true}}$ of the group that in reality contains that galaxy. Comparison of these two therefore reflects the amount of fragmentation or merging of groups in the recovered group catalogue. The galaxies shown in purple are members of (real) richer groups that have been fragmented, while those in yellow are "interlopers" that have been incorrectly merged into larger structures in the recovered group catalogue. 

Table~\ref{Table_galaxy_bases} gives the median true halo mass, the $16 \%$ and $84 \%$ scatter in halo mass, and the RMS of the ratio between $N_{\mathrm{rec}}$ and $N_{\mathrm{true}}$ for bins of different $N_{\mathrm{rec}}$. We will discuss how these are affected by different mass-selection $M_{\mathrm{stellar}}$ and sampling rates $s$ in detail later in this paper. 

\begin{table*}
	\centering
	\caption{As function of $N_{\mathrm{rec}}$: The median halo masses $M_{H,m}$, the RMS scatter in halo masses and the quotients of $N_{\mathrm{rec}}$ and $N_{\mathrm{true}}$, on a galaxy-basis (see also Fig.~\ref{M_halo_gal}).}
	\label{Table_galaxy_bases}
	\setlength{\tabcolsep}{6.pt}
	\begin{tabular}{@{}l||*{16}{c}r@{}} % four columns, alignment for each 
	\centering
	    $N_{\mathrm{rec}} =$ & 1 & 2 & 3 & 4 & 5 & 6-7 & 8-10 & 11-14 & 15-20 & 21-28 & 29-39 & 40- & mean & $\geq 2$\\
	    \hline
		log$(M_{H,m}/M_{\odot})$ & 11.88 & 12.3 & 12.62 & 12.8 & 12.92 & 13.11 & 13.2 & 13.28 & 13.58 & 13.72 & 13.5 & 13.79 & - & -\\
		\hline
		RMS(log($M_H/M_{H,m})$) & 0.42 & 0.42 & 0.45 & 0.48 & 0.49 & 0.52 & 0.56 & 0.72 & 0.71 & 0.78 & 0.68 & 0.85 & 0.59 & 0.49\\
		\hline
	    RMS(log$(N_{\mathrm{rec}}/N_{\mathrm{true}}$)) & 0.19 & 0.21 & 0.24 & 0.27 & 0.3 & 0.31 & 0.39 & 0.52 & 0.51 & 0.61 & 0.69 & 0.86 & 0.42 & 0.3\\
		\hline
	\end{tabular}
\end{table*}

We therefore introduce the first two metrics of the recovered group catalogue as follows (based on Fig.~\ref{M_halo_gal}):
\begin{itemize}
\item Metric 1: The median true halo mass ${^1}M_{H,m}$ of galaxies recovered as singletons (by definition on a galaxy basis). This may be thought of as giving the minimum halo masses for which there is any information about galaxy evolution.
\item Metric 2: The median true halo mass $^{2}M_{H,m}$ of galaxies in recovered 2-member groups, again on a galaxy basis. This may be thought of as giving the halo mass for which there is some non-trivial environmental information.  The overall $M_{H,m}(N_{\mathrm{rec}})$ relation for richer groups scales quite closely with this value.
\end{itemize}

\subsection{The recovered 2-way matched groups}\label{2-way_basis}

In addition to looking at the statistics of the dark matter haloes assigned on a galaxy-basis, we can also look at the groups directly. We have to restrict this analysis to recovered groups which have a 2-way match because only for these is the allocation of a true dark matter halo to a given recovered group unambiguous.  This comparison also really only makes sense for the 2-way matched groups since it is only for these that there is a reasonable correspondence (to within at most a factor of two) between the true and recovered membership.   Unfortunately, it is not possible for an observer to know (without having the "mock" reality) which of the recovered groups are 2-way matched and which are not, and so we also need to consider what fraction of the recovered groups are two-way matched. 

It is important to note that, unlike most of the metrics in this section, 2-way matching only makes sense in terms of a comparison with a "resampled" reality. Because it involves a comparison with the number of real members in the real group, any galaxies that are excluded by random sampling must not be included.  The real group membership is therefore defined here in terms of the real galaxy population after any random sampling of the galaxies (i.e. as given by our parameter $s$) has taken place. 

\begin{figure}
\includegraphics[width=\columnwidth]{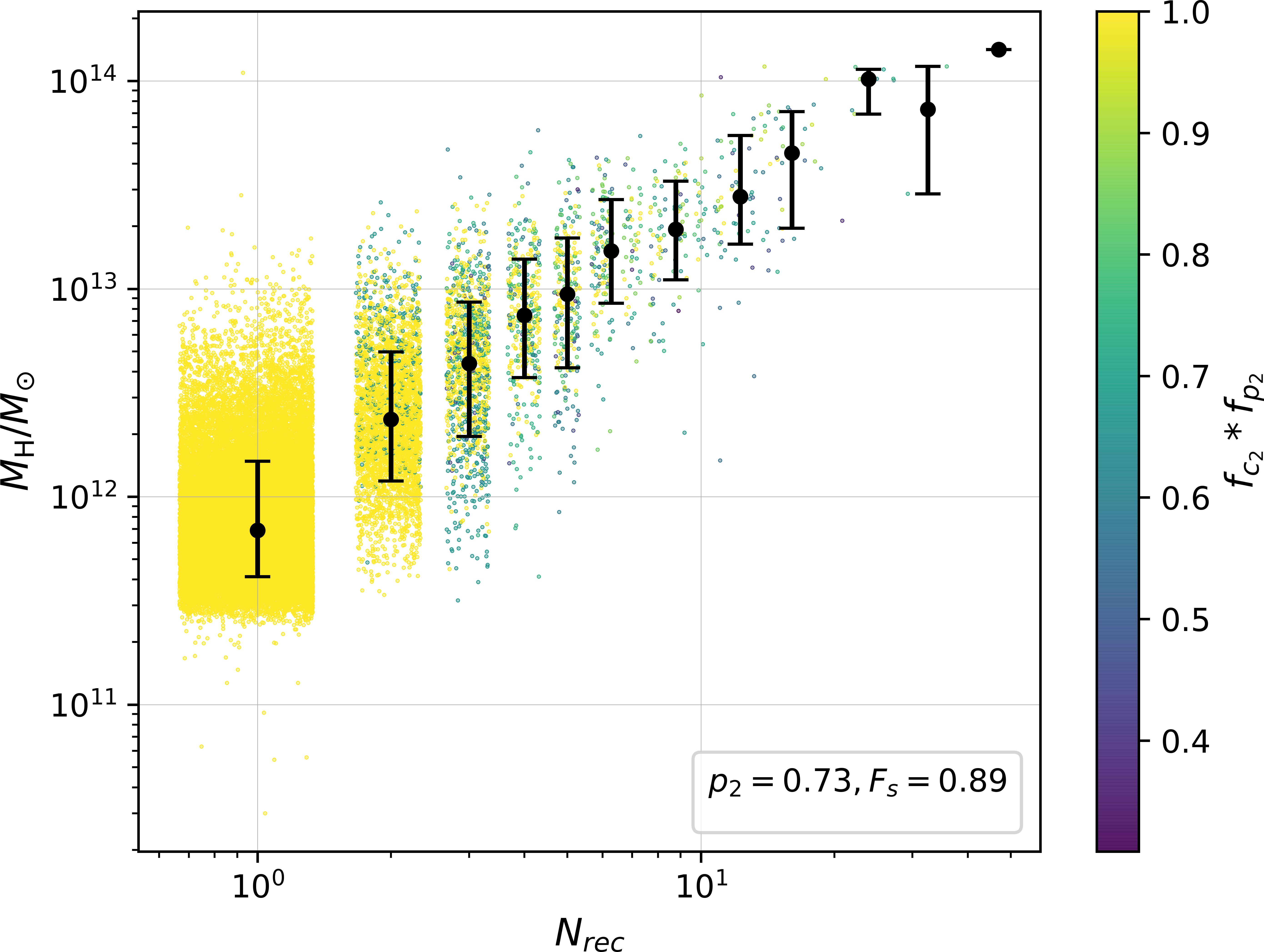}
\caption{The true halo mass of recovered 2-way matched groups as a function of the recovered richness, plotted on a group basis. Each colored dot represents a single 2-way matched group.  The error bars indicate the median and $16 \%$ and $84 \%$ halo mass dispersion in each recovered richness bin. $F_s$ is the fraction of recovered singletons which are truly singletons, only these are plotted at $N_{\mathrm{rec}} = 1$.}
\label{M_halo_gr}
\end{figure}

Fig.~\ref{M_halo_gr} shows the distribution of true halo mass for the 2-way matched recovered groups, including, with $N_{\mathrm{rec}} = 1$, those galaxies that are singletons in both the ("mock") true and recovered catalogues. Hence, each dot now represents a single recovered group (or singleton).  The points in the integer richness bins are again dispersed horizontally for clarity. In the representative recovered catalogue considered in this section, $73 \%$ of all recovered groups in the catalogue (with a richness of at least 2) have a 2-way match, while $89 \%$ of all recovered singletons are truly singletons (and thus two-way matched).  The color-coding shows the product $f_{c_2} f_{p_2}$, which indicates how "good" the 2-way match is. It gives the product of the fractions of true group members that are found in the recovered group and visa versa.  Straightforwardly from the definition of 2-way matched groups, this must be at least 0.25, i.e. $0.5 \times 0.5$.

Table~\ref{Table_Group_bases} gives the median true halo mass, the $16 \%$ and $84 \%$ scatter in true halo mass and the mean of  $f_{c_2} f_{p_2}$ for recovered groups of different richness. 

\begin{table*}
	\centering
	\caption{As function of $N_{\mathrm{rec}}$: The median halo masses $M_{H,m}$, the RMS scatter in halo masses and the mean of the product $f_{c_2} f_{p_2}$ for the 2-way matched groups and the recovered singletons which are truly singletons. (see also Fig.~\ref{M_halo_gr}).}
	\label{Table_Group_bases}
	\setlength{\tabcolsep}{6.pt}
	\begin{tabular}{@{}l||*{16}{c}r@{}} % four columns, alignment for each 
	\centering
	    $N_{\mathrm{rec}} =$ & 1 & 2 & 3 & 4 & 5 & 6-7 & 8-10 & 11-14 & 15-20 & 21-28 & 29-39 & 40- & mean & $\geq 2$\\
	    \hline
		log$(M_{H,m}/M_{\odot})$ & 11.84 & 12.37 & 12.64 & 12.87 & 12.97 & 13.18 & 13.29 & 13.44 & 13.65 & 14.01 & 13.86 & 14.15 & - & -\\
		\hline
		RMS(log($M_{\mathrm{H}}/M_{H,m})$) & 0.29 & 0.31 & 0.33 & 0.3 & 0.32 & 0.25 & 0.26 & 0.31 & 0.24 & 0.24 & 0.32 & 0.0 & 0.26 & 0.31\\
		\hline
	    mean($f_{c_2} f_{p_2}$) & - & 0.97 & 0.86 & 0.86 & 0.81 & 0.8 & 0.78 & 0.71 & 0.71 & 0.7 & 0.75 & 0.63 & 0.79 & 0.92\\
		\hline
	\end{tabular}
\end{table*}

Based on Fig.~\ref{M_halo_gr} and Table \ref{Table_Group_bases} we can construct the following three further metrics:

\begin{itemize}

\item Metric 3: The median true halo mass $_{2w}^{\ \ 2}M_{H,m}$ of recovered 2-way matched groups with two members. This metric is quite similar to Metric 2.  Again, this sets a scaling for the overall $_{2w}M_{H,m}(N_{\mathrm{rec}})$ relation.

\item Metric 4: The 2-way purity $p_2$ of all recovered groups of two or more members. This metric simply tells us the fraction of recovered groups that have a 2-way match to a real group (of at least two members, by construction). 

\item Metric 5: The mean product of ($f_{c_2} f_{p_2}$) for the 2-way matched groups, averaged over all 2-way matched recovered groups. This metric, which must by definition take values between 0.25 to 1.0, assesses the "quality" of the 2-way matches.  A value of 1.0 would represent perfection, a value of 0.25 would indicate that every 2-way matched group had only just scraped into that class.   
\end{itemize}

\subsection{Scatter in $M_{\mathrm{H}}$}
\begin{figure}
\centering
\includegraphics[width=\columnwidth]{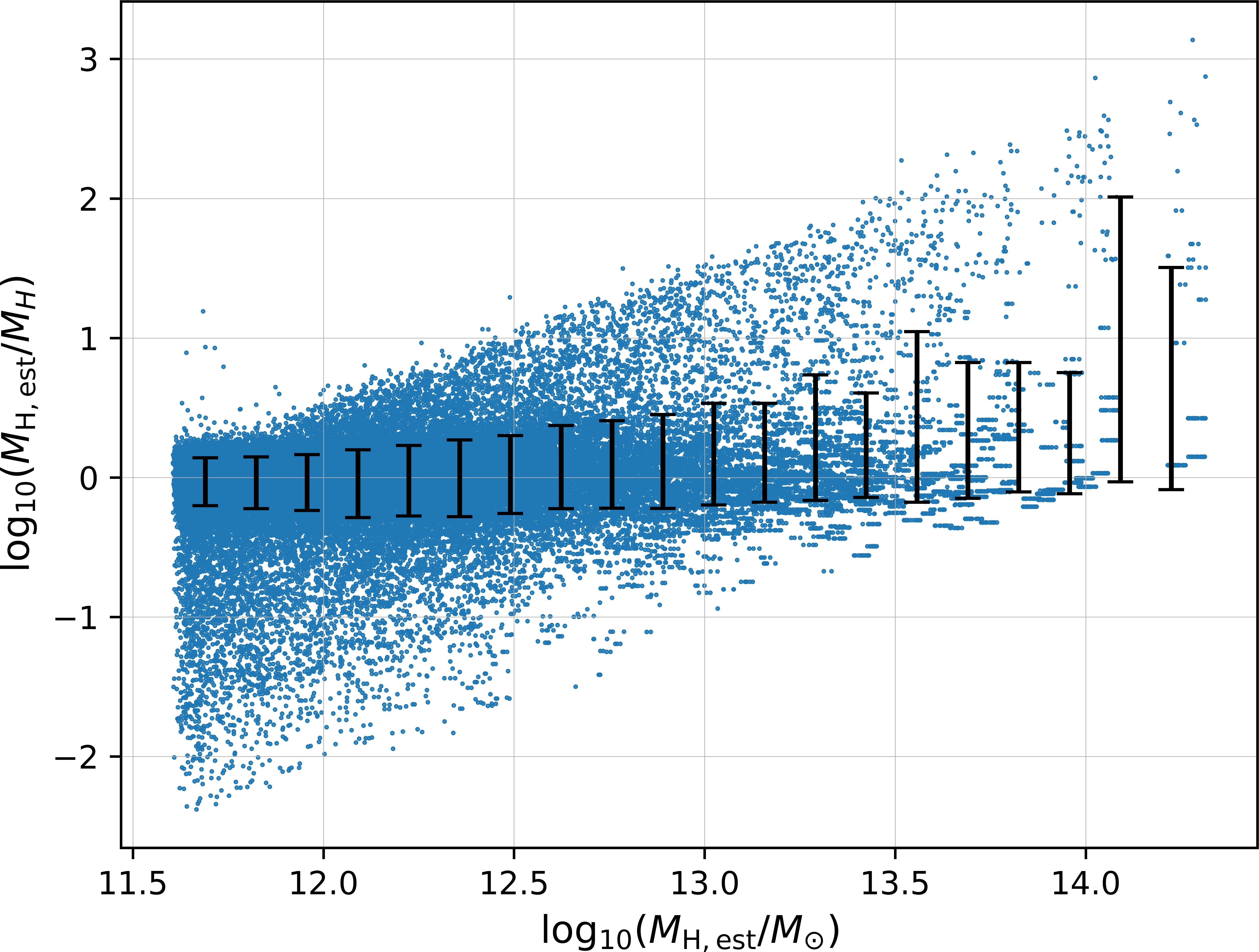}
\caption{The logarithmic difference between the true dark matter halo mass $M_{\mathrm{H}}$ and the estimated (recovered) halo mass $M_{\mathrm{H,est}}$, plotted for all groups on a galaxy basis as a function of $M_{\mathrm{H,est}}$.}\label{M_H_sc_gal}
\end{figure}

\begin{figure}
\centering
\includegraphics[width=\columnwidth]{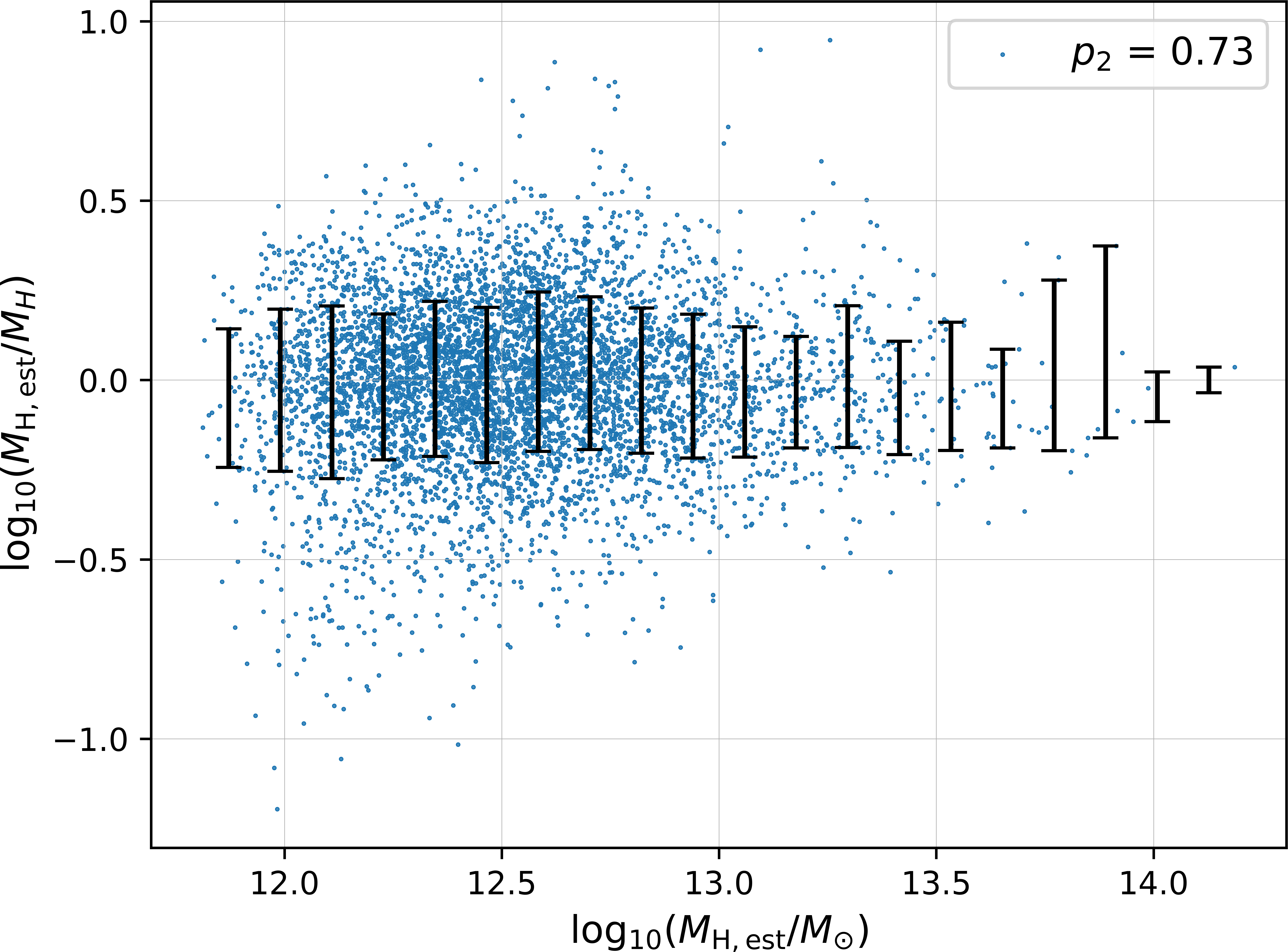}
\caption{The logarithmic difference between the true dark matter halo mass $M_{\mathrm{H}}$ and the estimated (recovered) halo mass $M_{\mathrm{H,est}}$, plotted for all 2-way matched groups, excluding singletons.}
\label{M_H_sc_gr}
\end{figure}

The $M_{\mathrm{H}}$-estimator based on $\sum{M_{\mathrm{stellar}}}$ and $\sigma_v$, presented in Section \ref{M_halo_est}, is superior to the richness $N_{\mathrm{rec}}$ as a halo mass estimator.  
Fig.~\ref{M_H_sc_gal} and Fig.~\ref{M_H_sc_gr} show the difference between the recovered (estimated) mass and the true mass on the galaxy-basis and the group-basis, respectively. As throughout this section, this is for the recovered group catalogue obtained with a mass cut of $\textrm{log}_{10}(M_{\mathrm{stellar}}/M_{\odot}) > 9.5$ and a sampling rate of $ s = 0.8$ in the $1.2 < z < 1.7$ redshift range. While in Fig.~\ref{M_H_sc_gal} all galaxies (also singletons) are included, in Fig.~\ref{M_H_sc_gr} singletons are excluded and furthermore only the 2-way matched groups are considered, as in the previous subsection. In both panels, the black uncertainty bars show the $16 \%$ and $84 \%$ percentile scatter in bins of recovered (estimated) halo mass.  
For the galaxy basis, the prominent locus extending up diagonally to the right arises from very low mass (true) groups that, mostly singletons, that are spuriously over-merged into much larger haloes. The discrepancy between apparent and true halo masses therefore increases towards larger (apparent) halo masses, causing a noticeable increase in the scatter at high masses.  In order to capture this effect, we will define the overall scatter to be the average of the individual error bars (spaced in log apparent halo mass) in Fig.~\ref{M_H_sc_gal}, rather than simply averaging over all galaxies, which would have been dominated by the lowest mass haloes which have the smallest dispersions.  The average uncertainty calculated in this way is $0.40$ dex.  This effect is not present, for obvious reasons, in the group-basis analysis in Fig.~\ref{M_H_sc_gr}, but we adopt the same approach to characterising the scatter in halo masses for consistency.

For the 2-way matched groups, the $M_{\mathrm{H}}$-scatter averages to $0.19$ dex. The large discrepancy between these two dispersions is mainly due to the interloper galaxies falsely merged into larger groups, for which, as discussed above, the true halo masses are significantly overestimated.

The $M_{\mathrm{H}}$-estimator and its application to the recovered group catalogue therefore defines two further metrics:

\begin{itemize}
\item Metric 6: The scatter in halo mass $\sigma_{\mathrm{gal}} \left(M_{\mathrm{H}} \right)$, as given by average of the $16 \%$ and $84 \%$ percentiles, averaged over (logarithmic) bins of recovered estimated halo mass ($M_{\mathrm{H,est}}$) and calculated within each bin on a galaxy basis.
This metric reflects both infidelities of the group-finder (in terms of over-merging and fragmentation, spurious groups and undetected groups), and also the underlying scatter in the halo mass estimator itself.

\item Metric 7: The scatter in halo mass $\sigma_{\mathrm{gr}} \left(M_{\mathrm{H}} \right)$ averaged over all logarithmic $M_{\mathrm{H,est}}$ bins, given by the $16 \%$ and $84 \%$ percentiles for each bin, and calculated within each for 2-way matched groups, 
but excluding the singletons.
This gives a better estimate of the underlying scatter in halo mass, as the 2-way matched avoid, by construction, the worst failures of the group reconstruction.
\end{itemize}

Uncertainties in the estimated stellar masses of galaxies (e.g. due to intrinsic mass-to-light ratio variations in the Universe, or observational uncertainties) may contribute an uncertainty in the halo mass estimates that use, in part, those stellar masses. We have looked at this effect by adding scatter randomly to the stellar masses of the galaxies  in Fig.~\ref{M_H_sc_gal} (Metric 6). Using a 0.1 dex ($1 \sigma$) scatter (see discussion in Section \ref{Introduction}), the increase in scatter in halo mass is undetectable.  Even using a much larger 0.5 dex (random) scatter in galaxy masses the additional scatter in halo mass remains small: the overall scatter increases from 0.40 dex to 0.42 dex.

\subsection{Metrics for Central/Satellite classification}

Whether a galaxy is the central galaxy or a satellite galaxy in its parent halo is thought to be a major driver of its evolution, both in the real universe, and certainly in the L-Galaxies model. Centrals exhibit different physical characteristics than satellites. 
The performance of the group-finder in correctly identifying centrals and satellites is important in galaxy evolution studies.  For both the true groups and the recovered groups, we define for simplicity the central to be the most massive (stellar mass) galaxy in the group and all others to be satellites. Singleton galaxies (whether in the real or recovered catalogues) are therefore always centrals.

It is then a well defined question whether a recovered central (including singletons) is really the central galaxy in its (true) halo and whether a recovered satellite is truly a satellite in its (true) halo.  In short, did the group-finder correctly classify centrals and satellites, accepting singletons as stand-alone centrals.  We define an accuracy $A$ to be the fraction of objects that were correctly classified, i.e. that their classification in the recovered catalogue matches their true classification. 

We stress here that the "true" central/satellite classification of the galaxies is done {\it before} the random sampling of the galaxy sample is done (c.f. the issue of 2-way matched groups discussed in Subsection \ref{2-way_basis}, where the 2-way matching is done by comparing with the "re-sampled membership" of the real group).  Galaxies are therefore classified as real centrals or real satellites based on their stellar-mass ranking amongst all the members of the full real group, not just those that were randomly selected for observation.

\begin{figure}
\includegraphics[width=\columnwidth]{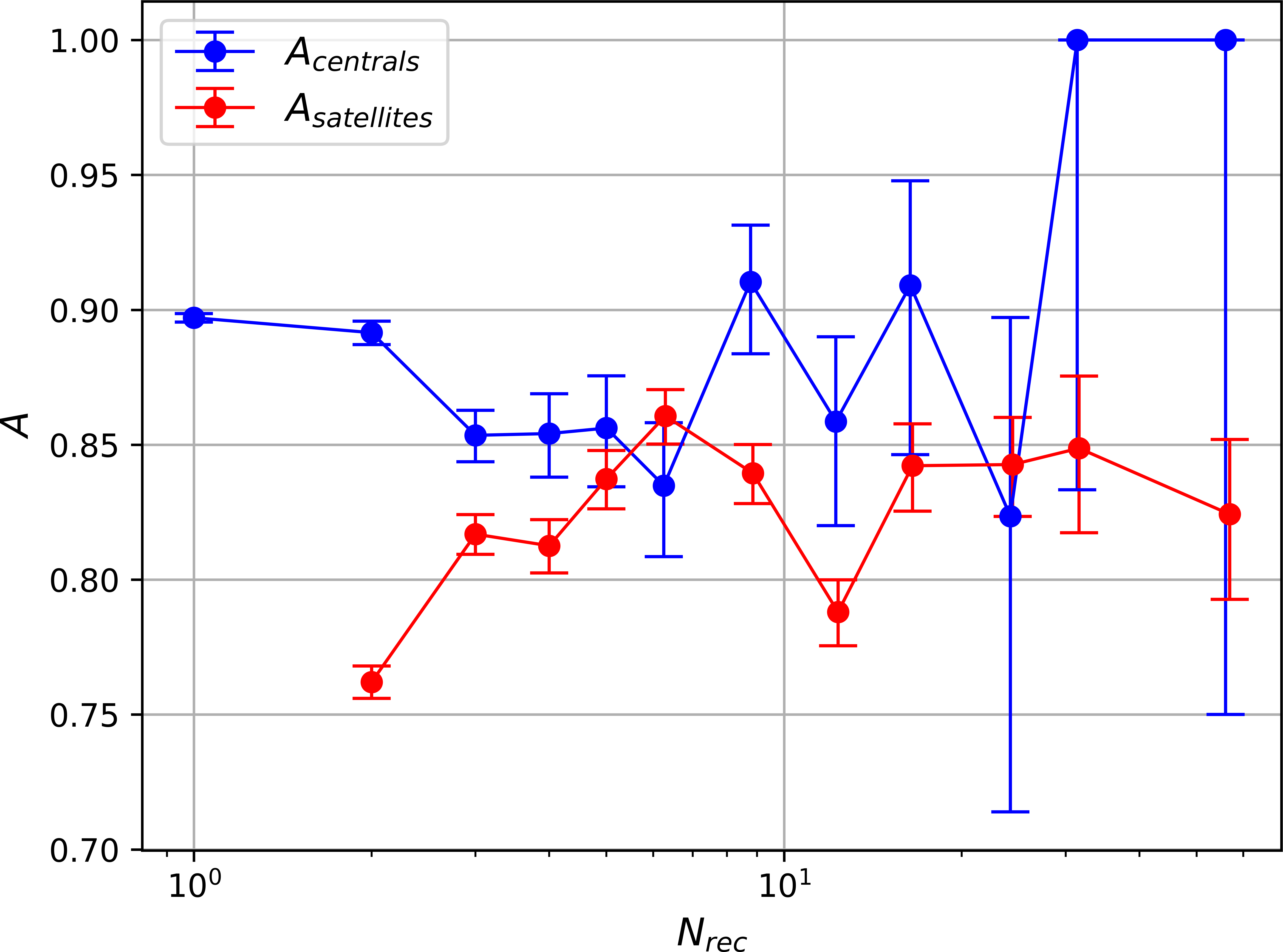}
\caption{The accuracy in classifying centrals and satellites, i.e. the fraction of recovered centrals and satellites that are truly centrals and satellites, as a function of the recovered richness.  Note that $N_{\mathrm{rec}}=1$ therefore refers to recovered singleton galaxies.}
\label{F_M_full}
\end{figure}

Fig.~\ref{F_M_full} shows $A$ for the recovered group catalogue used in this section as a function of the recovered richness  $N_{\mathrm{rec}}$. There is a weak overall trend of decreasing $A_{\mathrm{centrals}}$ with $N_{\mathrm{rec}}$. Singletons ($N_{\mathrm{rec}}=1$) are correctly identified as centrals in $90 \%$ of cases, the accuracy of the central classification in groups of at least two members goes down marginally to $88 \%$ and is even lower for richer groups. At first sight, both of these numbers could be surprisingly high: one might naively expect an accuracy of at most the sampling rate $s$ (in this case is $80 \%$) since this defines the chance that a given central is even observed in the first place.  However, this logic only works for high $N_{\mathrm{rec}}$, since $s$ will then reflect the chance that the true central was actually observed, and, if it wasn't, then for sure a true satellite will be wrongly recovered as the central.  At low $N_{\mathrm{rec}}$ however, this simple logic fails, as shown in Appendix \ref{A}. 

Turning to the satellites, the accuracy in recovering them lies around $80 \%$.  Since a recovered satellite by definition has a more massive galaxy nearby, it can only have been a true central because of infidelities in the group reconstruction, i.e. a real central was wrongly brought into the group through "over-merging".

Because these measurements of $A$ vary little with richness for  $N_{\mathrm{rec}} \geq 2$, we simply average across all the galaxies in the sample and define three further metrics as:

\begin{itemize}
\item Metric 8: The accuracy of recovered centrals for $N_{\mathrm{rec}} = 1$, ${^1}A_{\mathrm{centrals}}$, i.e. for singletons.  
This states the fraction of recovered singletons that are truly centrals (including singletons) rather than truly satellites.
\item Metric 9: The accuracy of centrals in recovered groups with richness $N_{\mathrm{rec}} \geq 2$. This states the fraction of recovered centrals (excluding the recovered singletons) which are indeed truly centrals (or singletons), rather than satellites, averaged across all the recovered centrals.
\item Metric 10: The accuracy of satellites  $A_{\mathrm{satellites}}$. This states the fraction of recovered satellites that are truly satellites (i.e. not truly centrals or singletons), averaged across all recovered satellites.
\end{itemize}

\subsection{Number of groups and multiplicity}

Finally, we come to two other important characteristics of the recovered group catalogue.  The first is simply the number of groups in the recovered catalogue. In our illustrative catalogue ($1.2 < z < 1.7$ redshift range with a mass cut of $M_{\mathrm{stellar}} > 9.5$ and a sampling rate of $ s = 0.8$ and covering $2\times2$ deg$^{2}$) there are 7793 recovered groups with at least two members.  This may also be defined, as desired, as per area of sky, or per comoving volume.

The second is the average multiplicity, by which we mean the average group richness in the catalogue, including singletons (with $N_{\mathrm{rec}} = 1$). The multiplicity tells us how many galaxies there are on average within a (recovered) individual dark matter halo. It therefore captures the amount of "information" in the group catalogue in the sense that the group catalogue loses its usefulness as the multiplicity reduces towards unity, i.e. as more and more galaxies become singletons. In the illustrative group catalogue considered here, the multiplicity is $1.34$.

Hence, we introduce the two last metrics:
\begin{itemize}
\item Metric 11: The number of groups $N_{\mathrm{grs}}$ in the recovered group catalogue. This assesses the size of the group catalogue and could usefully be expressed if desired as per comoving volume or surface area of sky.
\item Metric 12: The multiplicity of the recovered group catalogue. This measures the average richness (number of members) of the structures (including singletons) in the recovered group catalogue.
\end{itemize}

\section{Performance of "halo-based" and F\lowercase{o}F methods at low redshift} \label{Tinker_vs_FoF_main}

We now turn to compare the performance of the "halo-based" and the FoF group finding approaches. We do this at low redshift, using an L-Galaxies SAM light-cone in the redshift range $0.02<z<0.17$ with a flat stellar mass cut of $\textrm{log}_{10}(M_{\mathrm{stellar}}/M_{\odot})=9.0$ but varying the sampling rate in the range of $s=0.1-1.0$. This analysis is conducted at low redshift because the nearby Universe is well known, the SAM is probably most reliable, and the astrophysical knowledge used in the more refined "halo-based" approach is probably better established.  

We focus in this first analysis on the most basic performance metrics assessing the quality of the recovered group catalogues: completeness, purity and the fraction of interlopers $f_I$ in the catalogue, as well as the fraction of galaxies in true groups (of at least two members), $S_{\mathrm{Gal}}$, that are successfully put in a recovered group (of at least 2 members) by the group-finder. In our previous terminology, completeness and purity are "group-based" quantities, meaning that they assess the performance of the group-finder in recovering groups, while $f_I$ and $S_{\mathrm{Gal}}$ are "galaxy-based" quantities.   

\begin{figure}
\includegraphics[width=\columnwidth]{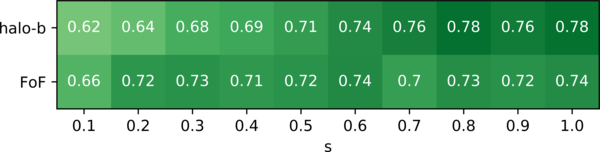}
\caption{Comparison of the 2-way purity $p_2$ in the "halo-based" and FoF recovered catalogues as a function of the sampling $s$. See text for details.}
\label{p_2_Tinker_FoF}
\end{figure}

Fig.~\ref{p_2_Tinker_FoF} shows the 2-way purity $p_2$ (Metric 4) of the "halo-based" and FoF recovered group catalogue over a wide range of random sampling rate $s=0.1-1.0$. There is an overall improvement in the purity with $s$, reaching up to $78 \%$ in the "halo-based" method and $74 \%$ in FoF.  The dependence is quite weak, and even at very low $s$, the purity remains at $62 \%$  for the "halo-based" method and $66 \%$  for FoF.  Overall the two group-finders perform very similarly, with only small differences: As might be expected for a more "refined" approached, the "halo-based" method performs better for complete samples (e.g. $4 \%$ better than FoF at $s=1.0$), while for seriously incomplete samples, $s \lesssim 0.6$, the less refined FoF yields better results. The "halo-based" method is the more sophisticated group-finder in the sense that it uses more information (knowledge of the size of a halo as a function of mass), the strength of the FoF group-finder possibly lies in its use of minimal information. Hence, while the "halo-based" method performs better for complete information, FoF performs better when the amount of available information is degraded by incomplete random sampling. 

\begin{figure}
\includegraphics[width=\columnwidth]{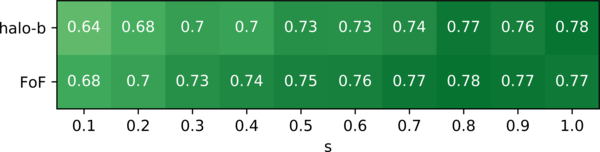}
\caption{2-way completeness $c_2$ of the "halo-based" and FoF methods.}
\label{c_2}
\end{figure}

For the 2-way completeness $c_2$ we observe the same trends, as shown in Fig.~\ref{c_2}: Again, both group-finder perform very similarly, but the "halo-based" method performs slightly better at high $s$ but worse at low $s$: At $s = 1.0$ the "halo-based" method has $c_2 = 78 \%$, while FoF reaches  $c_2 = 77 \%$ but FoF performs better at $s \lesssim 0.8$, declining down to $68 \%$ at $s=0.1$, while the "halo-based" method reduces more rapidly down to $64 \%$. Overall, the 2-way completeness is (perhaps) surprisingly rather stable over this wide range of sampling rate in both group finding approaches.

While $c_2$ and $p_2$ assess the quality of the recovered group catalogue on a group basis, $f_I$ and $S_{\mathrm{Gal}}$ assess the quality of the recovered group catalogue on a galaxy basis, considering galaxies and their environment individually: $f_I$ states the fraction of interlopers in groups, i.e. the fraction of galaxies which are put into a group of at least two members by the group finding algorithm, but are truly singletons; $S_{\mathrm{Gal}}$ states the fraction of galaxies truly populating groups of at least two members, which are recovered in groups of at least two members.

\begin{figure}
\includegraphics[width=\columnwidth]{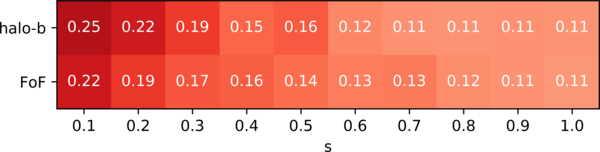}
\caption{Comparison of the fraction of interlopers $f_I$ in the "halo-based" and FoF methods, as a function of the sampling $s$.}
\label{f_I}
\end{figure}

Fig.~\ref{f_I} shows that the fraction of interlopers $f_I$ increases from $11 \%$ in the "halo-based" and FoF methods for $s=1.0$ full sampling, up to $22 \%$ for FoF and $25 \%$ for the "halo-based" method with a $s=0.1$ sampling rate. Again, while they perform comparably at high sampling rate, FoF performs slightly better for low sampling rate. Overall, the fraction of interlopers $f_I$ changes significantly with $s$.

\begin{figure}
\includegraphics[width=\columnwidth]{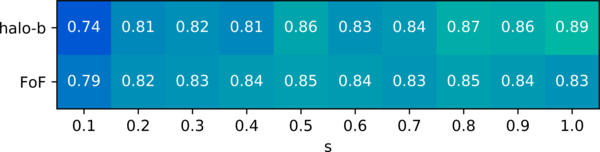}
\caption{Comparison of the fraction of galaxies correctly recovered in groups, $S_{\mathrm{Gal}}$, in the "halo-based" and FoF methods, as a function of the sampling $s$.}
\label{S_Gal}
\end{figure}

The fraction of true galaxies successfully recovered in groups is shown in Fig.~\ref{S_Gal}. Again, the "halo-based" method performs significantly better at $s=1.0$ with $S_{\mathrm{Gal}}=89 \%$, compared to  $S_{\mathrm{Gal}}=83 \%$ for FoF, while at low $s$ for "halo-based" method this  reduces to $74 \%$ while FoF declines only to $79 \%$. As for the other quality metrics, FoF performs better at low sampling rate, while the "halo-based" method performs better at high sampling rate.

The 12 science metrics can be found in Appendix \ref{12_science_metrics_Tinker_FoF}. We find that most metrics are virtually indistinguishable at the same sampling rate $s$, except the Metrics 6 (Fig.~\ref{sigma_gal_Tinker_FoF}) and Metric 7 (Fig.~\ref{sigma_gr_Tinker_FoF}) concerning the scatter in dark matter halo mass metrics on a galaxy-basis and on a group-basis, respectively. While the estimated dark matter halo mass in the "halo-based" catalogues is recovered by the built-in halo mass estimator in the group finding algorithm, for the FoF catalogues we use the halo mass estimator presented in Subsection \ref{M_halo_est} based on total stellar mass and velocity dispersion. As in the basic statistical metrics completeness and purity, FoF exhibits the slightly lower dark matter halo mass scatter $\sigma \left(M_{\mathrm{H}} \right)$ at low $s$ than the "halo-based" method, on both, galaxy-basis and group-basis. 

As a general conclusion, both of the two group-finders compared here perform rather similarly, despite their quite different conceptual approaches. In detail, the "halo-based" approach, using more external information (e.g. the halo size-mass relation), performs slightly better when the observational information level is high, from high sampling $s \approx 1$.  But as the completeness of the observational information reduces $s<1$, we find that the resulting degradation of group-finding performance is less for FoF, and it ends up performing slightly better than the "halo-based" approach at low sampling rates. 

For the rest of the paper we will focus on the expected performance of group finding algorithms at high redshift.  Since the sampling rate is unlikely to be extremely high, and because of the comparable overall performance of the two approaches established in this section, we will for simplicity henceforth focus on the FoF approach alone. 

\section{Performance of F\lowercase{o}F catalogues at high redshift in $\lowercase{s}/M_{\mathrm{\lowercase{stellar}}}$-space} \label{Science_metrics_M_s_space}

\subsection{The redshift range $1.2<z<1.7$}

In this section, we investigate how the fidelity of the recovered FoF group catalogues obtained from the L-Galaxies mock at high redshift depends on the lower stellar mass cuts, over the range $9.1 \leq \textrm{log}_{10}(M_{\mathrm{stellar}}/M_{\odot}) \leq 10.5$, and on the random sampling rate in the range $0.1 \leq s\leq 1.0$. The former produces a wide range in comoving density of the potential tracers, while the latter multiplies this to yield the actual final number density of tracers.  In the plots to follow, the thin red curves represent loci of constant tracer number density. 

We first focus on the 12 "science metrics" that we introduced in Section \ref{science_metrics} that are obtained in the redshift range $1.2<z<1.7$, before then briefly examining how the results differ with redshift, looking at two other redshift intervals by examining slightly lower $0.9<z<1.1$ and higher $2.0<z<2.6$ redshifts.

\begin{figure}
\centering
\includegraphics[width =\y\columnwidth]{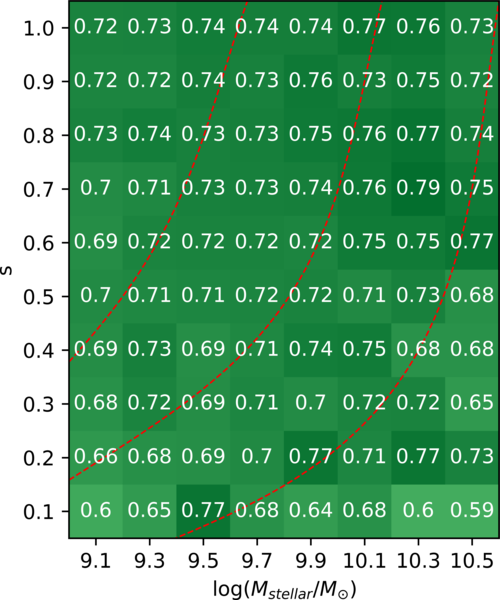}
\caption{Metric 4: The purity $p_2$. Changing $M_{\mathrm{stellar}}$ and $s$ does not significantly change $p_2$. See text for discussion. The red lines, reproduced from Fig.~\ref{N_obj_pp} later in this paper, show the constant (comoving) number density of available (post-sampling) tracers.} \label{p_2_1_2__1_7}
\end{figure}

Returning to $1.2<z<1.7$, Fig.~\ref{p_2_1_2__1_7} first presents the basic 2-way purity $p_2$ (Metric 4), the fraction of recovered groups with a 2-way match to a real group. It is therefore one of the most basic measures of the overall fidelity of the group catalogue.  It should however be remembered that it is also one of the more artificial metrics because it compares the recovered groups to the "resampled" reality obtained {\it after} the application of the random sampling $s$ to the full mock sample. Probably for this reason, $p_2$ is found to be largely independent of both $M_{\mathrm{stellar}}$ and $s$. Despite being quite noisy, values between $\approx 60-80 \%$ are found, with many close to 0.72, and there is no obvious trend across the diagram.   Once optimized, the FoF algorithm is evidently able to reconstruct the underlying "re-sampled" group population more or less independently of how much this has been degraded from the full underlying population by the incomplete sampling.  

It should be noted that this same characteristic value was also found in the SDSS-like comparison of the FoF and "halo-based" approaches (see Section \ref{Tinker_vs_FoF_main} above) and therefore reflects a fairly robust measure of the performance of optimized algorithms of different sorts.  That this $p_2$ purity Metric 4 nevertheless has a more or less uniform value that lies well below unity (i.e. perfection) is almost certainly a largely unavoidable consequence of the reduced phase-space information on galaxy positions (2 spatial positions and 1 velocity measure) that is available to the astronomer trying to use galaxy positions as tracers of underlying structure. 

\begin{figure}
\centering
\includegraphics[width=\y\columnwidth]{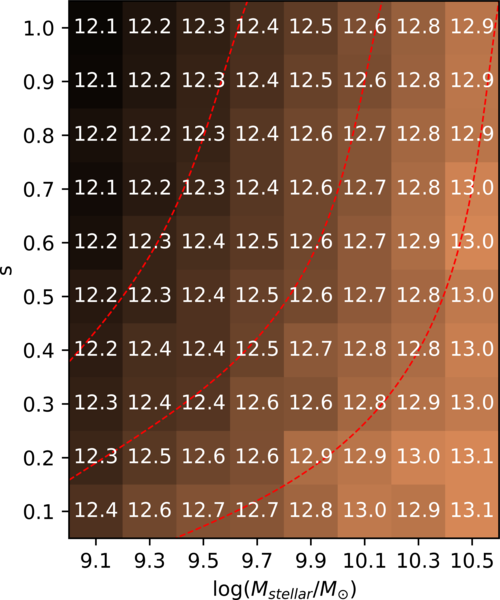}
\caption{Metric 2: The true median halo mass $^{2}M_{H,m}$ of galaxies which are recovered to be members of richness 2 groups depends primarily on the number density of tracers (indicated by the red lines).} 
\label{log_M_H_m_N_rec_2_galb}
\end{figure}

\begin{figure}
\centering
\includegraphics[width=\y\columnwidth]{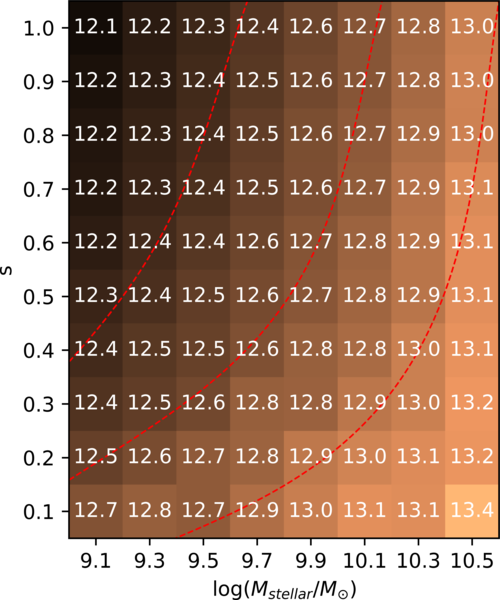}
\caption{Metric 3: The median halo mass $_{2w}^{\ \ 2}M_{H,m}$ of 2-way matched recovered groups of richness 2 depends mostly on the number density of tracers (indicated by the red lines).} \label{log_M_H_m_N_rec_2_grb}
\end{figure}

\begin{figure}
\centering
\includegraphics[width=\y\columnwidth]{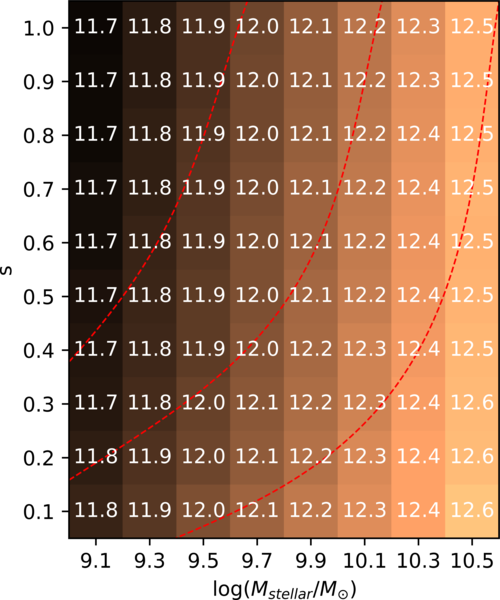}
\caption{Metric 1: The median halo mass ${^1}M_{H,m}$ of singletons depends mainly on the mass cut $M_{\mathrm{stellar}}$. This is due to a dominant fraction of singletons: Singletons after sampling were likely to be singletons in the original full catalogue.} \label{log_M_H_m_N_rec_1_galb}
\end{figure}

In Figures ~\ref{log_M_H_m_N_rec_2_galb}-\ref{log_M_H_m_N_rec_1_galb}, we show the changes in, respectively, the median halo mass for recovered $N_{\mathrm{rec}} = 2$ structures on a galaxy basis (Metric 2), for the 2-way matched recovered 2-member groups (Metric 3), and for haloes populated with only a recovered singleton (Metric 1). 

For the first two, the overall trend is for the mass of 2-member groups to decrease with both increasing sampling rate and with decreasing limiting stellar mass (towards the upper left of the figures). These two parameters determine the (comoving) number density of the tracers that are available for the group reconstruction (i.e. after the observational sampling has been applied to the tracer target population).  As noted above, the red lines in the figures, and in all later figures in this section, show the lines of constant (comoving) number density of the available (post-sampling) tracers:  Not surprisingly, increasing the number density of these tracers enables lower mass structures to be identified as multi-member groups.  

To first order, this gain is independent of whether the number density is increased by lowering $M_{\mathrm{stellar}}$ or by raising $s$. The fact that these metrics largely reflect the resulting final number density of tracers rather than the actual mass limit strongly suggests that these results will also be valid for other tracer selection criteria, e.g. selection by flux in some band. Any “fuzziness” of the mass-selection limit (at constant number density of tracers) due to uncertainties in the individual masses of galaxies is unlikely to have a large effect on these metrics once the number of density of tracers is fixed. 

The dependence of these two Metrics 2 and 3 on $s$ weakens at high $s$ due to a survivor effect. Most of the recovered 2-member groups, after the sampling, are truly 2-member groups and are just the "lucky survivors" in which both members were selected in the random sampling.

This "survivor-effect" is even more clearly seen for the singletons in Fig.~\ref{log_M_H_m_N_rec_1_galb}: Nearly all the recovered singletons, even after sampling, are truly singletons in the (full) mock catalogue, rather than being the result of incompletely sampled groups.  As $s$ is reduced, fewer galaxies (including singletons) are observed, but most of the recovered singletons would still have been been singletons with higher $s$.  This is why there is very little change in the median halo mass of singletons with $s$.   As a result, $^1 M_{H,m}$ is primarily affected by the stellar mass cut $M_{\mathrm{stellar}}$ alone and is found to be roughly proportional to $M_{\mathrm{stellar}}^{0.5}$, as expected from the cosmic halo-stellar mass relation in the mock (and real) universe.

\begin{figure}
\centering
\includegraphics[width =\y\columnwidth]{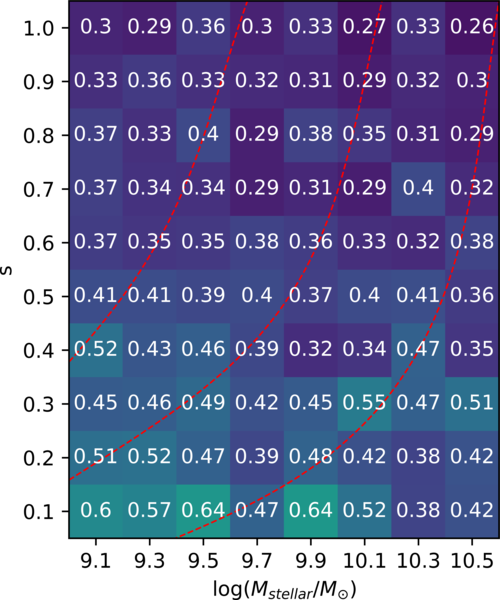}
\caption{Metric 6: This plot shows the average scatter in halo mass $\sigma_{\mathrm{gal}} \left(M_{\mathrm{H}} \right)$ of the galaxy groups (given by the average of the $16 \%$ and $84 \%$ percentiles, averaged across all the logarithmic bins in mass), considered on a galaxy basis.  This depends primarily on $s$.} \label{av_per_span_gal_otrr}
\end{figure}

\begin{figure}
\centering
\includegraphics[width =\y\columnwidth]{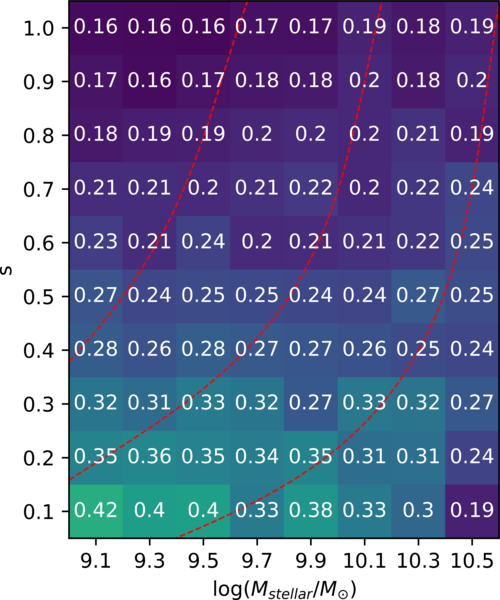}
\caption{Metric 7: The average scatter in halo mass $\sigma_{\mathrm{gr}} \left(M_{\mathrm{H}} \right)$ of the galaxy groups (given by the average of the $16 \%$ and $84 \%$ percentiles averaged across all the logarithmic bins in mass) of the 2-way matched groups. This also depends primarily on $s$.} \label{av_per_span_gr}
\end{figure}

\begin{figure}
\centering
\includegraphics[width =\y\columnwidth]{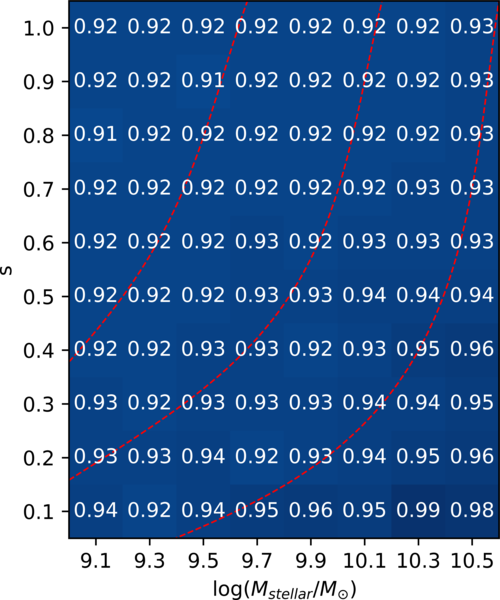}
\caption{Metric 5: The fidelity of the group-finder given as $f_{c_2} f_{p_2}$.  This is almost completely independent of both $s$ and the number of tracers.  However, it should be noted that this metric compares the recovered catalogue with the resampled true catalogue, i.e. it does not include any degradation that arises because of the random sampling.} \label{mean_f_f_group_z_1_2__1_7}
\end{figure}

Two further metrics focus on the uncertainty (i.e. scatter) in halo mass $\sigma \left(M_{\mathrm{H}} \right)$, averaged over bins in recovered halo mass. This scatter is calculated on a galaxy basis (Metric 6), see Fig.~\ref{av_per_span_gal_otrr}, and for 2-way matched groups on a group basis (Metric 7), see  Fig.~ \ref{av_per_span_gr}. 

As noted above, there will always be an intrinsic scatter in $M_{\mathrm{H}}$ in any recovered group catalogue, however good it is, because of underlying astrophysical or cosmological scatter in the galaxy content of haloes.  This underlying scatter is of course also present in the true "mock" catalogue. The uncertainty in halo mass in any recovered group catalogue is a result of (a) the almost irreducible scatter in the Universe and (b) of failures in the group-finding.

In order to assess the relative importance of these two effects we ran the halo mass estimator presented in Subsection \ref{M_halo_est} on perfectly sampled true group catalogues. We found that the intrinsic scatter varied from $0.15 \ \mathrm{dex}$ for a limiting stellar mass of $\textrm{log}_{10}(M_{\mathrm{stellar}}/M_{\odot}) = 9.1$ to $0.20 \ \mathrm{dex}$ for $\textrm{log}_{10}(M_{\mathrm{stellar}}/M_{\odot}) = 10.5$. By comparing these numbers to the top line of Fig.~ \ref{av_per_span_gal_otrr} one can see that the scatter on a galaxy basis is by a factor of $\approx2$ higher in the reconstructed catalogues compared to the true catalogues. Accordingly, the variance changes by a factor of $\approx4$. Hence, the failures in group-finding dominate over the intrinsic scatter.

For both metrics, 6 and 7, the scatter in the recovered halo mass shows very little change with $M_{\mathrm{stellar}}$, but a significant gradient with $s$. As $s$ increases from $0.5-1.0$, the scatter in halo mass decreases from $\approx 0.4-0.3 \ \mathrm{dex}$ on a galaxy basis, and from $\approx 0.25-0.18 \ \mathrm{dex}$ on a group basis. These changes are not negligible: the variance changes by a factor of $\approx 1.8$. The variance represents the sum of the contributions from (i) the astrophysical/cosmological scatter, (ii) the irreducible scatter due to the incomplete phase-space information even at full sampling, plus (iii) the additional scatter arising from the variation in the fidelity of the group catalogue with $s$. Note that we are here considering the fidelity of the recovered groups relative to the full true mock catalogue, not to the resampled mock catalogue that was discussed in connection with Metric 4. That the variance almost doubles when $s$ goes from $1.0$ to $0.5$ emphasizes the relative importance of the last term (iii) as $s$ is reduced.

In contrast, Fig.~ \ref{mean_f_f_group_z_1_2__1_7} shows $f_{c_2} f_{p_2}$ (Metric 5), which reflects the mean "quality" of the 2-way matches. The mean of $f_{c_2} f_{p_2}$ hardly changes at all with $s$ and $M_{\mathrm{stellar}}$ and has a more or less constant high value of $\approx 0.92$.  As with Metric 4 above (see Fig.~\ref{p_2_1_2__1_7}, this also reflects the robust performance of the FoF group-finder relative to the "resampled" reality, even as that degrades relative to the full unsampled reality.  Again, the difference with Metrics 6 and 7 is noticeable.

\begin{figure}
\centering
\includegraphics[width=\y\columnwidth]{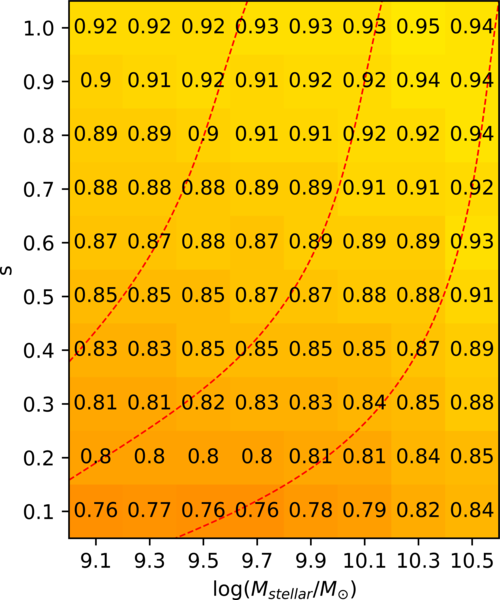}
\caption{Metric 8: The accuracy of central classification for singletons in the recovered group catalogue ${^1}A_{\mathrm{centrals}}$ for $N_{\mathrm{rec}} = 1$. The performance of the group-finder in recovering those improves with the sampling rate $s$ and slightly with a more restrictive (increasing) mass cut $M_{\mathrm{stellar}}$.} \label{A_singletons}
\end{figure}

\begin{figure}
\centering
\includegraphics[width = \y\columnwidth]{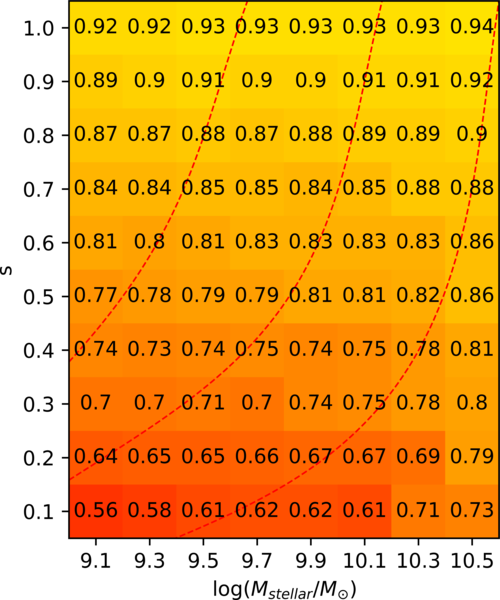}
\caption{Metric 9: The accuracy of the centrals $^{\geq2} A_{\mathrm{centrals}}$, i.e. the fraction of recovered centrals (excluding singletons), which are truly centrals. This improves quite significantly with sampling rate $s$.} \label{A_centrals}
\end{figure}

\begin{figure}
\centering
\includegraphics[width =\y\columnwidth]{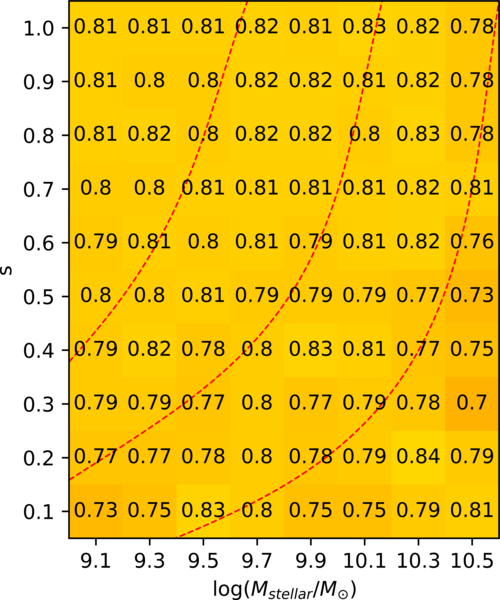}
\caption{Metric 10: The accuracy of the satellites $A_{\mathrm{satellites}}$ states the fraction of recovered satellites, which are truly satellites. The accuracy of the satellite classification is largely independent of both the mass cut $M_{\mathrm{stellar}}$ and the sampling rate $s$.} \label{A_satellites}
\end{figure}

\begin{figure}
\centering
\includegraphics[width =\y\columnwidth]{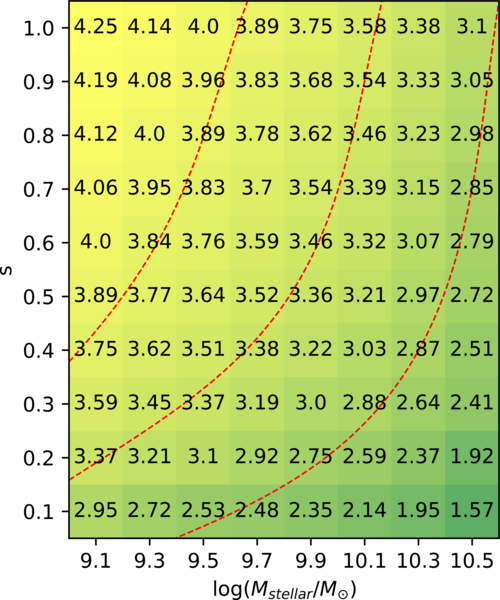}
\caption{Metric 11: Number of recovered groups $N_{\mathrm{grs}}$ (in log) of at least 2 members.} \label{Gr_N_l2_rec_pA_z_1_2__1_7}
\end{figure}

\begin{figure}
\centering
\includegraphics[width=\y\columnwidth]{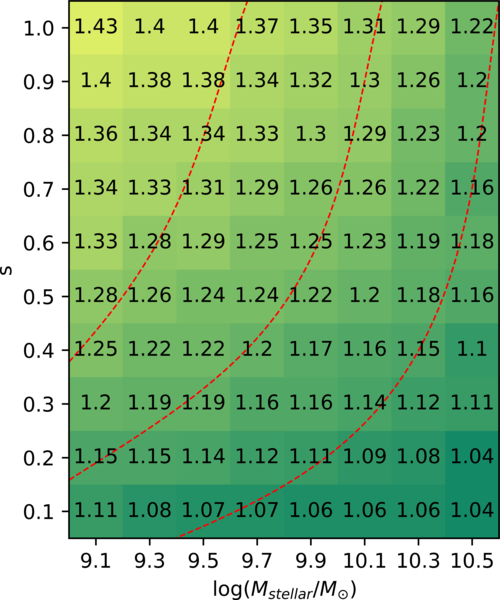}
\caption{Metric 12: The multiplicity $m$, which is the average richness of recovered galaxy groups (including singletons). The multiplicity decreases with decreasing sampling rate $s$ and increasing mass cut $M_{\mathrm{stellar}}$ and largely follows the number density of tracers (given by the red lines).} \label{Multiplicity}
\end{figure}

We next look at the performance of the group-finder in accurately classifying centrals (Fig.~\ref{A_centrals}), singletons  (Fig.~\ref{A_singletons}) and satellites (Fig.~\ref{A_satellites}) correctly.  For the centrals it will be recalled that we distinguish between centrals of recovered groups of at least 2 members (Metric 9), and the recovered singletons (Metric 8). 

We first note that the accuracy of the classification of centrals $^{\geq2} A_{\mathrm{centrals}}$ in the ($N_{\mathrm{rec}} \geqslant 2$)-structures improves very significantly with higher sampling rate $s$ (and much more weakly with $M_{\mathrm{stellar}}$), from a minimum of $ 56 \%$ at a sampling rate of $s = 0.1$ up to maximally $ 94 \%$ at $s = 1.0$. It is noticeable that the accuracy of classifying singletons ${^1}A_{\mathrm{centrals}}$ (i.e the centrals of ($N_{\mathrm{rec}} = 1$)-structures) is better than for ($N_{\mathrm{rec}} \geqslant 2$)-structures.  This is because, as noted above, most of the recovered singletons are the "lucky survivors" of the sampling and are true singletons in the original full mock catalogue, as opposed to being degraded 2- or more member groups structures, in which there is a danger of incorrect classification (the true unsampled group catalogue is dominated by small richness structures).  The parameter ${^1}A_{\mathrm{centrals}}$ goes from a minimum $76 \%$ at a sampling rate of $s = 0.1$ up to a maximum $95 \%$ at $s = 1.0$.  

It can be seen that the accuracy of central classification, both for $N_{\mathrm{rec}} = 1$ singletons and for the centrals of richer groups, improves slightly as $M_{\mathrm{stellar}}$ is increased.  This is due to the decreasing multiplicity of the overall group catalogue and the associated increase in the "survivor-effect", since more galaxies will be true singletons in the original (true) sample. 

In contrast to the centrals, the accuracy of the satellite classification $A_{\mathrm{satellites}}$ is remarkably constant at $\sim 80 \%$.  Since recovered satellites are by definition ranked below other more massive galaxies in the recovered groups, errors in satellite classification will only occur due to errors in assigning galaxies to groups, i.e. the basic underlying fidelity of the group reconstruction in terms of completeness and purity.  As shown for the Metrics 4 (Fig.~\ref{p_2_1_2__1_7}) and 5 (Fig.~\ref{mean_f_f_group_z_1_2__1_7}), this is only very weakly dependent on $s$ and $M_{\mathrm{stellar}}$.

Fig.~\ref{Gr_N_l2_rec_pA_z_1_2__1_7} shows the number of recovered groups of at least 2 members (Metric 11). This number is quite sensitive to $s$ and $M_{\mathrm{stellar}}$ and, as might be expected, follows quite closely the red lines of constant number of tracers, independent on whether this density is achieved through the initial selection of the tracers or their random sampling rate. 

Finally, we show in  Fig.~ \ref{Multiplicity} how the average group multiplicity (Metric 12) changes with $s$ and $M_{\mathrm{stellar}}$.  As might be expected, the overall trend is the same as for the number of recovered groups, and largely follows the number density of available tracers after the application random sampling.

\subsection{The variation with redshift}
\label{three_redshift_ranges}

\begin{figure*}
\centering
\hspace{\centpara mm}
\begin{subfigure}{\x \columnwidth}
\includegraphics[width=\columnwidth]{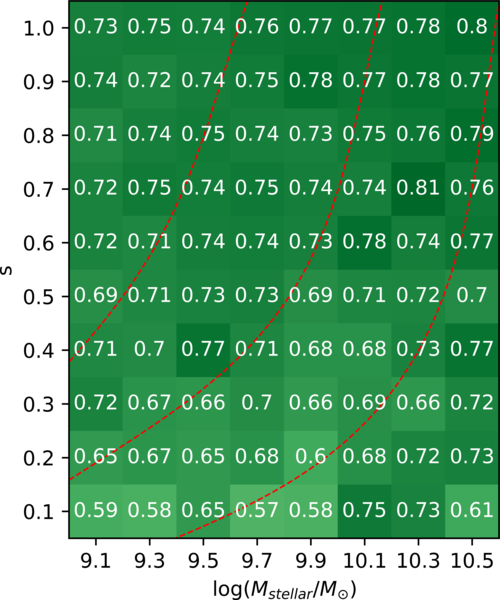}
\captionsetup{width= \columnwidth, skip= \z pt}
\caption{ $p_2$} 
\end{subfigure}
\hspace{\hdist mm}
\begin{subfigure}{\x \columnwidth}
\includegraphics[width=\columnwidth]{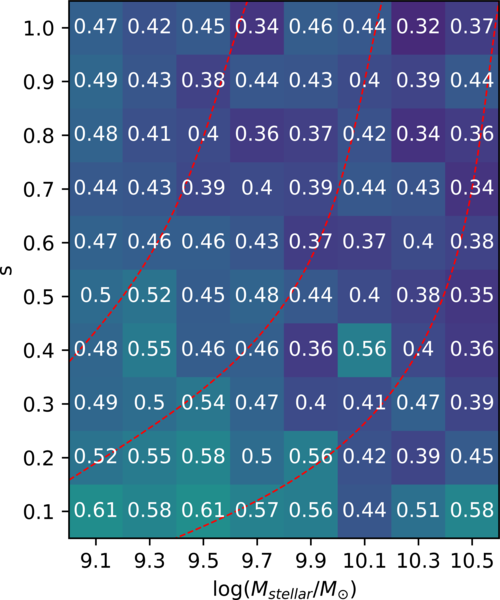}
\captionsetup{width=\columnwidth,skip = \z pt}
\caption{$\sigma_{\mathrm{gal}} \left(M_{\mathrm{H}} \right)$} 
\end{subfigure}
\hspace{\hdist mm}
\begin{subfigure}{\x \columnwidth}
\includegraphics[width=\columnwidth]{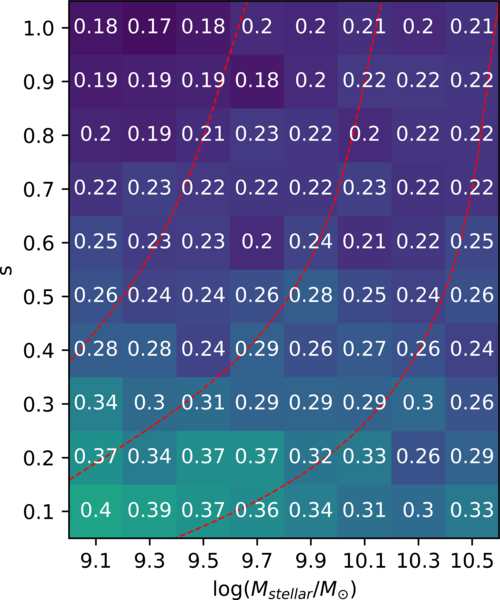}
\captionsetup{width=\columnwidth,skip = \z pt}
\caption{$\sigma_{\mathrm{gr}} \left(M_{\mathrm{H}} \right)$} 
\end{subfigure}
\vspace{\vdist mm}
\newline

\centering
\hspace{\centpara mm}
\begin{subfigure}{\x\columnwidth}
\includegraphics[width=\columnwidth]{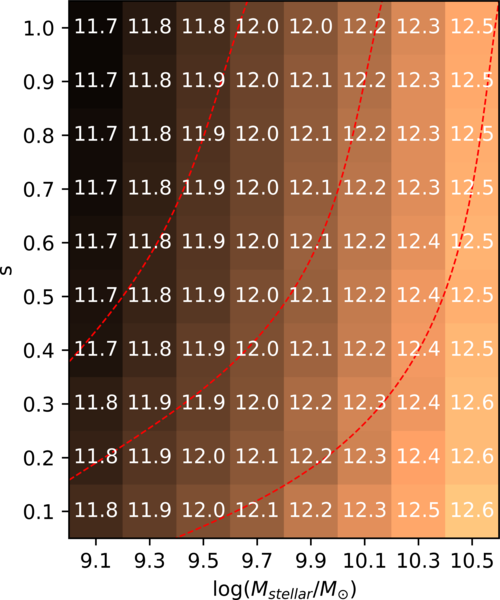}
\captionsetup{width=\columnwidth,skip = \z pt}
\caption{${^1}M_{H,m}$} 
\end{subfigure}
\hspace{\hdist mm}
\begin{subfigure}{\x \columnwidth}
\includegraphics[width=\columnwidth]{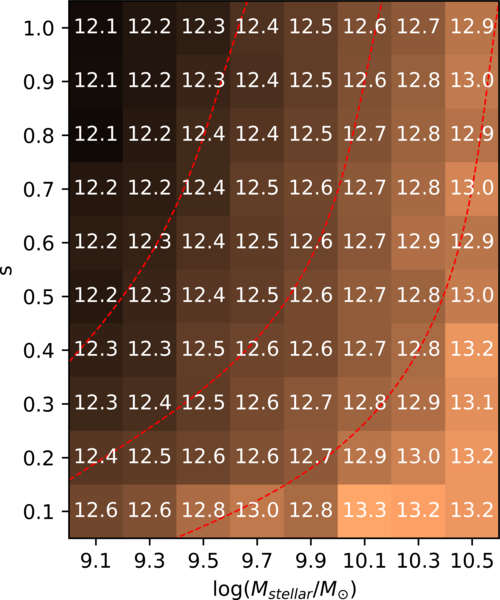}
\captionsetup{width=\columnwidth,skip = \z pt}
\caption{$^{2}M_{H,m}$} 
\end{subfigure}
\centering
\hspace{\hdist mm}
\begin{subfigure}{\x \columnwidth}
\includegraphics[width=\columnwidth]{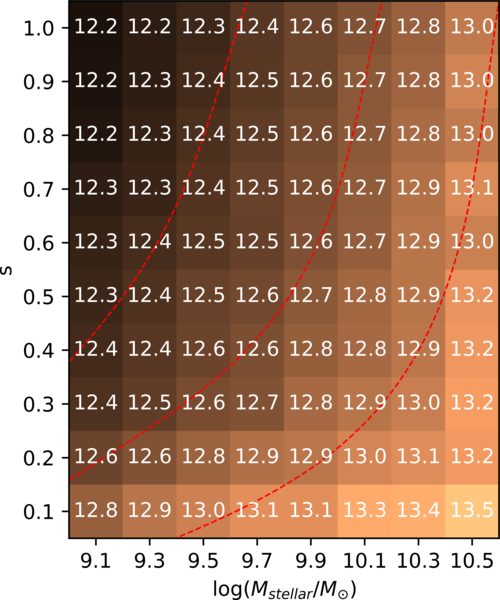}
\captionsetup{width=\columnwidth,skip = \z pt}
\caption{$_{2w}^{\ \ 2}M_{H,m}$} 
\end{subfigure}
\vspace{\vdist mm}
\newline

\centering
\hspace{\centpara mm}
\begin{subfigure}{\x \columnwidth}
\includegraphics[width=\columnwidth]{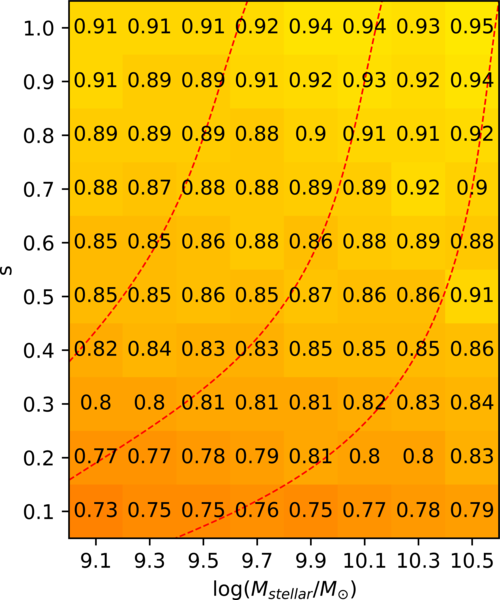}
%\vspace{\vdist mm}
\captionsetup{width=\columnwidth,skip = \z pt}
\caption{${^1}A_{\mathrm{centrals}}$} 
\end{subfigure}
\hspace{\hdist mm}
\begin{subfigure}{\x \columnwidth}
\includegraphics[width=\columnwidth]{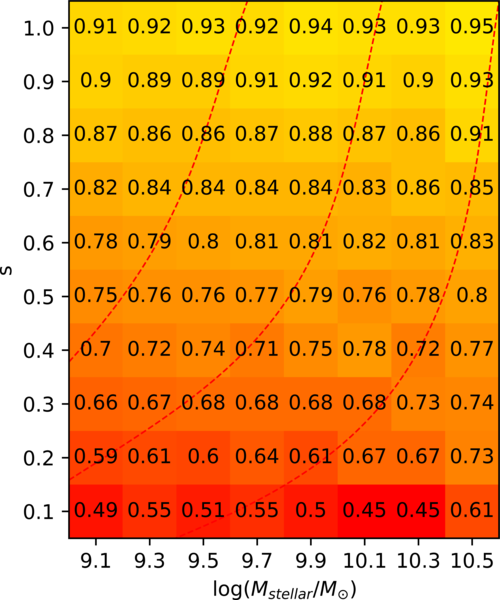}
\captionsetup{width=\columnwidth,skip = \z pt}
\caption{$^{\geq2} A_{\mathrm{centrals}}$} 
\end{subfigure}
\hspace{\hdist mm}
\begin{subfigure}{\x \columnwidth}
\includegraphics[width=\columnwidth]{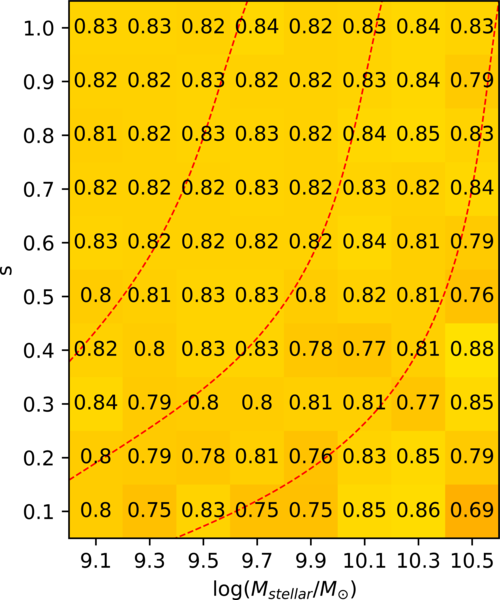}
\captionsetup{width=\columnwidth,skip = \z pt}
\caption{$A_{\mathrm{satellites}}$} 
\end{subfigure}
\vspace{\vdist mm}
\newline

\centering
\hspace{\centpara mm}
\begin{subfigure}{\x \columnwidth}
\includegraphics[width=\columnwidth]{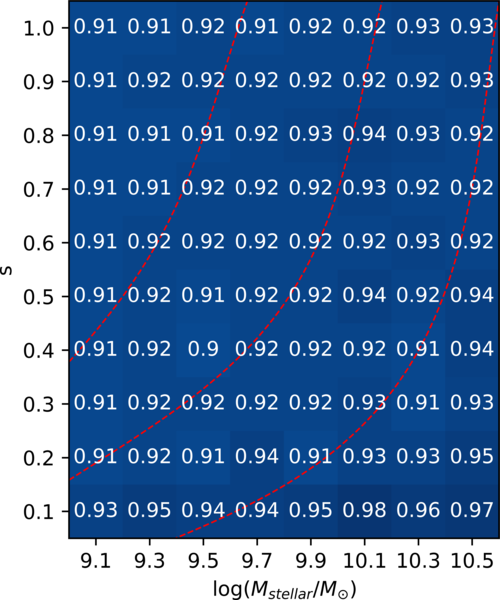}
\captionsetup{width=\columnwidth,skip = \z pt}
\caption{$f_{c_2} f_{p_2}$} 
\end{subfigure}
\hspace{\hdist mm}
\begin{subfigure}{\x \columnwidth}
\includegraphics[width=\columnwidth]{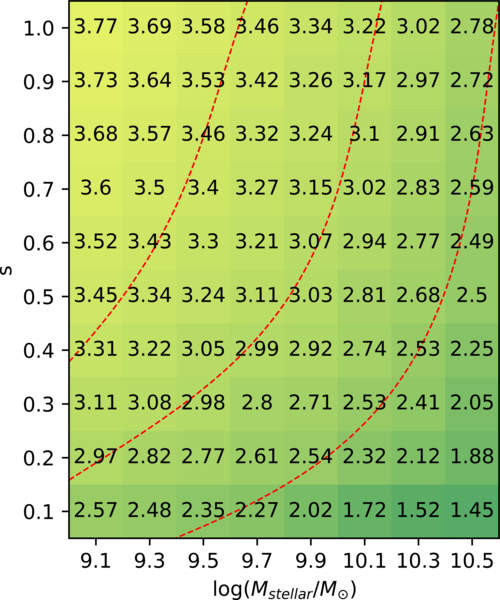}
\captionsetup{width=\columnwidth,skip = \z pt}
\caption{log$_{10}(N_{\mathrm{grs}})$} 
\end{subfigure}
\hspace{\hdist mm}
\begin{subfigure}{\x \columnwidth}
\includegraphics[width=\columnwidth]{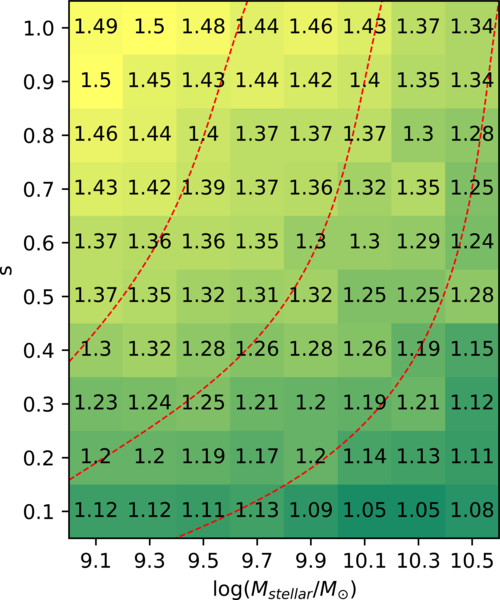}
\captionsetup{width=\columnwidth,skip = \z pt}
\caption{$m$} 
\end{subfigure}
\vspace{\vdist mm}
\newline
\centering
\caption{Metrics of the FoF group-finder for $0.9 < z < 1.1$.}
\label{09_11}
\end{figure*}

\begin{figure*}
\centering
\hspace{\centpara mm}
\begin{subfigure}{\x \columnwidth}
\includegraphics[width=\columnwidth]{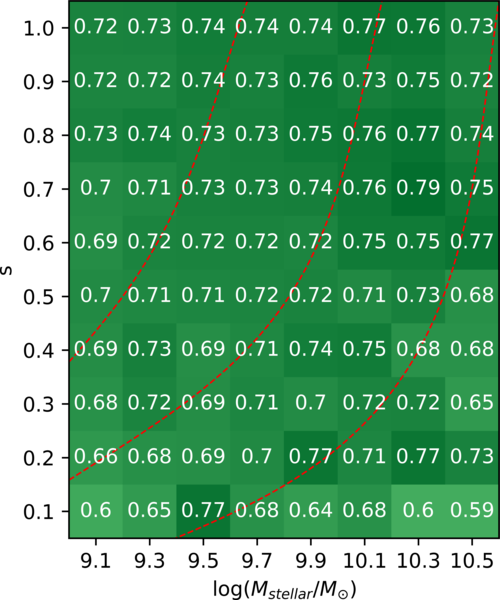}
\captionsetup{width=\columnwidth,skip = \z pt}
\caption{ $p_2$} 
\end{subfigure}
\hspace{\hdist mm}
\begin{subfigure}{\x \columnwidth}
\includegraphics[width=\columnwidth]{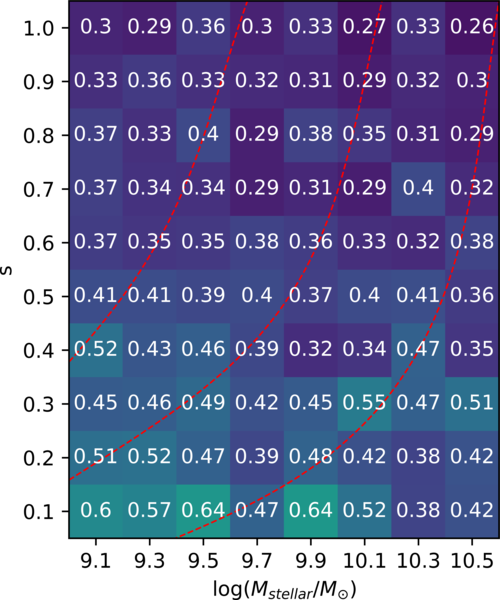}
\captionsetup{width=\columnwidth,skip = \z pt}
\caption{$\sigma_{\mathrm{gal}} \left(M_{\mathrm{H}} \right)$} 
\end{subfigure}
\hspace{\hdist mm}
\begin{subfigure}{\x \columnwidth}
\includegraphics[width=\columnwidth]{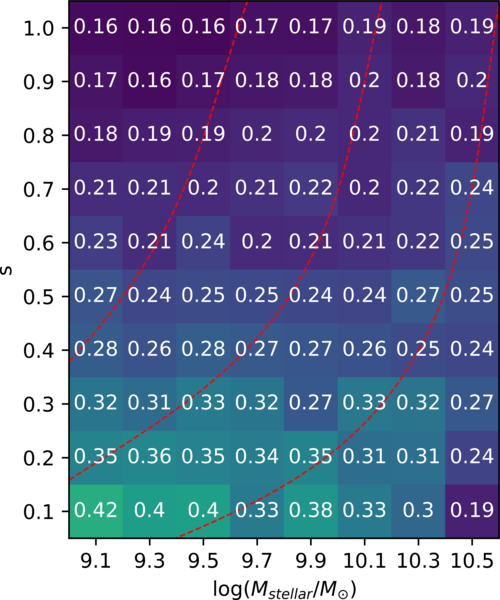}
\captionsetup{width=\columnwidth,skip = \z pt}
\caption{$\sigma_{\mathrm{gr}} \left(M_{\mathrm{H}} \right)$} 
\end{subfigure}
\vspace{\vdist mm}
\newline

\centering
\hspace{\centpara mm}
\begin{subfigure}{\x\columnwidth}
\includegraphics[width=\columnwidth]{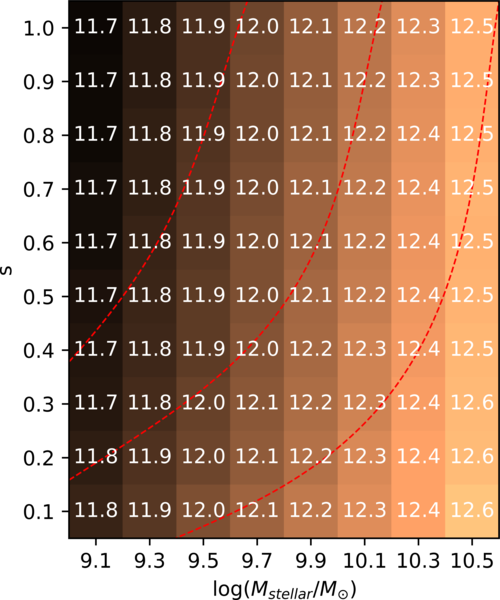}
\captionsetup{width=\columnwidth,skip = \z pt}
\caption{${^1}M_{H,m}$} 
\end{subfigure}
\hspace{\hdist mm}
\begin{subfigure}{\x \columnwidth}
\includegraphics[width=\columnwidth]{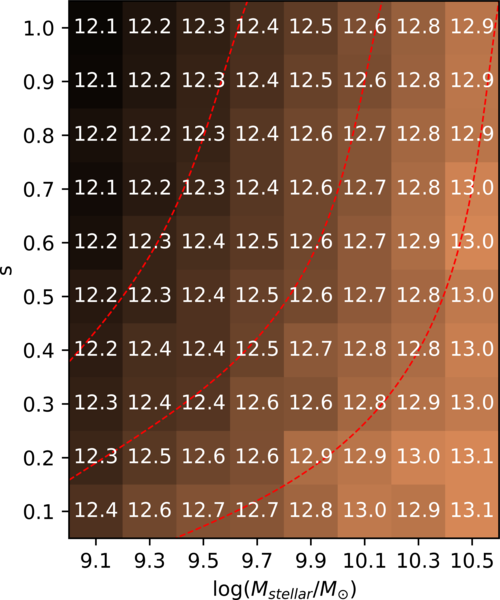}
\captionsetup{width=\columnwidth,skip = \z pt}
\caption{$^{2}M_{H,m}$} 
\end{subfigure}
\hspace{\hdist mm}
\begin{subfigure}{\x \columnwidth}
\includegraphics[width=\columnwidth]{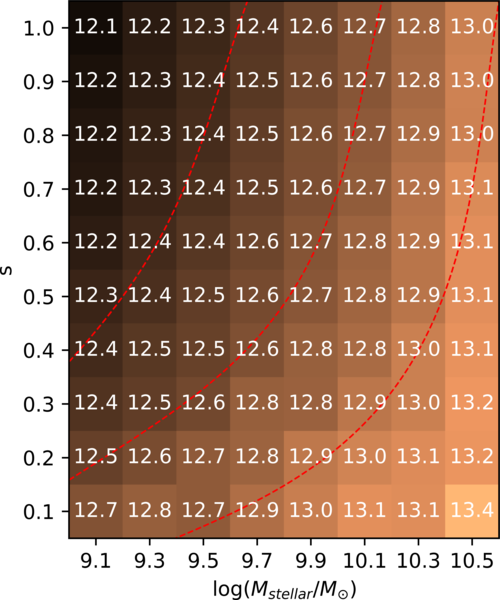}
\captionsetup{width=\columnwidth,skip = \z pt}
\caption{$_{2w}^{\ \ 2}M_{H,m}$} 
\end{subfigure}
\vspace{\vdist mm}
\newline

\centering
\hspace{\centpara mm}
\begin{subfigure}{\x \columnwidth}
\includegraphics[width=\columnwidth]{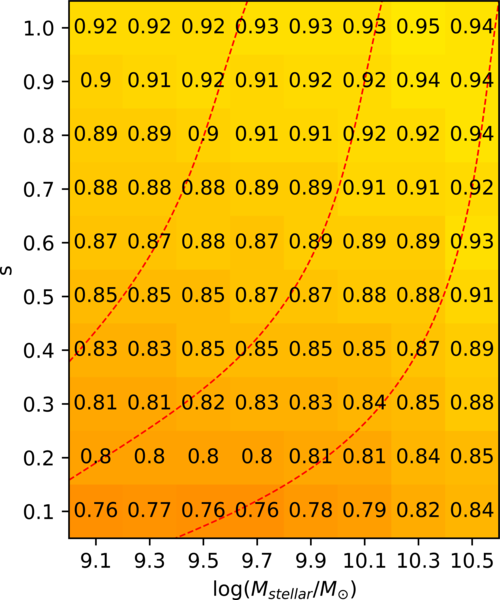}
\captionsetup{width=\columnwidth,skip = \z pt}
\caption{${^1}A_{\mathrm{centrals}}$} 
\end{subfigure}
\hspace{\hdist mm}
\begin{subfigure}{\x \columnwidth}
\includegraphics[width=\columnwidth]{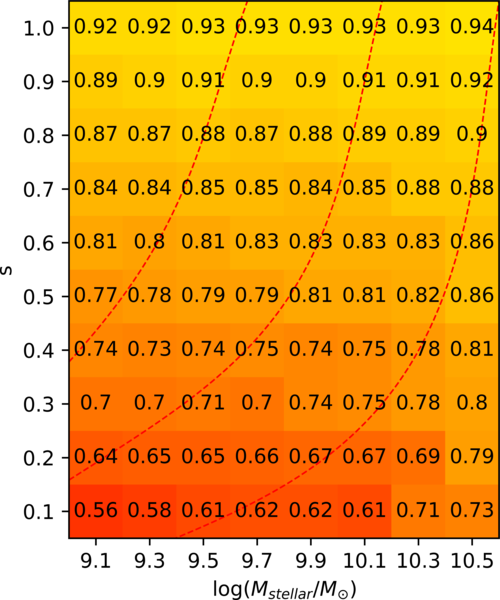}
\captionsetup{width=\columnwidth,skip = \z pt}
\caption{$^{\geq2} A_{\mathrm{centrals}}$} 
\end{subfigure}
\hspace{\hdist mm}
\begin{subfigure}{\x \columnwidth}
\includegraphics[width=\columnwidth]{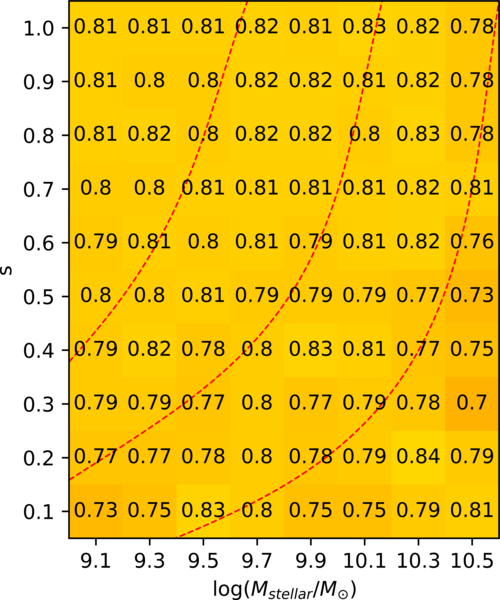}
\captionsetup{width=\columnwidth,skip = \z pt}
\caption{$A_{\mathrm{satellites}}$} 
\end{subfigure}
\vspace{\vdist mm}
\newline

\centering
\hspace{\centpara mm}
\begin{subfigure}{\x \columnwidth}
\includegraphics[width=\columnwidth]{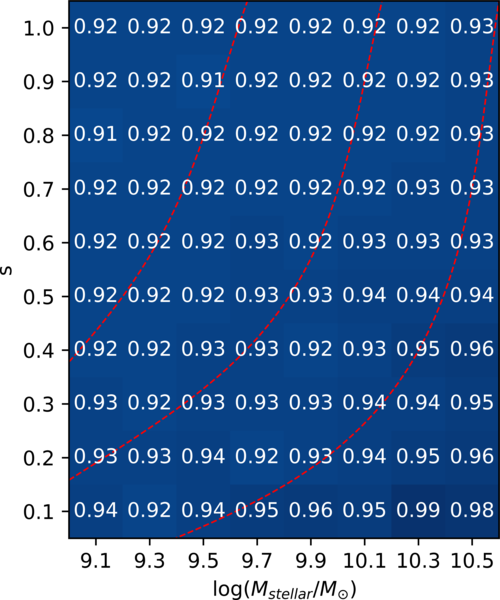}
\captionsetup{width=\columnwidth,skip = \z pt}
\caption{$f_{c_2} f_{p_2}$} 
\end{subfigure}
\hspace{\hdist mm}
\begin{subfigure}{\x \columnwidth}
\includegraphics[width=\columnwidth]{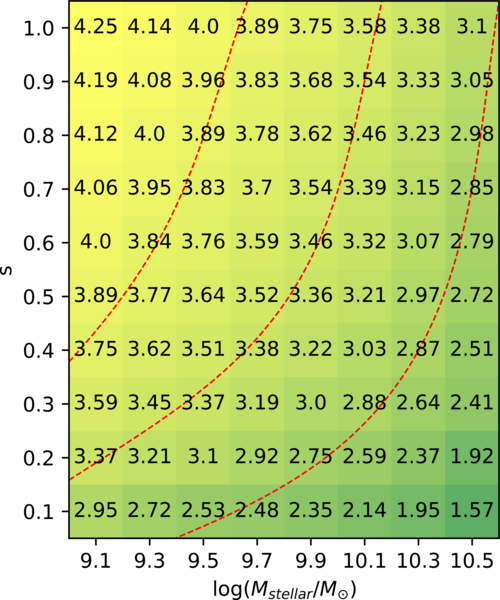}
\captionsetup{width=\columnwidth,skip = \z pt}
\caption{log$_{10}(N_{\mathrm{grs}})$} 
\end{subfigure}
\hspace{\hdist mm}
\begin{subfigure}{\x \columnwidth}
\includegraphics[width=\columnwidth]{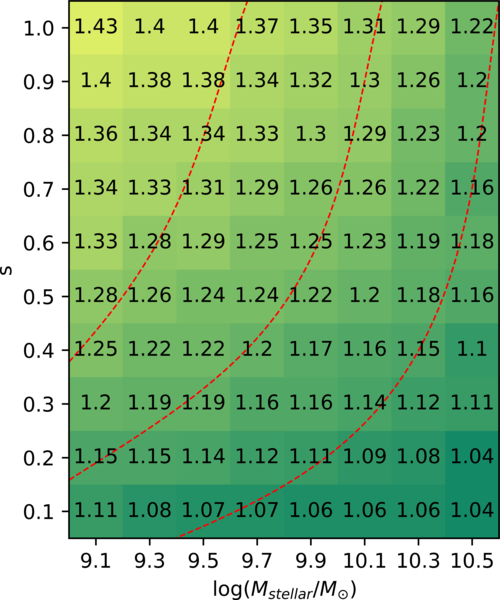}
\captionsetup{width=\columnwidth,skip = \z pt}
\caption{$m$} 
\end{subfigure}
\vspace{\vdist mm}
\newline
\caption{Metrics of the FoF group-finder for $1.2 < z < 1.7$.}
\label{12_17}
\end{figure*}

\begin{figure*}
\centering
\hspace{\centpara mm}
\begin{subfigure}{\x \columnwidth}
\includegraphics[width=\columnwidth]{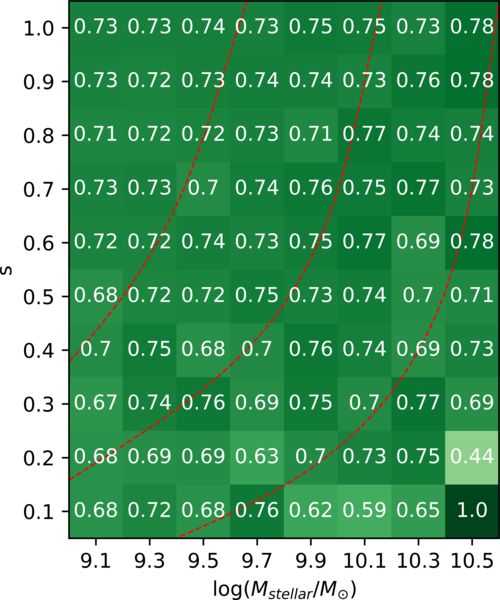}
\captionsetup{width=\columnwidth,skip = \z pt}
\caption{$p_2$} 
\end{subfigure}
\hspace{\hdist mm}
\begin{subfigure}{\x \columnwidth}
\includegraphics[width=\columnwidth]{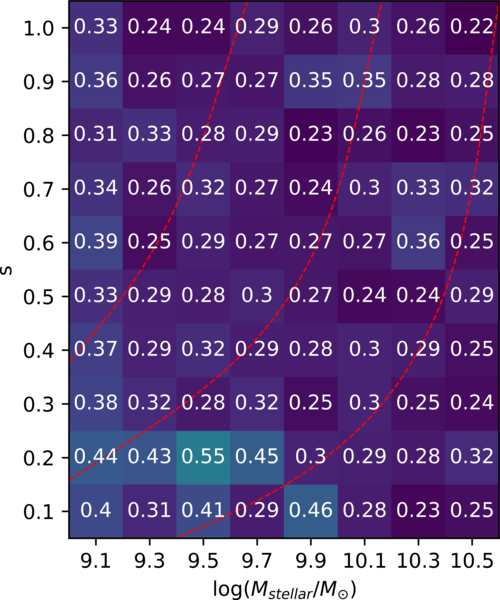}
\captionsetup{width=\columnwidth,skip = \z pt}
\caption{$\sigma_{\mathrm{gal}} \left(M_{\mathrm{H}} \right)$} 
\end{subfigure}
\hspace{\hdist mm}
\begin{subfigure}{\x \columnwidth}
\includegraphics[width=\columnwidth]{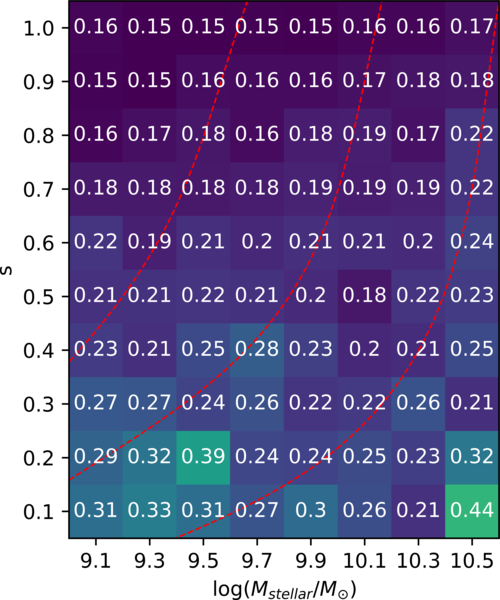}
\captionsetup{width=\columnwidth,skip = \z pt}
\caption{$\sigma_{\mathrm{gr}} \left(M_{\mathrm{H}} \right)$} 
\end{subfigure}
\vspace{\vdist mm}
\newline

\centering
\hspace{\centpara mm}
\begin{subfigure}{\x\columnwidth}
\includegraphics[width=\columnwidth]{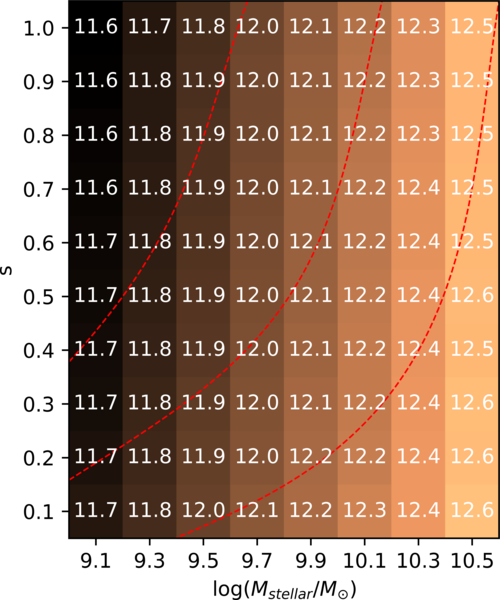}
\captionsetup{width=\columnwidth,skip = \z pt}
\caption{${^1}M_{H,m}$} 
\end{subfigure}
\hspace{\hdist mm}
\begin{subfigure}{\x \columnwidth}
\includegraphics[width=\columnwidth]{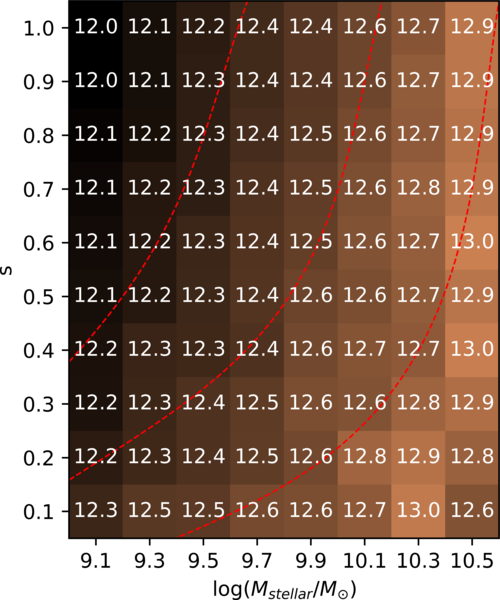}
\captionsetup{width=\columnwidth,skip = \z pt}
\caption{$^{2}M_{H,m}$} 
\end{subfigure}
\hspace{\hdist mm}
\begin{subfigure}{\x \columnwidth}
\includegraphics[width=\columnwidth]{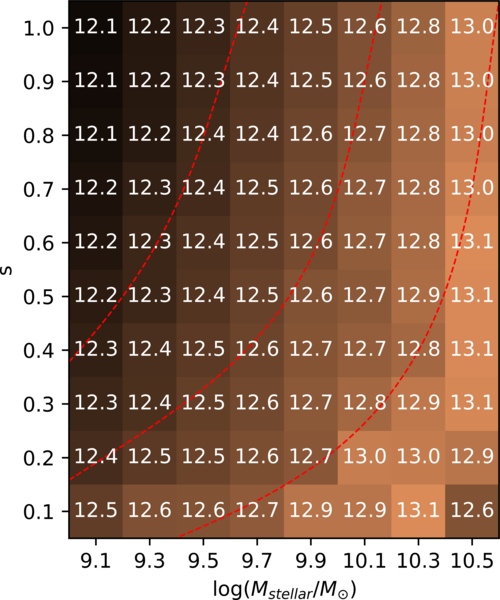}
\captionsetup{width=\columnwidth,skip = \z pt}
\caption{$_{2w}^{\ \ 2}M_{H,m}$} 
\end{subfigure}
\vspace{\vdist mm}
\newline

\centering
\hspace{\centpara mm}
\begin{subfigure}{\x \columnwidth}
\includegraphics[width=\columnwidth]{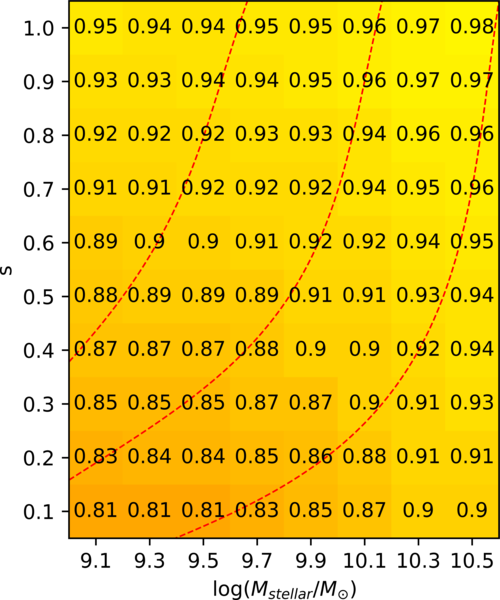}
\captionsetup{width=\columnwidth,skip = \z pt}
\caption{${^1}A_{\mathrm{centrals}}$} 
\end{subfigure}
\hspace{\hdist mm}
\begin{subfigure}{\x \columnwidth}
\includegraphics[width=\columnwidth]{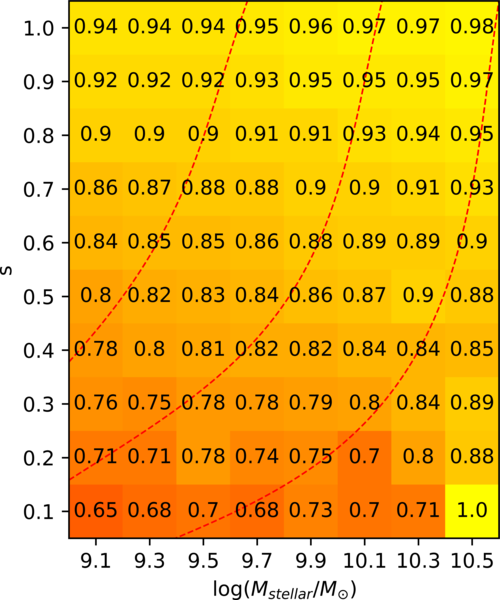}
\captionsetup{width=\columnwidth,skip = \z pt}
\caption{$^{\geq2} A_{\mathrm{centrals}}$} 
\end{subfigure}
\hspace{\hdist mm}
\begin{subfigure}{\x \columnwidth}
\includegraphics[width=\columnwidth]{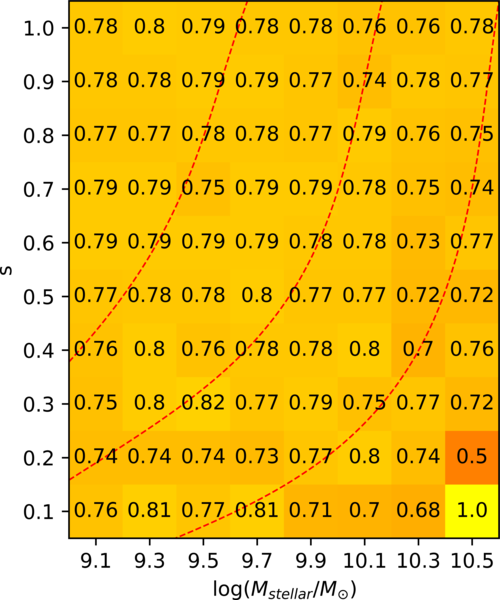}
\captionsetup{width=\columnwidth,skip = \z pt}
\caption{$A_{\mathrm{satellites}}$} 
\end{subfigure}
\vspace{\vdist mm}
\newline

\centering
\hspace{\centpara mm}
\begin{subfigure}{\x \columnwidth}
\includegraphics[width=\columnwidth]{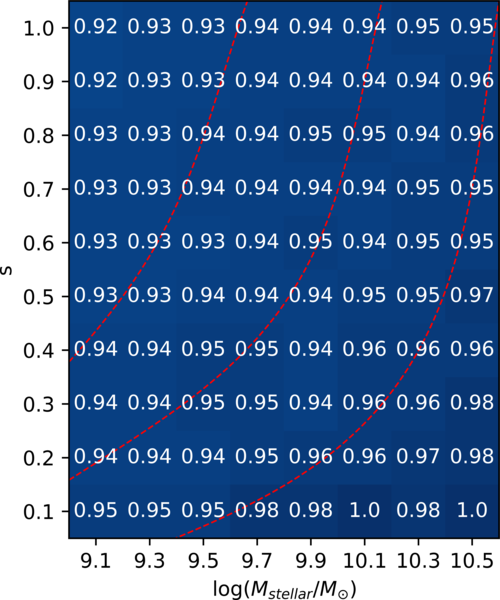}
\captionsetup{width=\columnwidth,skip = \z pt}
\caption{$f_{c_2} f_{p_2}$} 
\end{subfigure}
\hspace{\hdist mm}
\begin{subfigure}{\x \columnwidth}
\includegraphics[width=\columnwidth]{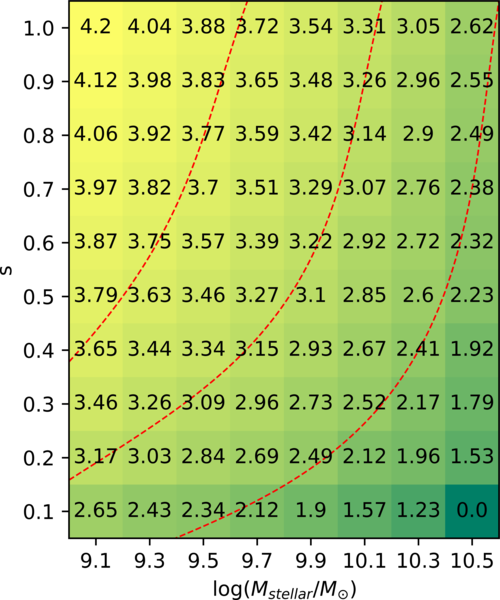}
\captionsetup{width=\columnwidth,skip = \z pt}
\caption{log$_{10}(N_{\mathrm{grs}})$} 
\end{subfigure}
\hspace{\hdist mm}
\begin{subfigure}{\x \columnwidth}
\includegraphics[width=\columnwidth]{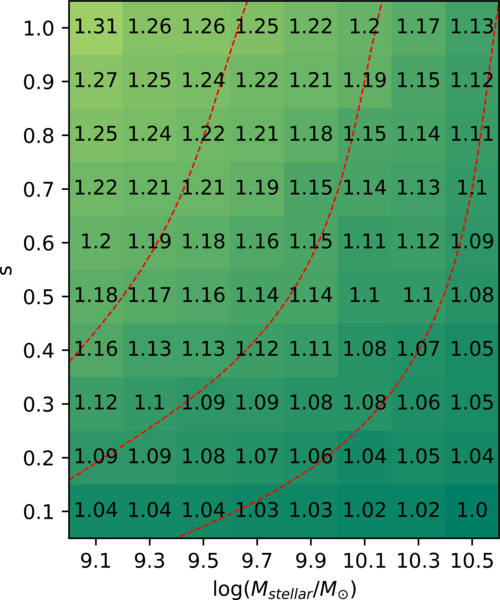}
\captionsetup{width=\columnwidth,skip = \z pt}
\caption{$m$} 
\end{subfigure}
\vspace{\vdist mm}
\newline
\caption{Metrics of the FoF group-finder for $2.0 < z < 2.6$.}
\label{20_27}
\end{figure*}

In this section, we simply and concisely compare all 12 performance metrics, and their dependence on $s$ and $M_{\mathrm{stellar}}$ over three almost contiguous redshift ranges $0.9 < z < 1.1$, $1.2 < z < 1.7$ and  $2.0 < z < 2.6$. Exactly the same procedures, discussed above in detail for the $1.2 < z < 1.7$ redshift range, are followed for the other two redshift ranges and all three are then presented together with the same color scales to allow easy visual comparison. Fig.~\ref{09_11} - \ref{20_27} present the 12 metrics in each redshift range. In order to compare the twelve metrics over all redshift ranges, the color-coding is set to be the same over all three figures. The goal here is to allow a simple visual comparison of the three redshift ranges.

The most noticeable change over the three redshift ranges lies in the changing multiplicity of the recovered structures (at fixed $s$ and $M_{\mathrm{stellar}}$).   This is simply due to the overall growth of structure over this redshift range in the mock (and real) universe.  Associated with this reduction in multiplicity is an improvement(!) with redshift in the accuracy of central classification, $A_{\mathrm{centrals}}$ (for ($N_{\mathrm{rec}} = 1$)-, as well as ($N_{\mathrm{rec}} \geqslant 2$)-structures) over the three redshift ranges at fixed $s$ and $M_{\mathrm{stellar}}$.   This is because the number of potentially contaminating satellites decreases at higher redshift because of the decrease in multiplicity. 

It is noticeable how the quality of the recovered group catalogue expressed by the 2-way purity $p_2$ and the fidelity of the 2-way matching represented by $f_{c_2} f_{p_2}$ is again strikingly independent of the redshift range.  Also the median halo masses $M_{H,m}$ of singletons, 2-member groups as well as 2-way matched 2-member groups (at fixed $s$ and $M_{\mathrm{stellar}}$) are largely independent of the redshift range, although this of course also reflects the details of the L-galaxies SAM.  

We also see a small decrease in the halo mass scatter $\sigma(M_{\mathrm{H}})$, especially on a galaxy basis, with increasing redshift. This effect is again due to the changing multiplicity with redshift due to the growth of structure. At lower redshift, the average $N_{\mathrm{rec}}$ is higher, and the recovered group catalogue is not quite so dominated by very low richness structures. Further, the scatter in dark matter halo mass increases with $N_{\mathrm{rec}}$ (and correlated with $\sum M_{\mathrm{stellar}}$), mostly due to the increasing number of interlopers, when calculated on a galaxy basis. Hence, the average scatter in halo mass over all $\sum M_{\mathrm{stellar}}$-bins, on a galaxy basis, is higher at low redshift for fixed $s$ and $M_{\mathrm{stellar}}$.

\subsection{Science example: Quenching fractions}
In this section, we discuss a single example of a possible science investigation that can be investigated with a galaxy group catalogue of the type discussed here. Based on a recovered galaxy group catalogue and a dark matter halo mass estimator, the fraction of central and satellite galaxies that are "quenched" (i.e. which have their sSFR reduced by a large amount relative to normal "Main Sequence" galaxies) can be investigated as a function of the parent dark matter halo mass. The true effect will however be perturbed by both errors in classifying centrals and satellites (see Metrics 8,9 and 10), in associating galaxies to haloes of a particular mass and in assigning a mass to that group (both of which are combined into Metric 6).

\begin{figure*}
\centering
\includegraphics[width=\textwidth]{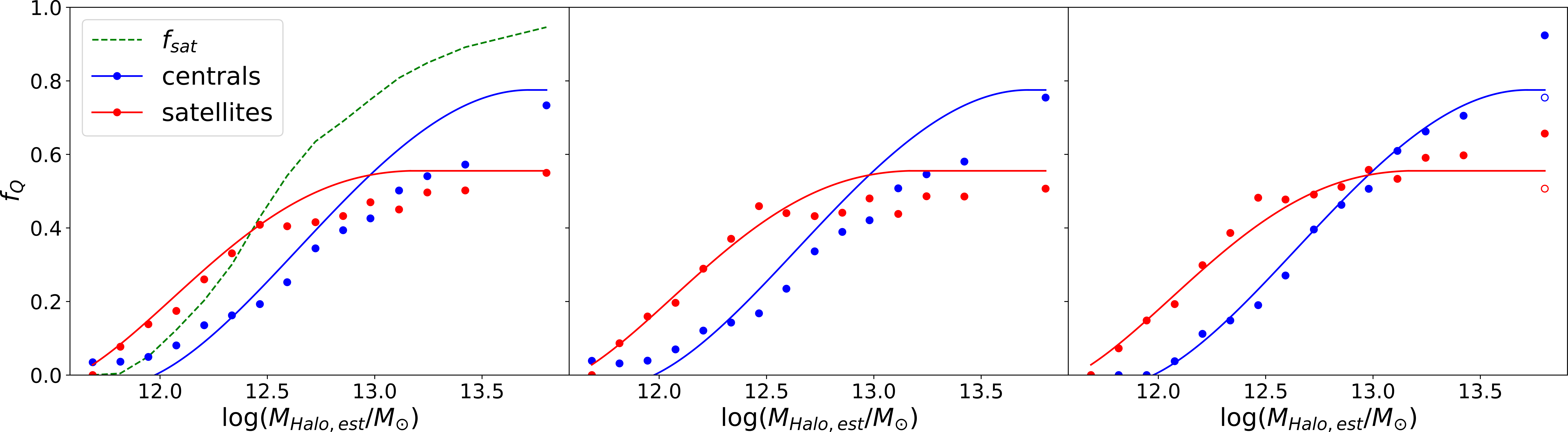}
\caption{The solid lines show the true (unsampled) quenched fractions, the dots the recovered quenched fractions, as a function of the true and recovered dark matter halo mass respectively, for centrals and satellites.  The recovered group catalogue was generated for a sampling rate of $80 \%$ and a stellar mass cut of $9.5$ solar masses (in log) in the $1.2<z<1.7$ redshift bin. Left panel: Uncorrected central and satellite quenched fractions. The recovered satellite fraction as a function of estimated dark matter halo mass is given by the green dashed line. Middle panel: Locally corrected central and satellite quenched fractions for global accuracies (see text for details). Right panel: Global accuracy and interloper corrected central and satellite quenched fractions (see text for details).}
\label{3_science_exmple_plots}
\end{figure*}

The quenching threshold is, in our example, set at a specific star formation rate (sSFR) of $0.22 \ Gyr^{-1}$, taken from the L-Galaxies SAM. The left panel of  Fig.~ \ref{3_science_exmple_plots} shows the true (unsampled) quenched fractions (solid lines), taken from the SAM, and the recovered quenched fraction (dots) that is constructed from the recovered group catalogue, again using for illustration a sampling rate of $80 \%$ and a stellar mass cut of $9.5$ solar masses (in log). Both are plotted as functions of the dark matter halo mass (true or estimated recovered, respectively).  The green dashed line shows the recovered satellite fraction as a function of estimated recovered halo mass.  While the recovered group catalogue reproduces the overall trends, significant inaccuracies are introduced due to the incomplete sampling and other failures in the group reconstruction. 

Information obtained from our different science metrics (about the accuracy in identifying centrals and satellites, and interloper fractions) can be used to try to correct the observed central and satellite quenched fractions for these shortcomings in the recovered galaxy group catalogue.  Of course, one could apply ever more sophisticated corrections to the observed quenching fraction using more and more information from the mock universe, until one ended up with a scheme that simply fully recovered the underlying mock.  This would be a meaningless exercise.  Rather, our goal here is to see how easily one can recover a good approximation to reality with very simple corrections.

The most straightforward (and traditional) correction would be simply to correct the raw observed quenched fractions of the two populations at a given mass, using the cross-contamination of centrals and satellites given by the accuracy Metrics 8,9 and 10. The middle panel of Fig.~ \ref{3_science_exmple_plots} shows the quenching fractions corrected for the (in)accuracy in identifying centrals and satellites by solving the simultaneous equations for $f_{\mathrm{Q,corr,cen/sat}}$:
\begin{multline}
f_{\mathrm{Q,obs,cen/sat}} = f_{\mathrm{Q,corr,cen/sat}} A_{cen/sat} \ +  \\ + f_{\mathrm{Q,obs,sat/cen}} (1-A_{\mathrm{cen/sat}}).
\end{multline}
The accuracies in recovering centrals, $A_{\mathrm{cen}}$, (here including singletons and higher richness centrals) and satellites, $A_{\mathrm{sat}}$, are simply taken from Metrics 8,9 and 10, i.e. they are the average "global" values across all richnesses, not those determined at that particular recovered dark matter halo mass. This is a justifiable simplification because there is little dependence on mass in Fig.~\ref{F_M_full}. 
On the other hand, the $f_{\mathrm{Q,obs,sat/cen}}$ is the uncorrected quenched fraction within that particular dark matter halo bin. We therefore refer to this as a "local correction". It can be seen that already this simple correction is much closer to the real (mock) quenched fractions.

Of course, interloper centrals or satellites are unlikely to have come from haloes of the same mass, and so another simple correction would be to instead use the global quenching fraction $f_{\mathrm{Q,obs,sat/cen,glb}}$ of the complement (centrals for satellites and vice versa), the global interloper fraction $f_{\mathrm{I,glb}}$ and the global singleton quenched fraction $f_{\mathrm{Q,s}}$.   This makes the simple assumption that the classification failures involve galaxies scattered in from anywhere on the recovered dark matter halo mass axis. We call this a "global" correction and therefore correct the quenching fractions by solving 
\begin{multline}
f_{\mathrm{Q,obs,cen}} = f_{\mathrm{Q,corr,cen}} A_{\mathrm{cen}}(1-f_{\mathrm{I,glb}}) \ +  \\ + f_{\mathrm{Q,s,glb}}A_{\mathrm{cen}}f_{\mathrm{I,glb}}  + f_{\mathrm{Q,obs,sat,glb}} (1-A_{\mathrm{cen}}) \label{cen_global}
\end{multline}
and 
\begin{equation}
f_{\mathrm{Q,obs,sat}} = f_{\mathrm{Q,corr,sat}} A_{\mathrm{sat}} +  f_{\mathrm{Q,obs,cen,glb}} (1-A_{\mathrm{sat}}) \label{sat_global}
\end{equation}
for $f_{\mathrm{Q,corr,cen/sat}}$. This correction is presented in the right panel of Fig.~ \ref{3_science_exmple_plots}.

This further improves the recovery of the true quenched fraction.  At the very high dark matter halo mass end, however, the corrected quenched fractions of both centrals and satellites are now over-corrected, and are noticeably worse than the "local" correction.  This is not too surprising.  For false centrals, it is probably unrealistic that satellites or interlopers from much lower dark matter haloes will have been falsely identified as the central of very high mass recovered groups: more likely is that a satellite in the same halo is misidentified as the central because the true central was not observed. For the satellites, it is likely that using the low global central quenched fraction $f_{\mathrm{Q,obs,cen,glb}}$ leads to an over-correction. 
Accordingly, we adopt the local correction at very high masses.  This combined correction, even though very simple, yields quenching fractions very close to the real mock quenched fractions.

In summary, this short science example provides a simple illustration of the recovery of accurate scientific information using only information from high level metrics in a very simple way.

\section{Survey costs analysis in $\lowercase{s}/M_{\mathrm{\lowercase{stellar}}}$-space} \label{cost_analysis}

In Section \ref{Science_metrics_M_s_space} of the paper we explored how the science return from a group catalogue constructed from a redshift-survey at high redshift depends on two key parameters of the survey design: the limiting mass of the tracer galaxy population (or more generally the number density of tracers) and the sampling rate of those tracers, i.e. what fraction of those tracers are observed and/or yield a usable redshift.
In this section, we analyze the "costs" of these different possible survey designs in terms of required observation time in order to better understand the trade-offs involved.

For a given amount of telescope observation time, there are clear trade-offs in the survey design parameters.  Going deeper (i.e. lowering the stellar mass limit $M_{\mathrm{stellar}}$ of the tracers) and achieving a high sampling rate $s$ are likely to  be expensive, in the sense of requiring more observing time to cover a given area, or equivalently permitting only a smaller area to be covered in a given total amount of observing time. The optimum choice of survey design may well be different for different science goals.  While these may to a certain degree be accommodated through a wedding-cake approach of different nested surveys of different depths and areas, the final detailed design requires the best possible quantitative understanding of the trade-offs between science performance and cost.  

This paper has aimed to provide a general framework for considering these trade-offs in the area of galaxy group science.   However, since some of the factors are undoubtedly instrument-specific, we will use for illustration in this section recent simulations of the multiplexing capabilities of the MOONS instrument (using the path analysis algorithm Faststar) kindly provided by co-workers on the MOONS project at IASF-MI \footnote{Andrea Belfiore, Paolo Franzetti, Bianca Garilli, Adriana Gargiulo et al.; https://www.iasf-milano.inaf.it/}.
However, since the ultimate performance of the MOONS instrument on the sky will not be known for some time, and since the efficiency of fibre positioning algorithms is still being developed, we will focus in this analysis primarily on relative effects.
This also helps to ensure that this analysis will also have validity for other (similar) observational instruments and survey designs. 

\subsection{Breakdown of survey costs}

In this section we present a generalized breakdown of the costs of a survey into four factors.   This should be applicable to any redshift survey carried out with a multiplexed spectrograph that can simultaneously obtain spectra of multiple sources.  Our approach is to construct a flexible formalism in which the increasing levels of detail can be easily incorporated as desired. 

We will assume for simplicity that the observations are broken down into units of some convenient integration time, during which the configuration of the instrument, e.g. the location of individual slits or fibres, is not changed.   It may well be that some sources may require much more integration time than others: an obvious example might be passive galaxies that will require much longer integration times than star-forming galaxies of the same stellar mass.  Another possibility would be very faint but rare very high redshift galaxies. We will however assume that these different needs can be accommodated by retaining some fibres or slits on those objects for subsequent configurations, while redeploying the others onto new "normal" targets.

The total telescope observation time $T_{\mathrm{obs}}$ required to complete the observations in a given area of sky, which we for convenience set equal to the patrol field of view (FoV) of the instrument, may then be written in quite general terms, as follows.  
We define the total number of the tracer target population within this given area to be $N_{\mathrm{obj}}$, so the number that is to be actually observed is $s \times N_{\mathrm{obj}}$.  In the context of the earlier sections of the paper,  $N_{\mathrm{obj}}$ will be a function of the limiting stellar mass $M_{\mathrm{stellar}}$, but this can be completely generalized for any selection function. 
We also define the ideal multiplexing capability of the instrument in a particular observing mode to be $N_{\mathrm{fibre}}$ (though the formalism applies equally well to slits).  It is then clear that $T_{\mathrm{obs}}$ may be written quite generally as:

\begin{multline}
T_{\mathrm{obs}} =\overline{T}_{\mathrm{exps}} \times \left( s \times N_{\mathrm{obj}}\left(M_{\mathrm{stellar}} \right) \right)\ \times \\ \times \frac{1}{N_{\mathrm{fibre}}} \times \epsilon^{-1} \left( s, N_{\mathrm{obj}} \right), \label{T_obs_four_fac}
\end{multline}

The first three terms are straightforward: the average exposure time per object, the total number of objects to be observed, and the number of objects that can, in principle, be observed simultaneously.   The final efficiency term is also conceptually simple, as it is just the fraction of the fibres that can in practice actually be usefully allocated to a target, averaged over all of the fibre configurations that are required to complete the observations in this one FoV.  

In practice, this efficiency term will likely be quite complicated and will depend on the details of the technical design of the instrument and on the mode of operation of it.   While it will generally be possible to use the full multiplexing capability when the number of potential targets is very much larger than $N_{\mathrm{fibre}}$, this will not be the case as the choice of targets becomes more constrained, whether through having a lower $N_{\mathrm{obj}}$, or by requiring high $s$, or through other constraints such as a possible need to locate "sky" fibres to allow nodding.  Given such constraints, not all fibres (slits) that are in principle available will actually be usable in practice. The efficiency will therefore drop.  However, it is important to appreciate that the $\epsilon$ in equation~(\ref{T_obs_four_fac}) is the efficiency {\it averaged} over all the fibre configurations that are needed to complete this field.  

Equation~(\ref{T_obs_four_fac}) is quite flexible and quite general. As a simple example, it might be decided that 10\% of the target objects required ten times the exposure time. In practice, this means that 47\% of the fibres should be allocated at any point in time to the majority objects needing only the shorter exposure time and 53\% to the more difficult minority objects.
One could then consider this in equation~(\ref{T_obs_four_fac}) in one of two ways.  One could either increasing the average exposure time $\overline{T}_{\mathrm{exps}}$ by a factor of 1.9, while keeping $N_{\mathrm{obj}}$ and $N_{\mathrm{fibre}}$ the same. Alternatively, one could conceptually "forget" about these more difficult objects, keep $\overline{T}_{\mathrm{exps}}$ the same, reduce $N_{\mathrm{obj}}$ by 10\%, but then also reduce the $N_{\mathrm{fibre}}$ to 0.47 of its original value.  Both approaches give exactly the same total observation time $T_{\mathrm{obs}}$.

All of the complications arising from practical limitations of fibre placement, etc., are put into the final efficiency term, $\epsilon \left( s, N_{\mathrm{obj}} \right)$.   This has the advantage that it may be calculated with whatever precision or sophistication is desired, as the performance of an instrument, including the fibre positioning etc., is further refined.   
The breakdown of $T_{\mathrm{obs}}$ in equation~(\ref{T_obs_four_fac}) is designed to separate the largely instrument-independent terms from the detailed instrument-specific technical constraints in the last term.

\subsection{Generic survey design considerations in $\lowercase{s}/M_{\mathrm{\lowercase{stellar}}}$-space}
\label{generic_design}

In this section we discuss the first three factors of  equation~(\ref{T_obs_four_fac}) which are largely instrument-independent apart from the "ideal" multiplexing factor $N_{\mathrm{fibre}}$. To first order, the average observation time per object $\overline{T}_{\mathrm{exps}}$ might be expected to vary as $M_{\mathrm{stellar}}^{-2}$ at a given redshift. This is certainly true for passive objects.  For star-forming galaxies, the combination of the weakly declining $sSFR(M_{\mathrm{stellar}})$ relation and the increasing effects of dust obscuration at higher stellar masses means that the luminosity of the emission lines on which a successful redshift determination depends may have a weaker dependence on mass (a point emphasized to us by Emanuele Daddi, private communication). For illustrative purposes, we assume a scaling as $M_{\mathrm{stellar}}^{-2}$.

The latest estimates for the MOONS suggest that an integration time of $2hr$ is needed for a star-forming object with a stellar mass of ${10^{9.5} M_{\odot}}$. Hence, we approximate
\begin{equation}
\overline{T}_{\mathrm{exps}} \approx  2 hr \times \left(\frac{10^{9.5} M_{\odot}}{M_{\mathrm{stellar}}} \right) ^2
\end{equation}

\begin{figure}
\centering
\includegraphics[width=\y\columnwidth]{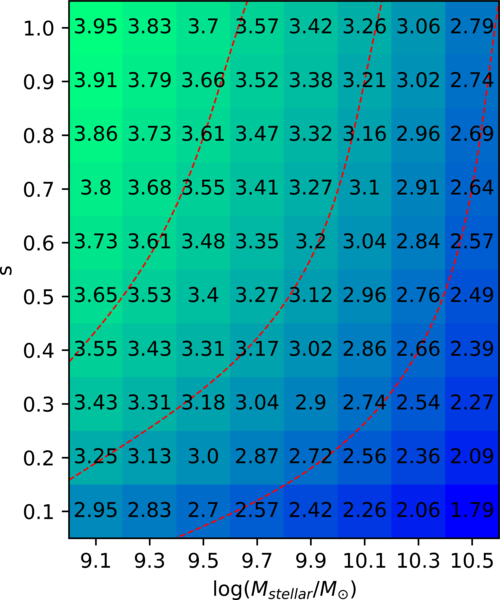} 
\caption{The product $\left( s \ N_{\mathrm{obj}} \left(M_{\mathrm{stellar}} \right) \right)$ indicates the (log) number of targets that need to be observed in a MOONS FoV to achieve the sampling rate $s$ of the tracer population down to a limiting stellar mass $M_{\mathrm{stellar}}$.  The red curves, representing the values $2.7,3.25$ and $3.65$, indicate the lines of constant number of tracers superposed on numerous other figures in this paper. The equal number density of tracers is of course equivalent to the constant number of targets (up to the appropriate spatial density factor).}
\label{N_obj_pp}
\end{figure}

Apart from this potential penalty, decreasing the limiting mass also has a large (beneficial) effect in increasing the  number density of the galaxy tracer population, but this also means that the number of objects to be observed in each field also increases proportionally in the second term in equation~(\ref{T_obs_four_fac}). Fig.~\ref{N_obj_pp} shows the number of objects that need to be observed in one MOONS FoV in order to reach a certain sampling rate $s$ at a given stellar mass cut $M_{\mathrm{stellar}}$, using the numbers from the L-Galaxies SAM light-cone, and summing over the three redshift ranges considered earlier in the paper.  The red lines in the figure represent lines of constant number of galaxies observed, and provide the lines that were superposed on the other figures in this paper. 

 The third term in equation~(\ref{T_obs_four_fac}) concerns the theoretical multiplexing capacity of the instrument, i.e. how many targets can be simultaneously observed in a given (fibre or slit) configuration.  This may depend on the choice between available observing modes. For instance, for a fibre-spectrograph like MOONS, a key decision is how to obtain spectra of blank sky for sky-subtraction of the target spectra.   For MOONS, three observing modes have been envisaged.  XSwitch is the most conservative and assigns a half of the fibres to target sources, paired with an equal number of fibres placed on blank regions of sky that are displaced by a fixed distance and direction from each target, enabling an ABBA nodding technique.  This extra constraint on the location of the sky fibres will inevitably degrade the fourth efficiency term in equation~(\ref{T_obs_four_fac}), but we should simply use the maximum number of allocatable fibres in the third term in equation~(\ref{T_obs_four_fac}).  STARE700 allocates $700$ fibres to target sources, with the remainder placed on random blank sky regions while STARE900 puts up to $900$ fibres simultaneously on targets.
 
%%%%%%%%%%%%%%%%%%%%%%%%%%%%%

\subsection{The efficiency factor $\epsilon$ of MOONS in $\lowercase{s}/M_{\mathrm{\lowercase{stellar}}}$-space}

We now turn to the instrument-specific fourth term in equation~(\ref{T_obs_four_fac}). As stated above, this term represents the average efficiency of fibre placement when averaged over all of the configurations required to complete the observations in a particular region of sky.  We will here apply several simplifying assumptions in this analysis.   A full treatment would require detailed simulations of the positioning of fibres on real targets, but these could be carried out as desired with arbitrary realism.  We are primarily interested in understanding relative effects.  

\begin{figure}
\includegraphics[width=\columnwidth]{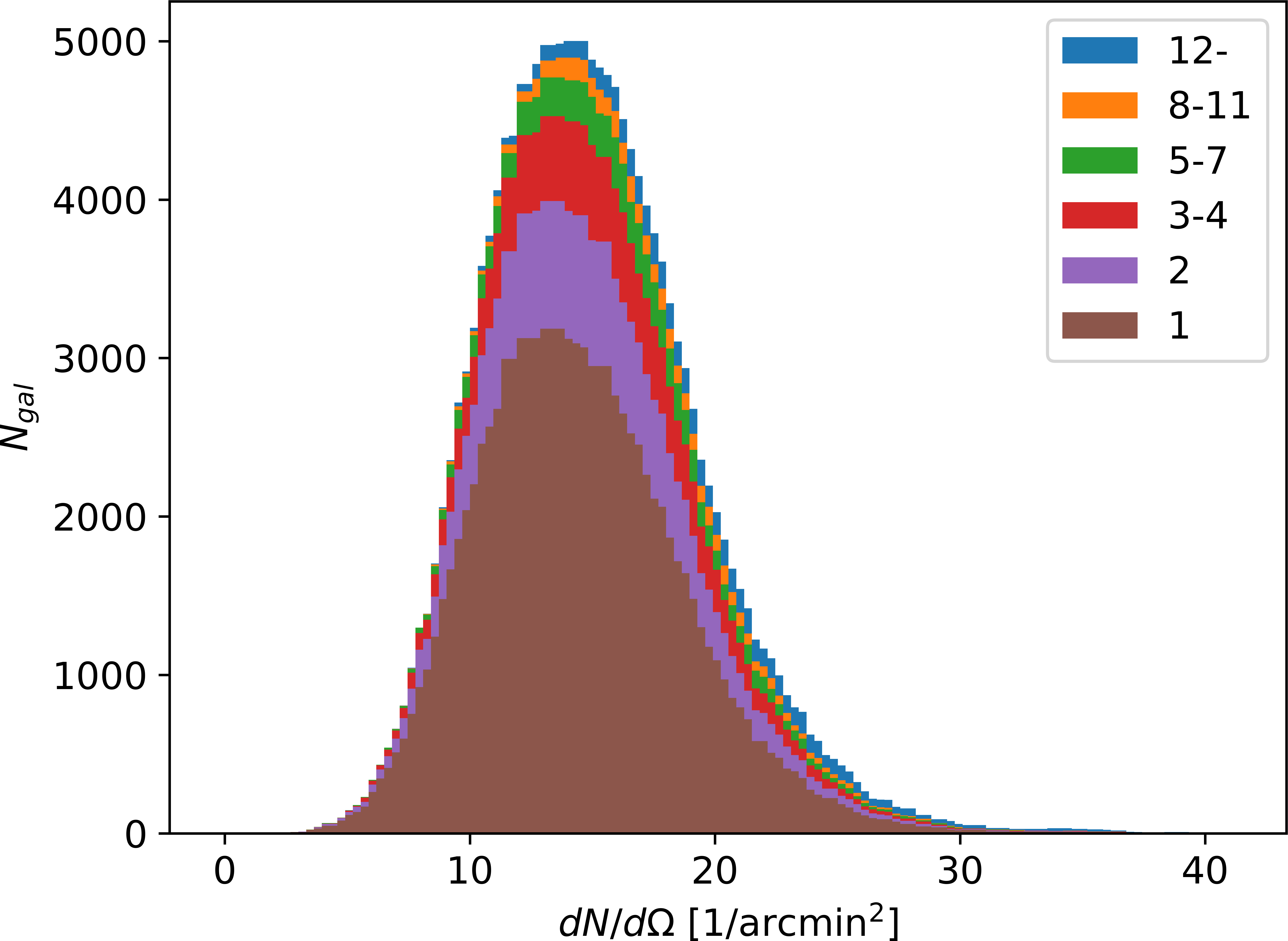}
\caption{The distribution of the (local) surface number density of tracers $dN/d\Omega$ for galaxies in different richness structures. The surface number density for each galaxy is calculated from counting the number of neighboring galaxies within a radius of $1 arcmin$.}
\label{N_gal_vs_Surface_N_density_of_Tracers_1.0_arcmin}
\end{figure}

\begin{figure*}
\centering
\includegraphics[width=\textwidth]{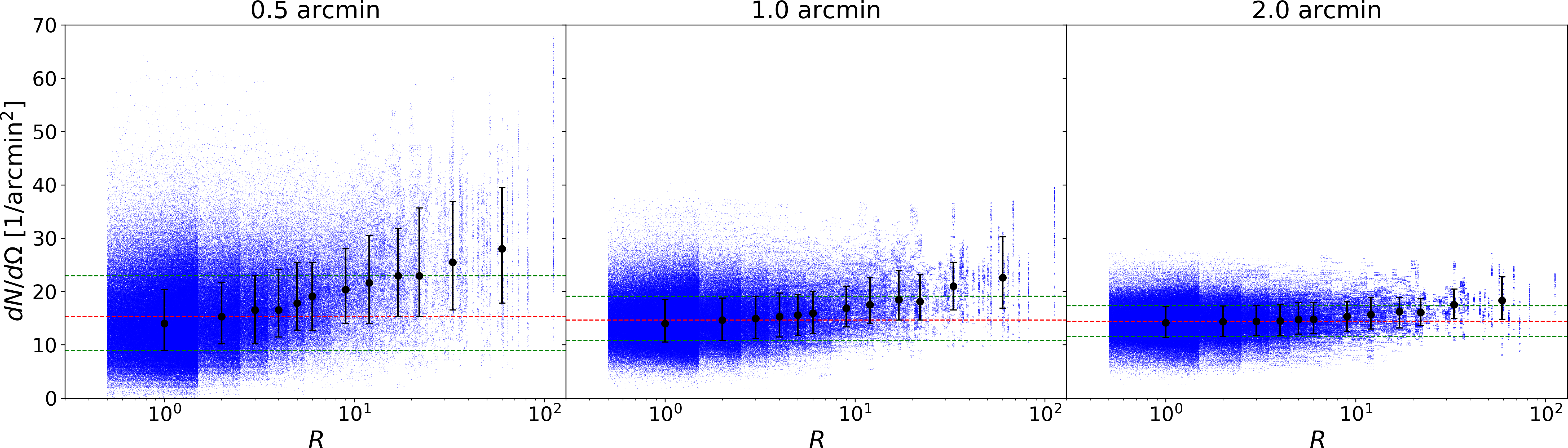}
\caption{Richness $R$ of groups versus surface number density of tracers $dN/d\Omega$, plotted on a galaxy basis. Each dot represents a galaxy within one of the three MOONS redshift ranges with $\textrm{log}_{10}(M_{\mathrm{stellar}}/M_{\odot}) > 9.5$ } and shows its true richness (same in all three panels) and the projected number density within three different radii (shown in the three panels).  In each panel, the black dots with error bars show the median $dN/d\Omega$ within a given richness bin, with the $16 \%$ and $84 \%$ percentiles.  The red and green dashed lines show the same for the overall sample.
\label{NdoT_wR_vs_R_linear_R_range}
\end{figure*}

We will therefore assume the following for simplicity.
\begin{itemize}
\item We will consider only the fibre placement for the "normal" shorter integration objects and include the effects of any objects requiring longer exposure times (such as passive galaxies or additional rare targets of special interest, such as very high redshift galaxies) by reducing $N_{\mathrm{fibre}}$, as in the example given previously in Subsection \ref{generic_design}. 

\item We further assume that all of these normal objects are observed for the same integration time, defined for simplicity by the nominal integration time required for the limiting stellar mass. 

\item For each successive fibre configuration in a given survey field, we assume that the fraction of "available" fibres that can be successfully placed on a target is a function only of the total number of targets remaining to be observed and the number of available fibres.  For each observing mode, we can estimate this function from the performance of the fibre positioner (see below).  

\item We might expect that the chance that a given target galaxy is observed depends on the surface density of all targets averaged within some solid angle of that galaxy.  For example, this solid angle could be set by the individual fibre patrol field. In this paper, we completely neglect any variation of the projected surface density of tracers $dN/d\Omega$ across the sky which may may have an impact on the (local) sampling rate $s$ via fibre-positioning issues. At these depths the variation in projected surface density is anyway modest. Furthermore, any correlation of the projected surface density with the richness of individual groups is washed out by the high number of foreground and background objects. This is shown in Fig.~\ref{N_gal_vs_Surface_N_density_of_Tracers_1.0_arcmin} and in Fig.~\ref{NdoT_wR_vs_R_linear_R_range} for three different solid angles that might be relevant for different instruments. Effectively, we have assumed in this paper that the random selection of targets (and hence the local sampling rate $s$) is independent of the local surface number density of tracers. Of course, this issue will be addressed in the final survey design of MOONRISE, which will use actual target lists and the final fibre allocation software, which is not yet available.
\end{itemize}


As stressed earlier, any or all of these assumptions can be avoided by undertaking completely realistic simulations of the fibre positioner on real catalogues of targets.  Our purpose here is only to explore the most general points.

The MOONS instrument is equipped with 1000 science fibres that can in principle be allocated to either a target object or to a blank sky position. In the XSwitch observing mode, the current implementation of the fibre-positioning Faststar algorithm allocates pairs of fibres (one fibre on target, one on the nearby sky) whenever possible, so the best case scenario would be to have 500 fibres on targets, 500 on the sky.  In the presence of passive objects and high-$z$ targets, requiring longer integrations, the number of fibres available for normal (star-forming) targets is likely to reduce in practice to of order $N_{\mathrm{fibre}} \approx 300$.  Furthermore, imposing the constraint that the remaining "sky" fibres must be placed with a fixed displacement from the target sources (to allow nodding), places large constraints the choice of targets. Given all the constraints (including the requirement of a constant offset to a blank region of sky), a general feature of such algorithms is that full fibre allocation is only achieved when the number of available (remaining) targets greatly exceeds the number of available fibres.   When the ratio of remaining targets to fibres reduces, the fraction of fibres that are usable in a given fibre configuration drops considerably.  While further optimization of the fibre positioning algorithm are no doubt likely, this drop off in efficiency as the number of targets is reduced is likely to remain a general feature of MOONS and similar instruments. 

Hence, in good first order approximation, the number of allocatable science fibres in each pointing to one instrument field of view depends only the number of targets available for selection, i.e. still not observed, and the number of available fibres. More precisely, the fraction $f=N_{\mathrm{fibre,all}}/N_{\mathrm{fibre}}$ of allocatable fibres are in good approximation a function of the ratio of the number of targets available and the number of available fibres,
\begin{equation}
f = f(N_{\mathrm{tar}}/N_{\mathrm{fibre}}).  
\end{equation}
The function $f$ we used in this work is based on the current (May 2020) Faststar algorithm for the MOONS instrument in the XSwitch mode that places a sky fibre at a fixed offset from each target fibre. As mentioned above, about $N_{\mathrm{fibre}} \approx 300$ science fibres are available for normal objects in this instrument mode.

For a given number density of tracers $N_{\mathrm{obj}}$ and a desired final sampling rate $s$, we can then progressively use the fraction $f$ of available fibres that are allocatable in the next pointing as a function of the "remaining targets" to compute the number of configurations $N_{\mathrm{config}}$ that are required to finally reach the desired $s$ for a particular $N_{\mathrm{obj}}$.  It is then straightforward to average the $f$ of these successive configurations to yield the efficiency factor $\epsilon^{-1}$ needed in equation~(\ref{T_obs_four_fac}).

\begin{figure}
\centering
\includegraphics[width=\y\columnwidth]{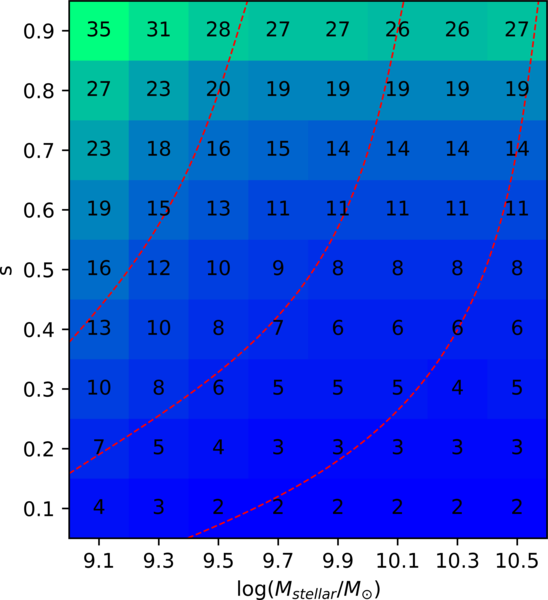}
\caption{The required number of configurations required to complete the observations in a given field, $N_{\mathrm{config}}$ as a function of $s$ and $M_{\mathrm{stellar}}$.} 
\label{N_point}
\end{figure}

Using for the sake of illustration the particular $f$ for the latest Faststar simulation, we show $N_{\mathrm{config}}$ as a function of $M_{\mathrm{stellar}}$ and $s$ in Fig.~\ref{N_point}. Full sampling, $ s = 1.0$ is not considered, because the efficiency in allocating fibres at very low target densities is the most uncertain and will be most dependent on the detailed spatial distribution of objects. Hence, we do not give an estimate for the costs of reaching $s = 1.0$ in Fig.~\ref{N_point}.

It should be noted that for typical survey designs in the middle of the ranges considered, $N_{\mathrm{config}}$ is large, of order 10 configurations or more. This makes it practical to repeatedly allocate some fibres to those objects requiring much longer exposure times, while giving "typical" objects many opportunities to be allocated a fibre, ensuring that constraints arising from the individual ''patrol-fields" of individual fibres (which will be present in each individual configuration) are washed out in the aggregate sampling map. 

At fixed $M_{\mathrm{stellar}}$, i.e. fixed $N_{\mathrm{obj}}$, the required number of configurations at first increases linearly with $s$ but then increases more steeply as more and more configurations are required in which the number of remaining targets falls below the optimum (large) multiple of the number of fibres. 

\begin{figure}
\centering
\includegraphics[width=\y\columnwidth]{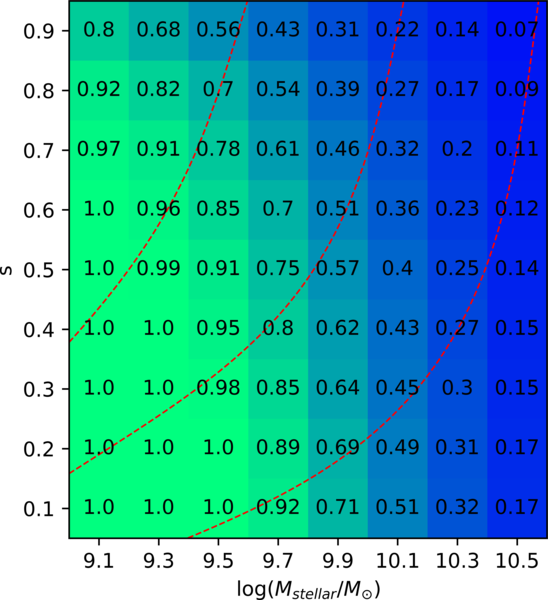}
\caption{The average fibre allocation efficiency $\epsilon$ as a function of the sampling rate $s$ and stellar mass cut $M_{\mathrm{stellar}}$.} \label{espilon_Faststar}
\end{figure}

A general point is that higher $s$ is more efficiently attained with a high target density of tracers $N_{\mathrm{obj}}$.  This is because the fraction of available targets to fibres, $N_{\mathrm{tar}}/N_{\mathrm{fibre}}$ stays high for longer as $s$ increases when the initial $N_{\mathrm{obj}}$ is larger.  For the particular example of the current implementation of Faststar shown in Fig.~\ref{N_point}, we show the efficiency $\epsilon$ across  $\lowercase{s}/M_{\mathrm{\lowercase{stellar}}}$-space in Fig.~\ref{espilon_Faststar}.  When going deeper, i.e. starting with a higher number density of targets, $\epsilon$ is overall high and drops only slightly with $s$ until a very high $s$ is desired. This is because many configurations are needed in the high density target-rich regime. Hence, the allocation of fibres stays efficient to quite high sampling rates.  In contrast, for a lower $N_{\mathrm{obj}}$, i.e. a higher stellar mass cut, the efficiency is overall lower and furthermore drops steeply with sampling rate at relatively low $s$.  The bottom line is that higher $s$ is overall much more expensive for a lower density of targets $N_{\mathrm{obj}}$ than for a high density of targets that require a large number of configurations per field.

\begin{figure}
\centering
\includegraphics[width=\y\columnwidth]{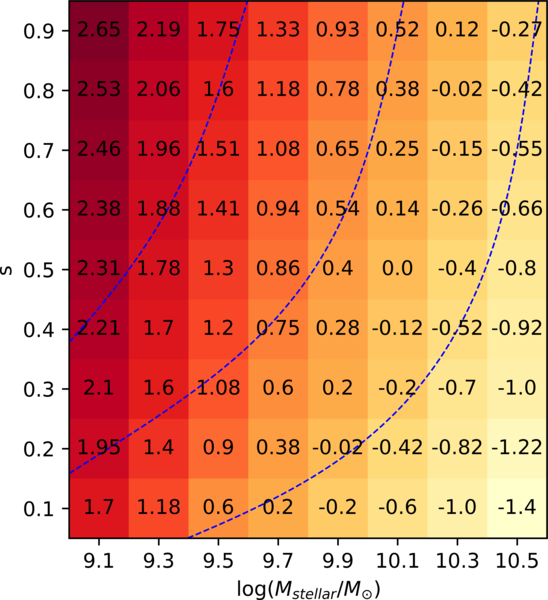}
\caption{The total observation time $T_{\mathrm{obs}}$ (in log hours) required to complete the observation in one MOONS FoV as a function of sampling rate $s$ and stellar mass cut $M_{\mathrm{stellar}}$.} \label{Obs_time_pp}
\end{figure}

The total observation time $T_{\mathrm{obs}}$ that is required to complete the observations in one instrument FoV can then be computed as a function of the sampling rate $s$ and the stellar mass cut $M_{\mathrm{stellar}}$ by multiplying all of the terms in equation~(\ref{T_obs_four_fac}).  This is presented in Fig.~\ref{Obs_time_pp}. It can be seen that $M_{\mathrm{stellar}}$, both through the (assumed) exposure time factor, $M_{\mathrm{stellar}}^{-2}$, and through the number of targets $N_{\mathrm{obj}}$ being proportional to $M_{\mathrm{stellar}}^{-3/4}$, is the dominant quantity in determining the total observing time $T_{\mathrm{obs}}$ required to complete the observation in one MOONS FoV, and thus, in a given amount of observing time, the total number of such FoV that can be surveyed.

\section{Summary and conclusion}
\label{summary}

In this paper, we have presented an in-depth study of the performance of group-finders, based on positional information of individual tracer galaxies, when they are applied to redshift surveys, especially at high redshift. This has been based on comparing the recovered group catalogues (including singleton galaxies) that are obtained by applying the group finding algorithms to realistic mock light-cones with the underlying "reality" from the L-galaxies SAM, in which the full halo information is known.

We have examined how the group-finding performance depends on the main design parameters of the survey, in particular on the limiting stellar mass (or to a certain degree equivalently, to the number density) of the galaxy tracers, and on the (assumed) random sampling rate of those tracers. We also examined how the observational cost of such a survey depends on those same parameters, choosing the future MOONS spectrograph on the VLT as an example.   This study should therefore be of use in designing future surveys with this and other instruments.  

We began with a comparison of two different group finding approaches, which we call the "halo-based" and the FoF group finding approaches, at low redshift, i.e. to SDSS-like surveys. 
This led to the conclusion that many of the performance and science metrics are almost indistinguishable between the group catalogues recovered by the two methods. Overall, the "halo-based" approach works slightly better at high $s$, while the FoF algorithm works slightly better at low $s$. This can for example be seen in the most basic statistical quantities of purity $p_2$ and completeness $c_2$. That the "halo-based" approach performs better at high $s$ might be expected, as it is a more "refined" approach in the sense that it assumes more information (about the sizes of dark matter haloes). The strength of FoF, on the other hand, lies in its minimal dependence on assumptions and it seems to be more robust as the fraction of objects for which positional information is known drops away. However, the "halo-based" and the FoF group-finders perform overall very similarly, and in the remainder of the paper we considered for simplicity only the FoF approach.

A total of 12 science metrics were then constructed which were designed to capture a range of physical information that might be relevant for a wide variety of science investigations.   We then examined how these different science metrics varied across ($s$/$M_{\mathrm{stellar}}$)-space when the FoF algorithm is applied at high redshift and identified the following main conclusions:

\begin{itemize}

\item The number of groups $N_{\mathrm{grs}}$ scales closely as the number of tracers observed $N_{\mathrm{tr}}$, given by the product of the sampling rate $s$ and the density of the underlying tracer population $N_{\mathrm{obj}}$, which is given in our analysis by the limiting stellar mass of the tracers $M_{\mathrm{stellar}}$.  There is a weak preference for $s$ over $M_{\mathrm{stellar}}$ (or $N_{\mathrm{obj}}$) to achieve a given $N_{\mathrm{tr}}$.

\item As would be expected, the median dark matter halo mass $M_{\mathrm{H,m}}$ for singleton galaxies scales as $M^{0.5}_{\mathrm{stellar}}$ and is more or less independent of $s$.  On the other hand, the median $M_{\mathrm{H,m}}$ for groups (containing more than one galaxy) scales quite closely with the number density of tracers $N_{\mathrm{tr}}$. 

\item The scatter in the estimated halo masses of recovered groups, $\sigma(M_{\mathrm{H}})$ reduces with $s$, more or less independent of $M_{\mathrm{stellar}}$, regardless of whether this scatter is calculated on what we call a galaxy-basis (considering the difference between the estimated parent halo mass and the true parent halo mass for each individual galaxy in the sample) or on what we called a group-basis, in which we compare the estimated and true masses of those recovered groups that are (uniquely) "2-way matched" with real groups.

\item The accuracy in correctly classifying galaxies as centrals or satellites is different for centrals and for satellites.  For centrals there is a dependency on $s$, especially for richer groups.  For singletons, the accuracy is high and depends only weakly on $s$. The accuracy of classifying satellites is very largely independent of $s$ (and also $M_{\mathrm{stellar}}$). is This because the vast majority of satellites are always seen as satellites.  Nevertheless, this classification accuracy is only about 80\%, because of infidelities in the group reconstruction.

\item Looking at how these results change with redshift, there is rather little change in almost all the metrics (at a given $(M_{\mathrm{stellar}},s)$) over the redshift range $0.9 < z < 2.6$.   The main effects are all related to the unavoidable reduction in the multiplicity with redshift, which is ultimately due to the cosmic growth of structure.  This reduction in multiplicity actually has some nominally beneficial effects, most notably an increase in the accuracy of central classification.
\end{itemize}

We then constructed a quite general and flexible formalism for considering the required cost in the observation time to complete the survey within each instrumental field of view, which translates (inversely) to the number of such fields that can be observed in a given amount of observing time.  This formalism puts all of the complexities of instrument operation, such as constraints on fibre positioning, into a single efficiency term, $\epsilon$.  Our analysis leads to the following main conclusions:

\begin{itemize}
\item The choice of the mass limit $M_{\mathrm{stellar}}$ has a strong effect:  Not only does this likely increase the typical observation time needed for each configuration (as ${M^{-2}_{\mathrm{stellar}}}$ for continuum objects, likely less for emission line objects), it also increases the number of targets that must be observed in each field so as to reach a give $s$, increasing as roughly $M^{-3/4}_{\mathrm{stellar}}$. 
%\item The efficiency $\epsilon$ does vary with $M_{\mathrm{stellar}}$ and $s$, but apparently it does not vary a lot from mode to mode. Hence, the effect of the choice of mode lies in $N_{\mathrm{fibre}}$.
\item Having a high density of target galaxies (e.g. by having a low $M_{\mathrm{stellar}}$) however produces high efficiency.  This is because the fraction of available fibres that can successfully be placed on a target is a strong function of the ratio of available targets to available fibres.  A survey at a low $N_{\mathrm{obj}}$ (i.e. high $M_{\mathrm{stellar}}$) is inefficient due to the low number of targets that are available in this regime. For the same reason, there is a cost associated with pursuing high $s$, although the decrease in average efficiency $\epsilon$ with $s$ is modest at high $N_{\mathrm{obj}}$, because more time is spent with high target density, but is much larger when one starts with a low $N_{\mathrm{obj}}$.
\end{itemize}

\section*{Acknowledgements}
We thank the anonymous referee for their highly insightful and constructive comments and suggestions, which have greatly benefited this paper. This work is based on an MSc thesis that was presented by TJL at the ETH Zürich. RM acknowledges support from the ERC Advanced Grant 695671 “QUENCH” and by the Science and Technology Facilities Council (STFC). This is a pre-copyedited, author-produced PDF of an article accepted for publication in MNRAS following peer review. %The version of record [insert complete citation information here] is available online at: xxxxxxx [insert URL that the author will receive upon publication here].

\section*{Data availability}
The L-Galaxies light-cone data used in this article are available on the L-Galaxies Munich Galaxy Formation Model website: \url{http://galformod.mpa-garching.mpg.de/public/LGalaxies/}.

%%%%%%%%%%%%%%%%%%%%%%%%%%%%%%%%%%%%%%%%%%%%%%%%%%



\bibliographystyle{mnras}
\bibliography{Bib_for_paper} % if your bibtex file is called example.bib

%%%%%%%%%%%%%%%%%%%%%%%%%%%%%%%%%%%%%%%%%%%%%%%%%%

%%%%%%%%%%%%%%%%% APPENDICES %%%%%%%%%%%%%%%%%%%%%

\appendix

\section{The science metrics for "halo-based" and F\lowercase{o}F methods}\label{12_science_metrics_Tinker_FoF}
The science metrics designed to capture a wide range of environmental information introduced in Section \ref{science_metrics} are here presented for the "halo-based" and FoF group finding algorithms in the $0.02<z<0.17$ redshift range, taking a constant stellar mass cut of $\textrm{log}_{10}(M_{\mathrm{stellar}}/M_{\odot}) = 9.0$ but exploring the full sampling rate range $s = 0.1-1.0$. As truth, we use light-cones produced by the L-Galaxies SAM.

\begin{figure*}
\centering
\begin{subfigure}{\m \columnwidth}
\includegraphics[width=\columnwidth]{Figures_for_paper_GF/Tinker_vs_FoF/resized-p_2.png}
\captionsetup{width=\columnwidth,skip = \z pt}
\caption{ $p_2$} 
\end{subfigure}
\hspace{\hdistTF mm}
\begin{subfigure}{\m\columnwidth}
\includegraphics[width=\columnwidth]{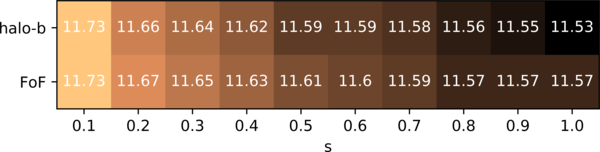}
\captionsetup{width=\columnwidth,skip = \z pt}
\caption{${^1}M_{H,m}$} 
\end{subfigure}
\vspace{\vdist mm}
\newline

\centering
\begin{subfigure}{\m \columnwidth}
\includegraphics[width=\columnwidth]{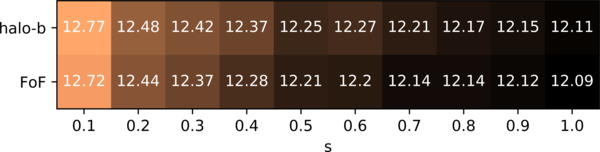}
\captionsetup{width=\columnwidth,skip = \z pt}
\caption{$^{2}M_{H,m}$} 
\end{subfigure}
\hspace{\hdistTF mm}
\begin{subfigure}{\m \columnwidth}
\includegraphics[width=\columnwidth]{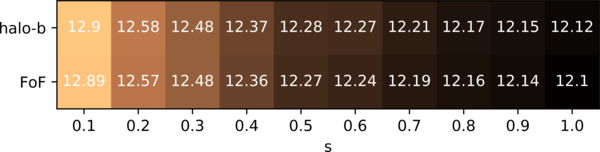}
\captionsetup{width=\columnwidth,skip = \z pt}
\caption{$_{2w}^{\ \ 2}M_{H,m}$} 
\end{subfigure}
\vspace{\vdist mm}
\newline

\centering
\begin{subfigure}{\m \columnwidth}
\includegraphics[width=\columnwidth]{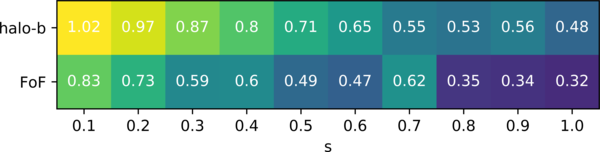}
\captionsetup{width=\columnwidth,skip = \z pt}
\caption{$\sigma_{\mathrm{gal}} \left(M_{\mathrm{H}} \right)$} \label{sigma_gal_Tinker_FoF}
\end{subfigure}
\hspace{\hdistTF mm}
\begin{subfigure}{\m \columnwidth}
\includegraphics[width=\columnwidth]{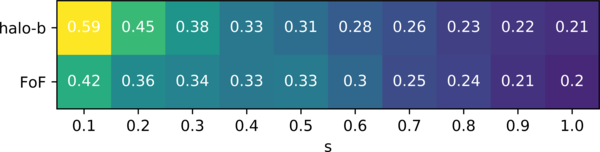}
\captionsetup{width=\columnwidth,skip = \z pt}
\caption{$\sigma_{\mathrm{gr}} \left(M_{\mathrm{H}} \right)$} \label{sigma_gr_Tinker_FoF}
\end{subfigure}
\vspace{\vdist mm}
\newline

\centering
\begin{subfigure}{\m \columnwidth}
\includegraphics[width=\columnwidth]{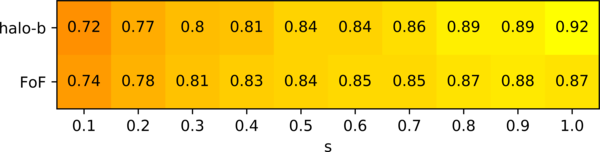}
\captionsetup{width=\columnwidth,skip = \z pt}
\caption{${^1}A_{\mathrm{centrals}}$} 
\end{subfigure}
\hspace{\hdistTF mm}
\begin{subfigure}{\m \columnwidth}
\includegraphics[width=\columnwidth]{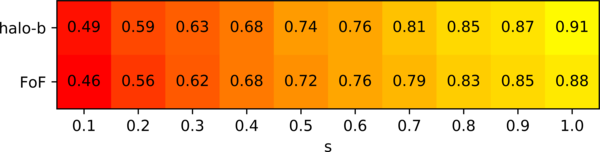}
\captionsetup{width=\columnwidth,skip = \z pt}
\caption{$^{\geq2} A_{\mathrm{centrals}}$} 
\end{subfigure}
\vspace{\vdist mm}
\newline

\centering
\begin{subfigure}{\m \columnwidth}
\includegraphics[width=\columnwidth]{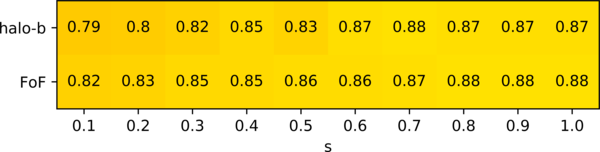}
\captionsetup{width=\columnwidth,skip = \z pt}
\caption{$A_{\mathrm{satellites}}$} 
\end{subfigure}
\hspace{\hdistTF mm}
\begin{subfigure}{\m \columnwidth}
\includegraphics[width=\columnwidth]{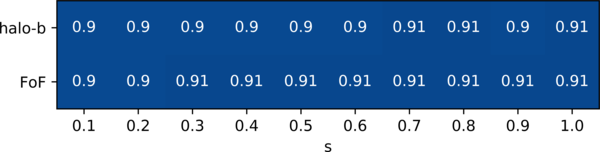}
\captionsetup{width=\columnwidth,skip = \z pt}
\caption{$f_{c_2} f_{p_2}$} 
\end{subfigure}
\vspace{\vdist mm}
\newline

\centering
\begin{subfigure}{\m \columnwidth}
\includegraphics[width=\columnwidth]{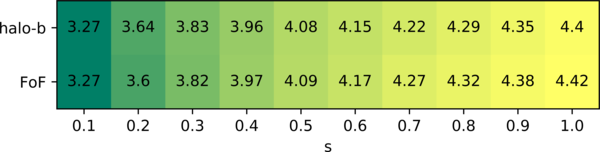}
\captionsetup{width=\columnwidth,skip = \z pt}
\caption{log$_{10}(N_{\mathrm{grs}})$} 
\end{subfigure}
\hspace{\hdistTF mm}
\begin{subfigure}{\m \columnwidth}
\includegraphics[width=\columnwidth]{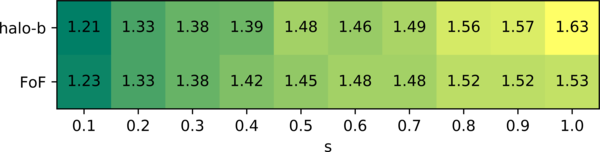}
\captionsetup{width=\columnwidth,skip = \z pt}
\caption{$m$} 
\end{subfigure}
\vspace{\vdist mm}
\newline

\caption{The science metrics designed to capture a wide range of environmental information for the "halo-based" and FoF group catalogues at low redshift in SDSS.}
\label{Tinker_vs_FoF}
\end{figure*}

\section{How the sampling rate affects the accuracy of centrals} \label{A}
We statistically analyse the effect of the sampling rate $s$ on the success in recovering centrals i.e. $A_{\mathrm{centrals}}$. Recovering centrals incorrectly, i.e. claiming that a galaxy is a central in the recovered group catalogue, even though it is a satellite in truth, can have two causes: This can happen (a) due to a failure of the group finding algorithm (fragmentation of the real group), making a central of a false group, or (b) due to the incomplete sampling of galaxies in the recovered group catalogue, the true central was not in fact observed and so a satellite was wrongly identified as the most massive group member. 

We present here a statistical analysis of case (b). For simplicity, we first consider a $50 \%$ sampling rate: When applying a $50 \%$ sampling on an underlying true group catalogue, half of the true singletons in the true group catalogue survive. Hence, obviously, there are singletons after the sampling that are truly singletons. For a group catalogue dominated by small richness structures - as it is expected to be the case - even a dominant fraction of them are. Hence, $A_{\mathrm{centrals}}$ of singletons is expected to be close to $1.0$ (in fact this is the case for wide range of sampling rates, even to quite low ones, due to the same reasoning). Let us next consider centrals of groups with at least two members. When applying a $50 \%$ sampling on a true two member group, there are four possible outcomes (which all have the same probability): (1) Both galaxies are sampled out (not observed), (2) the central or (3) the satellite is sampled out, or (4) both survive the sampling (and are observed). If small richness structures are dominant, most of the recovered two member groups are truly two member groups and hence must (case 4) have a correct detection of the central (unless there is a failure of the group-finder). Considering true groups of three members, there are in total eight possible outcomes, four of them are groups of at least two members, and in three of them the central is not sampled out. Hence, as long as there is no failure of the group-finder, the central is identified correctly three out of four times. Generalizing this to groups of richness $N$, there are $\sum_{i = 0}^{N-2} \binom{N}{i}$ possible outcomes that exhibit a group of at least two members after the sampling, out of which in $\sum_{i = 0}^{N-2} \binom{N-1}{i}$ cases the central is not sampled out. The ratio of these two sums goes to the sampling rate $0.5$ with increasing $N$, as it obviously has to. In summary, if the group catalogue is dominated by small richness structures, the success of recovering centrals  $A_{\mathrm{centrals}}$ is expected to lie well above the sampling rate of $50 \%$.

When generalizing this statistical analysis to any sampling rate $s$, we have to take into account that the possible outcomes of the sampling do not happen with the same probability any more. For any sampling rate $s$ the recovered two member groups which are truly two member groups (which is the dominant fraction down to low sampling rate) have - as long as there is no other failure of the group-finder - a correct central identification. For the true three member groups, the four possible and three successful outcomes (with a group of at least two members) have to be weighted by their relative probabilities; For example when applying a sampling of $80 \%$, in $86 \%$ of the cases the central is correctly identified. 

Generalizing this to any $N$ and $s$, the fraction of correctly recovered centrals (when the sampled group has at least two members) is

\begin{equation}
\frac{\sum_{i = 0}^{N-2} \binom{N-1}{i} s^{N-i} \left(1-s \right) ^i }{\sum_{i = 0}^{N-2} \binom{N}{i} s^{N-i} \left(1-s \right) ^i }.
\end{equation}
This fraction obviously converges to the sampling rate $s$ with increasing $N$. 
In conclusion, the accuracy of recovered centrals $A_{\mathrm{centrals}}$ (for singletons and richer structures) is expected to be always higher than the sampling rate $s$, provided (a) that the performance of the group-finder is otherwise good enough and (b) that the underlying true group catalogue is dominated by small richness structures, as it will be in any hierarchical universe such as our own.

%%%%%%%%%%%%%%%%%%%%%%%%%%%%%%%%%%%%%%%%%%%%%%%%%%

% Don't change these lines
\bsp	% typesetting comment
\label{lastpage}
\end{document}

% End of mnras_template.tex